\documentclass[12pt]{iopart}
\usepackage{setspace}
\usepackage{graphicx}
\usepackage{epsfig}
\usepackage{enumerate}
\usepackage{indentfirst}
\usepackage[figbotcap]{subfigure}

\usepackage[left=1.54cm,top=1.54cm,bottom=1.54cm,right=1.54cm]{geometry}
\renewcommand{\theequation}{\arabic{equation}}  
\date{\today}
\usepackage[usenames]{color}

\begin{document}
\bibliographystyle{unsrt}
\newcommand{\bm}[1]{\mbox{\boldmath{$#1$}}}

\title[Entangled Quantum Dynamics]{\begin{center}Emergent Time Scale in Entangled Quantum Dynamics\\ of Ultracold Molecules in Optical Lattices\end{center}}
\author{M. L. Wall and L. D. Carr}
\address{Department of Physics, Colorado School of Mines, Golden, Colorado 80401, USA}

\begin{abstract}
We derive a novel lattice Hamiltonian,  the \emph{Molecular Hubbard Hamiltonian} (MHH), which describes the essential many body physics of closed-shell ultracold heteronuclear molecules in their absolute ground state in a quasi-one-dimensional optical lattice.  The MHH is explicitly time-dependent, making a dynamic generalization of the concept of quantum phase transitions necessary.  Using the Time-Evolving Block Decimation (TEBD) algorithm to study entangled dynamics, we demonstrate that, in the case of hard core bosonic molecules at half filling, the MHH exhibits an emergent time scale over which spatial entanglement grows, crystalline order appears, and oscillations between rotational states self-damp into an asymptotic superposition.  We show that this time scale is a non-monotonic function of the physical parameters describing the lattice.  We also point out that experimental mapping of the static phase boundaries of the MHH can be used to measure the molecular polarizability tensor.
\end{abstract}

\section{Introduction}
\label{sec:introduction}

In recent years, ultracold atomic gases have provided near perfect realizations of condensed matter Hamiltonians, acting as \emph{quantum simulators}~\cite{feynmanRP1982,LewensteinMCMPAB} that allow the study of complex condensed matter phenomena in a clean and highly controllable environment.  Ultracold polar molecular gases, which have recently been brought to the edge of quantum degeneracy in their absolute ground state~\cite{niKK2008,langF2008}, offer additional features over atomic gases, such as a large internal Hilbert space and a greater susceptibility to external fields via a permanent electric dipole.  There have been a number of proposals on how to use ultracold molecular gases for mimicking well-known Hamiltonians such as spin-1 lattice models~\cite{1367-2630-9-5-138}.  Ultracold  molecules have also been suggested as a model system for the study of strongly correlated 2D quantum phases~\cite{buchler:060404} or for quantum information processing schemes~\cite{demilleD2002,goral2002,pupilloG2008}.  However, these proposals frequently involve complex and yet-to-be implemented experimental techniques.  In this article, we instead focus on the completely \emph{new} quantum many body physics which results naturally from the simplest quantum lattice experiments that can be performed in the immediate future with established techniques in ultracold molecular quantum gases.

Towards this end we derive a novel lattice Hamiltonian, which we refer to as the \emph{Molecular Hubbard Hamiltonian} (MHH).  The MHH describes the physics of an ultracold polar molecular gas in a 1D optical lattice that is oriented using a DC electric field, giving rise to a resonant dipole-dipole interaction, and is driven between rotational levels using a microwave AC field.  In particular, new aspects of our derivation include explicit dependence of hopping energy on the molecular polarizability tensor.  This in turn allows a determination of the tensor elements, an outstanding experimental issue, from the borders of the static phase diagram of the MHH, which are identical to those of the extended Bose-Hubbard Hamiltonian~\cite{kuhner2000} when a single molecular rotational level is occupied.

Beyond the statics, the MHH naturally has a dynamical component due to the AC driving fields, as well as an internal structure in terms of rotational modes which is inherently different from spinor atomic systems~\cite{dickerscheidDBM2008,higbie2005}.  We study this dynamical aspect with Time-Evolving Block Decimation (TEBD)~\cite{PhysRevLett.91.147902,PhysRevLett.93.040502}, a newly developed entangled quantum dynamics algorithm which takes spatial entanglement (specifically, Schmidt number~\cite{nielsenMA2000}) as a cut-off.  We find an emergent time scale in the case of half-filling for hard core bosonic molecules.  We emphasize that a \emph{quantum} lattice model requires low filling (average number of particles per site), in contrast to a mean field lattice model, for which the filling would typically be quite high.  Thus, although experiments can most easily access the mean field regime of hundreds of molecules per site with a single pair of counter-propagating laser beams, we look slightly ahead to the quantum regime, which will require two pairs of such beams in order to create an array of quasi-1D ``tubes.''  A third pair is then used to create the lattice in each tube.  This technique is already well established for ultracold atoms~\cite{kinoshita2006}.

Dynamical aspects of quantum phase transitions are just beginning to be considered~\cite{polkovnikov2002,polkovnikov2005b}, and have so far been a limited area of study restricted to mean field considerations, due to lack of numerical tools.  With the recent advent of entangled quantum dynamics algorithms, namely TEBD, dynamical properties of many-body systems are becoming amenable to numerical study.  For example, TEBD has been used to address key questions such as the dynamics of a quantum quench~\cite{manmanaSR2007,kollathC2007} or the speed at which correlations propagate in a lattice~\cite{laeuchliA2008}; these are not issues which can be studied with other dynamical methods such as dynamical mean field theory (DMFT)~\cite{footnoteTDFT}.  We give a brief review of TEBD in Sec. 3.  The reader interested in computational details can find them in Ref.~\cite{carr2008h}.

The first main contribution of this paper is to present a careful derivation of the Molecular Hubbard Hamiltonian.  This is done in Sec.~\ref{sec:mhh}, with some previously known aspects of molecular physics relegated to~\ref{sec:single}.  The second main contribution is to present an emergent time scale for half filling; although we treat the case of hard core bosons, the MHH can also be applied to fermionic molecules.  To this end, in Sec.~\ref{sec:methods} we first give a brief explanation of TEBD and the quantum measures we use.  Then, in Sec.~\ref{sec:results} we present and analyze our simulations, with an accompanying convergence study in~\ref{sec:convergence}.  Finally, in Sec.~\ref{sec:conclusions} we summarize.

\section{The Molecular Hubbard Hamiltonian}
\label{sec:mhh}

The Molecular Hubbard Hamiltonian (MHH) is
\begin{eqnarray}
\hat{H}&=&-\sum_{JJ'M}t_{JJ'M}\sum_{\langle i,i'\rangle}\left(\hat{a}_{i',J'M}^{\dagger}\hat{a}_{iJM}+\mbox{h.c.}\right)\nonumber\\
&&+\sum_{JM}E_{JM}\sum_i\hat{n}_{iJM}-\pi\sin\left(\omega t\right)\sum_{JM}\Omega_{JM}\sum_{i}\left(\hat{a}_{iJ,M}^{\dagger}\hat{a}_{iJ+1,M}+\mbox{h.c.}\right)\nonumber\\
&&+\frac{1}{2}\sum_{\tiny{\begin{array}{c} J_1,J_1',J_2,J_2'\\ M,M'\end{array}}}U_{dd}^{\tiny{\begin{array}{c} J_1,J_1',J_2,J_2'\\ M,M'\end{array}}}\sum_{\langle i,i'\rangle}\hat{a}_{iJ_1M}^{\dagger}\hat{a}_{iJ_1'M}\hat{a}_{i'J_2M'}^{\dagger}\hat{a}_{i'J_2'M'}.
\label{eqn:discreteHamiltonian}
\end{eqnarray}
where $\hat{a}_{iJM}$ destroys a bosonic or fermionic molecule in the $|\mathcal{E};JM\rangle$ state (defined below) on the $i^{th}$ lattice site, and the bracket notation $\langle\dots\rangle$ denotes that the sum is taken over nearest neighbors.  The first term in Eq.~(\ref{eqn:discreteHamiltonian}) corresponds to hopping both between sites and molecular rotational states with quantum numbers $J$, $M$.  The second term represents the rotational energy along with rotational state-dependent energy differences due to a DC electric field.  The third term corresponds to an AC electric field, making this a driven system.  The fourth term corresponds to electric dipole-dipole interactions.  In the following subsections and~\ref{sec:single} we justify Eq.~(\ref{eqn:discreteHamiltonian}) with a careful derivation and present the energy scales of each term.

\subsection{Derivation of the Molecular Hubbard Hamiltonian}
\label{ssec:derivation}

\noindent The full molecular Hamiltonian in second quantization is
\begin{eqnarray}
\nonumber \hat{H}&=&\int \!d^3r\, \hat{\psi}^{\dagger}\left(\mathbf{r}\right)\left[\hat{H}_{\mathrm{kin}}+\hat{H}_{\mathrm{rot}}+\hat{H}_{\mathrm{DC}}+\hat{H}_{\mathrm{AC}}\left(t\right)+\hat{H}_{\mathrm{\mathrm{opt}}}\left(\mathbf{r}\right)\right]\hat{\psi}\left(\mathbf{r}\right)\\
&&+\int \! d^3r d^3r'\, \hat{\psi}^{\dagger}\left(\mathbf{r}\right)\hat{\psi}^{\dagger}\left(\mathbf{r}'\right)\hat{H}_{\mathrm{dd}}\left(\left|\mathbf{r}-\mathbf{r}'\right|\right)\hat{\psi}\left(\mathbf{r}'\right)\hat{\psi}\left(\mathbf{r}\right).
\label{eqn:fullHamiltonian}
\end{eqnarray}
The terms on the first line correspond to single-molecule effects: kinetic energy, rotation, the DC electric field which orients the dipole, the AC microwave field which drives transitions between rotational levels, and the far off-resonant optical lattice potential, respectively.  The second line is the two-molecule resonant dipolar energy.  The field operators $\hat{\psi}$ can be either bosonic or fermionic.  We focus on the bosonic case for brevity.  There are five key assumptions underlying our derivation, as follows.  We consider all five assumptions to be reasonable for present and near-future experiments.

\begin{enumerate}
\item We consider ultracold closed-shell polar heteronuclear diatomic molecules, characterized by permanent dipole moment $d$ and rotational constant $B$.  The most experimentally relevant bosonic species in this category are SrO, RbCs, and LiCs~\cite{buchler:060404}.  The individual molecules are assumed to be in their electronic and vibrational ground states, and it is assumed that none of these degrees of freedom can be excited at the large intermolecular separations and low temperatures/relative energies that we consider.

\item The molecule is assumed to have a $^1\Sigma$ ground state.  The characteristic trapping potential length is chosen large enough compared to the internuclear axis to assume spherical symmetry, i.e. a locally constant potential.

\item We neglect any intramolecular interactions (e.g., hyperfine structure), as they are typically very small for $^1\Sigma$ molecules~\cite{RSODM}.

\item We consider only the lowest three rotational levels.  All AC fields will be sufficiently weak to allow this assumption.

\item We work in the ``hard-core" limit where at most one molecule is allowed per site.  This is enforced by strong dipole-dipole interactions on-site.  We consider the lattice spacing large enough to include only nearest-neighbor dipole-dipole interactions.  Other short-range interactions such as exchange or chemical reactions or long range interactions such as dispersion and quadrupole-quadrupole interactions are not considered.
\end{enumerate}

We proceed to follow the usual procedure~\cite{lewensteinM2007} of expanding the field operators of our second-quantized Hamiltonian in a Wannier basis of single-molecule states centered at a particular discrete position $\mathbf{r}_i$:
\begin{eqnarray}
\hat{\psi}&=\sum_{i} \hat{a}_{i}w\left(\mathbf{r}-\mathbf{r}_i\right)\,,
\end{eqnarray}
where $i$ is a site index and the sum is over all lattice sites.
For our Wannier Basis we choose the single-molecule basis that diagonalizes the rotational and DC electric field Hamiltonians, spanned by kets $|\mathcal{E};JM\rangle$.  In this basis, which we refer to as the ``dressed basis'' (the DC field ``dresses'' the rotational basis) we have the field operator expansion
\begin{eqnarray}
\label{Wannierbasis} \hat{\psi}_{JM}&=\sum_{i}\hat{a}_{iJM}w_{JM}\left(\mathbf{r}-\mathbf{r}_i\right)\equiv\sum_{i} \hat{a}_{iJM}|\mathcal{E};JM\rangle_i\,.
\end{eqnarray}
We note that such a basis, while highly efficient for the hard core limit we consider, becomes progressively worse for higher filling factors, till in the mean field limit the single-molecule basis, whether dressed or not, is so poor that many bands must be considered.  Here we do not include a band index for simplicity, although the generalization of Eq.~(\ref{eqn:discreteHamiltonian}) to include multiple bands is straightforward.

This choice of Wannier basis associates the terms in Eq.~(\ref{eqn:discreteHamiltonian}) to the terms in Eq.~(\ref{eqn:fullHamiltonian}) as follows:
\begin{eqnarray}
\fl\label{TunnelingEnergies} t_{J,J',M}&\equiv-\int \! d\mathbf{r}\,w^{\star}_{JM}\left(\mathbf{r}-\mathbf{r}_{i}\right)\left[H_{\mathrm{kin}}
+H_{\mathrm{\mathrm{opt}}}\right]w_{J'M}\left(\mathbf{r}-\mathbf{r}_{i+1}\right),\\
\fl E_{JM}&\equiv \int \! d\mathbf{r}\,w^{\star}_{JM}\left(\mathbf{r}-\mathbf{r}_{i}\right)
\left[H_{\mathrm{rot}}+H_{\mathrm{DC}}\right]w_{JM}\left(\mathbf{r}-\mathbf{r}_{i}\right),\\
\fl-\pi\Omega_{JM}\sin\left(\omega t\right)&\equiv \int \! d\mathbf{r}\,w^{\star}_{JM}\left(\mathbf{r}-\mathbf{r}_{i}\right)\left[H_{\mathrm{AC}}\right]
w_{J+1,M}\left(\mathbf{r}-\mathbf{r}_{i}\right),\\
\fl U_{dd}^{\tiny{\begin{array}{c} J_1,J_1',J_2,J_2'\\ M,M'\end{array}}}&\equiv\int d\mathbf{r}d\mathbf{r}'\, w_{J_1'M}^{\star}\left(\mathbf{r}-\mathbf{r}_i\right)w_{J_2'M'}^{\star}\left(\mathbf{r}'-\mathbf{r}_{i+1}\right)H_{dd}\left(\mathbf{r}-\mathbf{r}'\right)w_{J_1M}\left(\mathbf{r}-\mathbf{r}_i\right)w_{J_2M'}\left(\mathbf{r}'-\mathbf{r}_{i+1}\right)\,,
\end{eqnarray}
where the operators $H_{\mathrm{kin}}$, $H_{\mathrm{opt}}$, etc., are taken to be in position space representation.
For the derivation of the single-molecule terms (rotational, DC electric field, and AC electric field) and discussion of the properties of our Wannier basis we refer the reader to ~\ref{sec:single}.  In the following sections we present the derivation of the tunneling (hopping) and dipole-dipole terms, which have new aspects not heretofore appearing in the literature~\cite{micheliA2007}.

\subsection{Tunneling}
\label{ssec:tunneling}

The tunneling term represents the sum of the molecular kinetic energy with the potential energy of the lattice.  After expanding in the Wannier basis of Eq.~(\ref{Wannierbasis}), we find the effective tunneling Hamiltonian
\begin{eqnarray}
\label{DressedTunneling}  \hat{H}_t^{\mathrm{eff}}&=-\sum_{J,J',M}t_{JJ'M}\sum_{\langle i,i'\rangle} \left(\hat{a}^{\dagger}_{i,J'M}\hat{a}_{i',JM}+\mbox{h.c.}\right)
\end{eqnarray}
where $t_{J,J',M}$ was defined in Eq.~(\ref{TunnelingEnergies}).  To understand why this operator mixes states of different $J$, we note that the kinetic energy and (far off-resonant) optical lattice potential do not mix rotational eigenstates.  Because our Wannier basis states are dressed and therefore superpositions of rotational eigenstates with different $J$,  the tunneling operator in the dressed basis will mix $J$.  Although the dressed basis makes the tunneling more complex to analyze, it simplifies other terms in the MHH, such as the DC term, and is in any case a more standard basis for analysis of the diatomic molecules we study here.  Comparable basis changes are sometimes made in other quantum many body systems, where, for instance, particles and holes are mixed, or particles are paired.  Note that, because we assume $z$-polarized fields, $M$ is still a good quantum number.  To discuss the actual form of the tunneling energies $\left\{t_{J,J',M}\right\}$ we must first examine the interaction of a diatomic molecule with the optical lattice.

\subsection{Interaction with an Optical Lattice}
\label{ssec:lattice}

The charge redistribution that occurs when a molecule is subjected to a static, spatially uniform electric field $\mathbf{E}$ is reflected in its dipole moment $\mathbf{d}$ via the polarizability series
\begin{eqnarray}
d_{j}&=d_{j}^{\left(0\right)}+\alpha_{jk}E_{k}+\frac{1}{2!}\beta_{jkl}E_{k}E_{l}
+\frac{1}{3!}\Gamma_{jklm}E_{k}E_{l}E_{m}+\ldots
\label{eqn:tensorSeries}
\end{eqnarray}
where the first, second, and third order coefficients $\alpha_{jk}$, $\beta_{jkl}$, and $\Gamma_{jklm}$  are elements of the polarizability, hyperpolarizability, and second hyperpolarizability tensors, respectively.  The polarizability tensor is a symmetric rank-two tensor with no more than six independent elements (less if molecular symmetry is greater), and characterizes the lowest order dipole moment induced by an applied electric field.  From this tensor we can form the scalar invariants
\begin{eqnarray}
\bar{\alpha}&\equiv \frac{1}{3}\mbox{Tr} \tilde{\alpha}\,,\\
\left(\Delta \alpha\right)^2&\equiv\frac{1}{2}\left[3\mbox{Tr} (\tilde{\alpha}^2)-\left(\mbox{Tr} \tilde{\alpha}\right)^2\right]\,,
\end{eqnarray}
referred to as the polarizability and the polarizability anisotropy, respectively.  Note that we use the tilde to clarify that $\tilde{\alpha}$ with elements $\alpha_{jk}$ is a tensor, not a scalar -- we reserve the accent circumflex (the ``hat'' symbol) for quantum operators. In linear molecules, such as diatomic molecules, the presence of only two distinct moments of inertia allows for the classification of $\tilde{\alpha}$ according to its components along and perpendicular to the internuclear axis, denoted $\alpha_{\parallel}$ and $\alpha_{\perp}$, respectively.  In the presence of AC electric fields with frequency $\omega$ we speak of the dynamic polarizability tensor $\tilde{\alpha}\left(\omega\right)$, with the series of Eq.~(\ref{eqn:tensorSeries}) being the zero frequency limit.  The tensor $\tilde{\alpha}\left(\omega\right)$ is, in general, complex, with the real part inducing a dipole moment and the imaginary part accounting for power absorption by the dipole and out-of-phase dipole oscillation.  In the case of $\Sigma$ diatomic molecules in their electronic and vibrational ground states~\cite{micheli:043604}
\begin{eqnarray}
\tilde{\alpha}\left(\omega\right)&\equiv\alpha_{\parallel}\left(\omega\right)\mathbf{e}_0'\otimes \mathbf{e}_0'+\alpha_{\perp}\left(\omega\right)\sum_{\Lambda=\pm 1}\left(-1\right)^{\Lambda}\mathbf{e}_{\Lambda}'\otimes \mathbf{e}_{-\Lambda}'\,,
\end{eqnarray}
where the $\mathbf{e}_q'$ are molecule-fixed spherical basis vectors.  The parallel and perpendicular dynamic polarizabilities are
\begin{eqnarray}
\alpha_{\parallel}&=\sum_{\pm}\sum_{\nu, v}\frac{\left|d_{\nu\Sigma\left(v\right)-X\Sigma\left(0\right)}\right|^2}{E_{\nu\Sigma\left(v\right)}-E_{X\Sigma\left(0\right)}\mp \hbar \omega}\,,\\
\alpha_{\perp}&=\sum_{\pm}\sum_{\nu, v}\frac{\left|d_{\nu\Pi\left(v\right)-X\Sigma\left(0\right)}\right|^2}{E_{\nu\Pi\left(v\right)}-E_{X\Sigma\left(0\right)}\mp \hbar \omega}\,,
\end{eqnarray}
respectively.  In these expressions $d_{\nu\Lambda\left(v\right)-X\Sigma\left(0\right)}$ is the transition dipole moment from the ground state to the $\nu\Lambda\left(v\right)$ state (following the usual diatomic molecular notation, $\Lambda\in\{\Sigma,\Pi\}\equiv\{0,1\}$ is the quantum number associated with the projection of the total electronic orbital angular
momentum along the internuclear axis, i.e., in the molecule-fixed basis) and the sum over $\mp$ accounts for the near-resonant and typically far off-resonant terms.

Transforming $\tilde{\alpha}$ from the molecule-fixed basis to the space-fixed basis using the transformation discussed in~\ref{sec:single}, we find
\begin{eqnarray}
\nonumber\tilde{\alpha}'\left(\omega_L\right)&=\sum_{p_1p_2}\sum_{j=0,2}\sum_{m=-j}^{j}\left(2j+1\right)\left(\begin{array}{ccc} 1&1&j\\ p_1&p_2&m\end{array}\right)\sqrt{\frac{1}{\left(2-j\right)!\left(3+j\right)!}}\\
\label{polarizability}&\times\left[\alpha_{\parallel}\left(j+2\right)\left(j-1\right)-4\alpha_{\perp}\right]
C^{\left(j\right)}_{m}\mathbf{e}_{p_1}\otimes\mathbf{e}_{p_2}\,,
\end{eqnarray}
where $C_{m}^{\left(j\right)}$ is an unnormalized spherical harmonic, $\left(\dots\right)$ denotes the Wigner 3-$j$ coefficient~\cite{ZareNote}, and the $\mathbf{e}_p$ are space-fixed spherical basis vectors.

The interaction of the lattice with the molecule is represented by the Hamiltonian
\begin{eqnarray}
H_{\mathrm{\mathrm{opt}}}\left(\mathbf{x}\right)
=-\mathbf{E}_{\mathrm{opt}}^{\star}\left(\mathbf{r}\right)\cdot \tilde{\alpha}'\left(\omega_L\right)\cdot\mathbf{E}_{\mathrm{opt}}\left(\mathbf{r}\right)\,.
\end{eqnarray}
If the electric field has polarization $p$  in the space-fixed spherical basis then we find
\begin{eqnarray}
H_{\mathrm{\mathrm{opt}}}\left(\mathbf{x}\right)&=-\frac{\left|\mathbf{E}_{\mathrm{\mathrm{opt}}}\left(\mathbf{r}\right)\right|^2}{3}\left[\left(\alpha_{\parallel}+2\alpha_{\perp}\right)C_{0}^{\left(0\right)}+\left(-1\right)^{p}\frac{2}{\left(1-p\right)!\left(1+p\right)!}
\left(\alpha_{\parallel}-\alpha_{\perp}\right)C_0^{\left(2\right)}\right]\,.
\end{eqnarray}
For light linearly polarized in the $\hat{x}$-direction we obtain
\begin{eqnarray}
 H_{\mathrm{\mathrm{opt}}}&=-\frac{\left|\mathbf{E}_{\mathrm{\mathrm{opt}}}\left(\mathbf{r}\right)\right|^2}{6}\left[2\left(\alpha_{\parallel}+2\alpha_{\perp}\right)C_0^{\left(0\right)}+\left(\alpha_{\parallel}-\alpha_{\perp}\right)
 \left(\sqrt{6}C_{-2}^{\left(2\right)}-2C_{0}^{\left(2\right)}+\sqrt{6}C_2^{\left(2\right)}\right)\right]\,,
\end{eqnarray}
whereas for light linearly polarized in the $\hat{y}$-direction we find
\begin{eqnarray}
 H_{\mathrm{\mathrm{opt}}}&=\frac{\left|\mathbf{E}_{\mathrm{\mathrm{opt}}}
 \left(\mathbf{r}\right)\right|^2}{6}\left[-2\left(\alpha_{\parallel}+2\alpha_{\perp}\right)
 C_0^{\left(0\right)}+\left(\alpha_{\parallel}-\alpha_{\perp}\right)\left(\sqrt{6}C_{-2}^{\left(2\right)}
 +2C_{0}^{\left(2\right)}+\sqrt{6}C_2^{\left(2\right)}\right)\right]\,.
\end{eqnarray}
Since $C_0^{\left(0\right)}=1$, these terms give a state-independent energy shift.  The $C_q^{\left(2\right)}$ terms produce a tensor shift. Because the depth (in energy) of a typical optical lattice is much smaller than the energy of transitions between rotational levels (of order $B$, as defined in~\ref{ssec:rot}), we can ignore far off-resonant Raman coupling between different $J$ manifolds and use only the diagonal matrix elements. The $C_2^{\left(2\right)}$ term and the $C_{-2}^{\left(2\right)}$ will both mix $M$ in the $J\ge 2$ manifolds, but do not affect the lowest two rotational levels, again, because we neglect Raman couplings.  Thus $x$, $y$, and $z$ polarizations all have the same Hamiltonian in this approximation.  We can calculate the matrix elements of $C_0^{\left(2\right)}$ in the field free basis using the Wigner-Eckart theorem to find
\begin{eqnarray}
\nonumber \langle J'M'|H_{\mathrm{opt}}\left(\mathbf{r}\right)|JM\rangle&=
-\frac{\left|\mathbf{E}_{\mathrm{opt}}\left(\mathbf{r}\right)\right|^2}{3}
\Big[\left(\alpha_{\parallel}+2\alpha_{\perp}\right)\\
&+\left(-1\right)^{p}\frac{2}{\left(1-p\right)
!\left(1+p\right)!}\left(\alpha_{\parallel}-\alpha_{\perp}\right)
\frac{J\left(J+1\right)-3M^2}{\left(2J-1\right)\left(2J+3\right)}\Big]\delta_{JJ'}\delta_{MM'}.
\end{eqnarray}
In our effective Hamiltonian we choose right circular polarization for the $z$ lattice, $x$ polarization for the $x$ lattice, and $y$ polarization for the $y$ lattice, where each ``lattice'' refers to a pair of counter-propagating laser beams used to create a standing wave.

We consider the fields making up the optical lattice to have sinusoidal spatial profiles, resulting in sine-squared intensity profiles.  In addition, we assume that the $y$ and $z$ lattices are tight, meaning that the molecules are strongly confined at the potential minimum (for a red-detuned trap).  This tight confinement allows us to approximate them via a Taylor series, e.g., $\sin^2\left(k_zz\right)\simeq k_z^2z^2$ in the vicinity of the molecule.  Using the above results, the matrix elements of the Hamiltonian for the optical lattice can be written
\begin{eqnarray}
\fl \nonumber \langle J'M'|H_{\mathrm{\mathrm{opt}}}\left(\mathbf{r}\right)|JM\rangle&=&-\frac{\left|\mathbf{E}_{\mathrm{opt}}\left(\mathbf{y}\right)\right|^2k_y^2y^2+\left|\mathbf{E}_{\mathrm{opt}}\left(\mathbf{x}\right)\right|^2\sin^2\left(k_xx\right)}{3}\left[\bar{\alpha}+2\Delta\alpha\frac{J\left(J+1\right)-3M^2}{\left(2J-1\right)\left(2J+3\right)}\right]\delta_{JJ'}\delta_{MM'}\\
&&-\frac{\left|\mathbf{E}_{\mathrm{opt}}\left(\mathbf{z}\right)\right|^2k_z^2z^2}{3}\left[\bar{\alpha}-\Delta\alpha\frac{J\left(J+1\right)-3M^2}{\left(2J-1\right)\left(2J+3\right)}\right]\delta_{JJ'}\delta_{MM'}
\end{eqnarray}
or, more compactly, as
\begin{eqnarray}
\fl \nonumber \langle J'M'|H_{\mathrm{\mathrm{opt}}}\left(\mathbf{r}\right)|JM\rangle&=&\left[-\alpha_{JM}^{\left(t\right)}\left|\mathbf{E}_{\mathrm{opt}}\left(\mathbf{y}\right)\right|^2k_y^2y^2-\alpha_{JM}^{\left(t\right)}\left|\mathbf{E}_{\mathrm{opt}}\left(\mathbf{x}\right)\right|^2\sin^2\left(k_xx\right)\right]\delta_{JJ'}\delta_{MM'}\\
&&-\left|\mathbf{E}_{\mathrm{opt}}\left(\mathbf{z}\right)\right|^2 \alpha_{JM}^{\left(z\right)} k_z^2z^2\delta_{JJ'}\delta_{MM'}
\end{eqnarray}
by defining
\begin{eqnarray}
\label{atazdef}\alpha_{JM}^{\left(t\right)}&\equiv&\frac{1}{3}\left[\bar{\alpha}+2\Delta\alpha\frac{J\left(J+1\right)-3M^2}{\left(2J-1\right)\left(2J+3\right)}\right],\\ \label{atazdef2}\alpha_{JM}^{\left(z\right)}&\equiv&\frac{1}{3}\left[\bar{\alpha}-\Delta\alpha\frac{J\left(J+1\right)-3M^2}{\left(2J-1\right)\left(2J+3\right)}\right]\,.
\end{eqnarray}
We now define, as is customary, the ``lattice heights" in the $x$, $y$, and $z$ directions, respectively, as
\begin{eqnarray}
V_{x}^{\left(JM\right)}&\equiv&-\left|\mathbf{E}_{\mathrm{opt}}\left(\mathbf{x}\right)\right|^2\alpha_{JM}^{\left(t\right)},\\
V_{y}^{\left(JM\right)}&\equiv&-\left|\mathbf{E}_{\mathrm{opt}}\left(\mathbf{y}\right)\right|^2\alpha_{JM}^{\left(t\right)},\\
V_{z}^{\left(JM\right)}&\equiv&-\left|\mathbf{E}_{\mathrm{opt}}\left(\mathbf{z}\right)\right|^2\alpha_{JM}^{\left(z\right)}\,.
\end{eqnarray}
The tight confinement in the transverse ($y$ and $z$) directions strongly suppresses tunneling in these directions, making the overall lattice effectively 1D along x.

From Eqs.~(\ref{atazdef})-(\ref{atazdef2}), it is apparent that different rotational levels experience different trapping frequencies and different tunneling energies.  To make this clearer, we parse our full field-free tunneling matrix element as
\begin{eqnarray}
\nonumber &&t_{JM}\equiv-\int d\mathbf{r}\,w^{\star}_{JM}\left(\mathbf{r}-\mathbf{r}_{i}\right)\left[H_{\mathrm{kin}}+H_{\mathrm{\mathrm{opt}}}\right]w_{JM}\left(\mathbf{r}-\mathbf{r}_{i+1}\right)\\
\nonumber &=&\int d\mathbf{r}\, w_{JM}^{\star}\left(\mathbf{r}-\mathbf{r}_i\right)\left[-H_{\mathrm{kin}}+V_x^{\left(JM\right)}\sin^2\left(k_xx^2\right)\right]w_{JM}\left(\mathbf{r}-\mathbf{r}_{i+1}\right)\\
\nonumber &&+\int d\mathbf{r}\, w_{JM}^{\star}\left(\mathbf{r}-\mathbf{r}_i\right)\left[V_y^{\left(JM\right)}k_y^2y^2+V_z^{\left(JM\right)}k_z^2z^2\right]w_{JM}\left(\mathbf{r}-\mathbf{r}_{i+1}\right)\,.
\end{eqnarray}
Defining
\begin{eqnarray}
\label{eqn:hoppingnaught}t_{JM}^{\left(0\right)}&\equiv&\int d\mathbf{r}\, w_{JM}^{\star}\left(\mathbf{r}-\mathbf{r}_i\right)\left[-H_{\mathrm{kin}}+V_x^{\left(JM\right)}\sin^2\left(k_xx^2\right)\right]w_{JM}\left(\mathbf{r}-\mathbf{r}_{i+1}\right)\,,\\
\label{eqn:hoppingQuad}t_{JM}^{\left(\mathrm{trans}\right)}&\equiv&\int d\mathbf{r}\, w_{JM}^{\star}\left(\mathbf{r}-\mathbf{r}_i\right)\left[V_y^{\left(JM\right)}k_y^2y^2+V_z^{\left(JM\right)}k_z^2z^2\right]w_{JM}\left(\mathbf{r}-\mathbf{r}_{i+1}\right)\,,
\end{eqnarray}
we proceed to compute each piece separately.

In the evaluation of the first integral, Eq.~(\ref{eqn:hoppingnaught}) we assume that the Bloch function of a molecule in the sinusoidal optical lattice is a Mathieu function along $x$.  This may seem to contradict our assumption of spherical symmetry in the above derivation.  However, the assumption of spherical symmetry (i.e. a locally constant potential) need only hold on the order of an internuclear axis ($\sim 5\AA$) near the molecule.  In contrast, on the order of the characteristic lattice length $\sqrt{\hbar/\mu\omega_{\mathrm{\mathrm{opt}}}}$ the rigid-rotor molecule is indistinguishable from a point particle (such as an alkali atom), and so spherical symmetry is not required.  With this understanding, we recognize $t^{\left(0\right)}_{JM}$ as the expression for the hopping energy for point particles in optical lattices~\cite{rey-thesis} with the additional feature that the lattice height along the quasi-1D direction $V_0=V_x^{\left(JM\right)}$  is dependent on $J$ through the polarizability tensor.  Thus, altering the expression from the theory of point particles in optical lattices, we obtain the result
\begin{eqnarray}
\label{eqn:JayTunn}\frac{t_{JM}^{\left(0\right)}}{E_R}&\approx& A\left(\frac{V_x^{\left(JM\right)}}{E_R}\right)^{B}\exp\left(-C\sqrt{\frac{V_x^{\left(JM\right)}}{E_R}}\right)\,,
\end{eqnarray}
where $A=1.397$, $B=1.051$, $C=2.121$, and \begin{equation}E_R\equiv\hbar^2 k_x^2/2m\end{equation} is the recoil energy.

\begin{table}[t]
\begin{center}
\begin{tabular}{|c|c|c|}
\hline $|JM\rangle$&$3\alpha_{JM}^{\left(t\right)}/\bar{\alpha}$&$3\alpha_{JM}^{\left(z\right)}/\bar{\alpha}$\\
\hline \hline $|00\rangle$&1&1\\
\hline $|10\rangle$&1.715&0.642\\
\hline $|1\pm 1\rangle$&0.642&1.178\\
\hline $|20\rangle$&1.511&0.744\\
\hline $|2\pm 1\rangle$&1.255&0.872\\
\hline $|2\pm 2\rangle$&0.488&1.255\\
\hline
\end{tabular}
  \caption{Values of the polarizabilities for LiCs in different rotational states $|JM\rangle$.}
  \label{table:1}
\end{center}
\end{table}
For the second integral, Eq.~(\ref{eqn:hoppingQuad}), we approximate the Wannier functions with the ground state of a simple harmonic oscillator
\begin{eqnarray}
w\left(y\right)&\approx \left(l_{\mathrm{ho,y}}^{\left(JM\right)}\right)^{-1/2}\pi^{-1/4}\exp\left(-y^2/2\left(l_{\mathrm{ho,y}}^{\left(JM\right)}\right)^2\right)\,,\\
w\left(z\right)&\approx \left(l_{\mathrm{ho,z}}^{\left(JM\right)}\right)^{-1/2}\pi^{-1/4}\exp\left(-z^2/2\left(l_{\mathrm{ho,z}}^{\left(JM\right)}\right)^2\right)\,,
\end{eqnarray}
where the harmonic oscillator lengths are given by
\begin{eqnarray}
&\left(l_{\mathrm{ho,y}}^{\left(JM\right)}\right)^2\equiv \frac{\hbar^2}{2m\sqrt{V_y^{\left(JM\right)}E_R}}\,,\;\;\;\;\;\left(l_{\mathrm{ho,z}}^{\left(JM\right)}\right)^2\equiv \frac{\hbar^2}{2m\sqrt{V_z^{\left(JM\right)}E_R}}\,.
\end{eqnarray}
Then
\begin{eqnarray}
t_{JM}^{\left(\mathrm{trans}\right)}&\propto \exp\left[-\frac{\lambda^2}{4\left(l_{\mathrm{ho,y}}^{\left(JM\right)}\right)^2}\right]+\frac{\alpha_{JM}^{\left(z\right)}}{\alpha_{JM}^{\left(t\right)}}\exp\left[-\frac{\lambda^2}{4\left(l_{\mathrm{ho,z}}^{\left(JM\right)}\right)^2}\right]\,,
\end{eqnarray}
where  $\lambda$ is the wavelength of the optical lattice.  Because we consider tight traps such that the lattice height in the $y$ and $z$ directions is much greater than the lattice height in the $x$ direction, $V_y\sim V_z\gg V_x$, this contribution is exponentially suppressed compared to $t_{JM}^{\left(0\right)}$, and so we neglect it.  Thus,
\begin{eqnarray}
\frac{t_{JM}}{E_R}\approx\frac{t_{JM}^{\left(0\right)}}{E_R}&\approx& A\left(\frac{\left|\mathbf{E}_{\mathrm{\mathrm{opt}}}\right|^2\alpha_{JM}^{\left(t\right)}}{E_R}\right)^{B}\exp\left(-C\sqrt{\frac{\left|\mathbf{E}_{\mathrm{\mathrm{opt}}}\right|^2\alpha_{JM}^{\left(t\right)}}{E_R}}\right)\,.
\end{eqnarray}
This is equivalent to the array of tubes we discussed in Sec.~\ref{sec:introduction}, where each tube is isolated from its neighbors.

\begin{figure}[t]
 \begin{center}
  \epsfig{figure=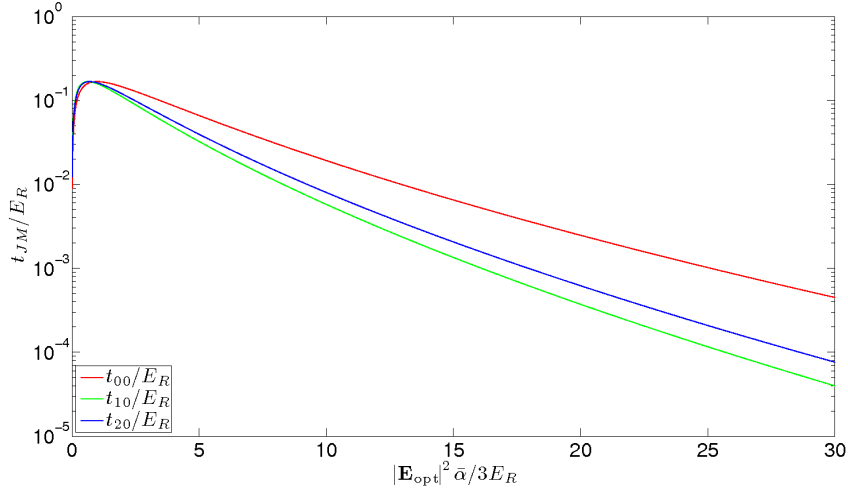, scale=0.3}
\caption{Dependence of the field-free tunneling (hopping) coefficient on rotational state and lattice height.}
\label{tunnelingmath}
 \end{center}
\end{figure}

Using tabulated values of the polarizabilities for LiCs\cite{deiglmayr:064309} as given in Table~\ref{table:1}, we find that, for a reasonable lattice height $V_x^{\left(00\right)}/E_R\simeq 10$, the tunneling term for the $|11\rangle$ state is only about $20\%$ of that in the $|00\rangle$ state, as shown in Fig.~\ref{tunnelingmath}.
For LiCs in a red-detuned optical lattice of wavelength $\lambda=985$nm,  $E_R=2\pi\times 1.46\hbar $ kHZ.  Typical values of the lattice heights are $V_{x}\sim10E_{R}$, $V_{y},V_{z}\sim 25E_{R}$~\cite{PhysRevLett.81.3108}.

We reiterate that the above matrix elements and tunneling energies $\left\{t_{JM}\right\}$ have been computed in the field-free basis for simplicity.  To transform to the dressed basis, we use the unitary matrix with dressed eigenvectors as columns, recovering Eq.~(\ref{DressedTunneling}), where the tunneling matrix element is no longer diagonal in $J$.

\subsection{Dipole-Dipole Interactions}
\label{ssec:dipole}

The induced dipoles from the DC field give rise to a resonant dipole-dipole interaction.  The  Hamiltonian for this interaction in the two-site dressed basis spanned by $|\mathcal{E};J_1M_1J_2M_2\rangle$ is
\begin{eqnarray}
\hat{H}_{\mathrm{dd}}&=\frac{1}{2}\sum_{\tiny{\begin{array}{c} J_1,J_1',J_2,J_2'\\ M,M'\end{array}}}U_{dd}^{\tiny{\begin{array}{c} J_1,J_1',J_2,J_2'\\ M,M'\end{array}}}\sum_{\langle i,i'\rangle}\hat{a}_{iJ_1M}^{\dagger}\hat{a}_{iJ_1'M}\hat{a}_{i'J_2M'}^{\dagger}\hat{a}_{i'J_2'M'}\,,
\end{eqnarray}
where we have defined
\begin{eqnarray}
\fl U_{dd}^{\tiny{\begin{array}{c} J_1,J_1',J_2,J_2'\\ M,M'\end{array}}}&\equiv\int d\mathbf{r}d\mathbf{r}'\, w_{J_1'M}^{\star}\left(\mathbf{r}-\mathbf{r}_i\right)w_{J_2'M'}^{\star}
\left(\mathbf{r}'-\mathbf{r}_{i+1}\right)H_{dd}\left(\mathbf{r}-\mathbf{r}'\right)
w_{J_1M}\left(\mathbf{r}-\mathbf{r}_i\right)w_{J_2M'}\left(\mathbf{r}'-\mathbf{r}_{i+1}\right)\,,
\end{eqnarray}
and for notational simplicity we have suppressed the $\mathcal{E}$ subscripts.  Note that because of our choice of polarizations of the optical lattice and AC and DC electric fields, $M_1=M_2\equiv M$ and $M_1'=M_2'\equiv M'$.

The resonant dipole-dipole interaction between two permanent dipoles $d_1$ and $d_2$ whose respective centers of mass are separated by a vector $\mathbf{R}$ in the space-fixed frame is
\begin{eqnarray}
\label{dipole}\hat{H}_{dd}&=\frac{{\mathbf{\hat{d}_1}\cdot \mathbf{\hat{d}_2}}-3\left(\mathbf{e}_R\cdot \mathbf{\hat{d}_1}\right)\left(\mathbf{\hat{d}_2}\cdot\mathbf{e}_R\right)}{R^3}\, ,
\end{eqnarray}
where $\mathbf{e}_R$ is a unit vector in the direction of $\mathbf{R}$.  Using standard angular momentum recoupling we recast this in spherical tensor notation as
\begin{eqnarray}
 \label{dipnosph}H_{dd}&=&-\frac{\sqrt{6}}{R^3}\sum_{\mu}\left(-1\right)^{\mu}C^{\left(2\right)}_{-\mu}\left(\mathbf{R}\right)\left[\hat{\mathbf{d}}_1\otimes \hat{\mathbf{d}}_2\right]_{\mu}^{\left(2\right)}\, ,
\end{eqnarray}
where $\left(T\right)_q^{\left(k\right)}$ denotes the component of the rank-$k$ spherical tensor $T$ that has projection $q$ along $\mathbf{R}$, $C_{m}^{\left(j\right)}\left(\mathbf{R}\right)$ is an unnormalized spherical harmonic in the polar coordinates defined with respect to $\mathbf{R}$, and we have defined the tensor product of the vector operators $\hat{\mathbf{d}}_1$ and $\hat{\mathbf{d}}_2$ as
\begin{eqnarray}
 \left[\hat{\mathbf{d}}_1\otimes \hat{\mathbf{d}}_2\right]_{q}^{\left(k\right)}\equiv \sum_{m}\langle 1,m,1,q-m|kq\rangle \left(\hat{\mathbf{d}}_1\right)^{\left(1\right)}_m\left(\hat{\mathbf{d}}_2\right)^{\left(1\right)}_{q-m}\, .
\end{eqnarray}
In the last line, $\langle j_1,m_1,j_2,m_2|J,M\rangle$ is a Clebsch-Gordan coefficient.  We now take matrix elements of Eq.~(\ref{dipnosph}) in the two dressed-molecule basis $|\mathcal{E};J_1M_1,J_2M_2\rangle$, where molecule $1$ is on site $i$ and molecule $2$ is on site $i+1$, yielding
\begin{eqnarray}
 \nonumber&\langle\mathcal{E};J_1'M_1',J_2'M_2'|\hat{H}_{dd}|\mathcal{E};J_1M_1,J_2M_2\rangle=-\frac{\sqrt{6}}{R^3}\sum_{\mu}\left(-1\right)^{\mu}C^{\left(2\right)}_{-\mu}\left(\mathbf{R}\right)\\
&\times \sum_{m}\langle 1,m,1,\mu-m|2\mu\rangle \langle \mathcal{E};J_1'M_1'|\left(\hat{\mathbf{d}}_1\right)^{\left(1\right)}_m|\mathcal{E};J_1M_1\rangle\langle\mathcal{E};J_2'M_2'|\left(\hat{\mathbf{d}}_2\right)^{\left(1\right)}_{\mu-m}|\mathcal{E};J_2M_2\rangle\,.
\end{eqnarray}
Because our DC field is polarized along $z$, only $(\hat{\mathbf{d}}_1)^{(1)}_0$ and $(\hat{\mathbf{d}}_2)^{(1)}_0$ matrix elements are nonzero, enforcing $\mu=0$, $m=0$.  With this in mind, the interaction takes the particularly simple form
\begin{eqnarray}
 \nonumber&\langle\mathcal{E};J_1'M_1',J_2'M_2'|\hat{H}_{dd}|\mathcal{E};J_1M_1,J_2M_2\rangle=-\frac{\sqrt{6}}{R^3}C^{\left(2\right)}_{0}\left(\mathbf{R}\right)\\
&\times \langle 1,0,1,0|20\rangle \langle \mathcal{E};J_1'M_1|\left(\hat{\mathbf{d}}_1\right)^{\left(1\right)}_0|\mathcal{E};J_1M_1\rangle\langle\mathcal{E};J_2'M_2|\left(\hat{\mathbf{d}}_2\right)^{\left(1\right)}_{0}|\mathcal{E};J_2M_2\rangle\\
 \label{DipDip}
&=\langle \mathcal{E};J_1'M_1|\left(\hat{\mathbf{d}}_1\right)^{\left(1\right)}_0|\mathcal{E};J_1M_1\rangle\langle\mathcal{E};J_2'M_2|\left(\hat{\mathbf{d}}_2\right)^{\left(1\right)}_{0}|\mathcal{E};J_2M_2\rangle
\left(\frac{1-3\cos^2\theta}{R^3}\right)\,.
\end{eqnarray}
The intermolecular axis plays a crucial role in the sign of the interaction.  Two molecules oriented along the intermolecular axis attract if their dipoles are parallel and repel if their dipoles are antiparallel.  Two molecules oriented perpendicular to the intermolecular axis, on the other hand, repel if their dipoles are parallel and attract if their dipoles are antiparallel.  The DC field that orients the molecules in our setup is polarized along $z$, perpendicular to the intermolecular quasi-1D axis $x$. This gives rise to repulsive interactions for positive dipole matrix elements.  With this geometry the dipole potential becomes
\begin{eqnarray}
\langle\mathcal{E};J_1'M_1',J_2'M_2'|\hat{H}_{dd}|\mathcal{E};J_1M_1,J_2M_2\rangle=\nonumber\\
 \frac{1}{R^3}\langle \mathcal{E};J_1'M_1|\left(\hat{\mathbf{d}}_1\right)^{\left(1\right)}_0|\mathcal{E};J_1M_1\rangle\langle\mathcal{E};J_2'M_2|\left(\hat{\mathbf{d}}_2\right)^{\left(1\right)}_{0}|\mathcal{E};J_2M_2\rangle\,,
\end{eqnarray}
yielding
\begin{eqnarray}
U_{dd}^{\tiny{\begin{array}{c} J_1,J_1',J_2,J_2'\\ M,M'\end{array}}}&=&\frac{8}{\lambda^3}\langle \mathcal{E};J_1'M_1|\left(\hat{\mathbf{d}}_1\right)^{\left(1\right)}_0|\mathcal{E};J_1M_1\rangle\langle\mathcal{E};J_2'M_2|\left(\hat{\mathbf{d}}_2\right)^{\left(1\right)}_{0}|\mathcal{E};J_2M_2\rangle\,,
\end{eqnarray}
where $\lambda$ is the wavelength of the optical lattice.

\subsection{Energy Scales}
\label{ssec:scales}

We proceed to clarify the energy scales associated with each term in Eq.~(\ref{eqn:discreteHamiltonian}).  Between previous discussion in Sec.~\ref{sec:mhh} and that of~\ref{sec:single}, all terms in Eq.~(\ref{eqn:discreteHamiltonian}) are now clearly defined.
The energy scales of the dressed basis are $B$, the rotational constant, which is roughly $60\hbar$ GHz, and $d\mathcal{E}_{\mathrm{DC}}$, which is of order $1-10 B$.  The DC term has no length scale associated with it because the field is uniform, and the length scale of the rotational term is the internuclear separation, on the order of angstroms.  The relative contribution of the DC electric field and rotational terms in Eq.~(\ref{eqn:discreteHamiltonian}) are expressed through the dimensionless parameter
\begin{eqnarray}
\beta_{\mathrm{DC}}\equiv d\mathcal{E}_{\mathrm{DC}}/B,
\end{eqnarray}
the ratio of the DC field energy to the rotational level splitting.

The energy scales of the AC term are $\hbar\omega$, where $\omega$ is the angular frequency of the driving field, and $d\mathcal{E}_{\mathrm{AC}}$.  The scale $\hbar\omega$ is of order $2B$ for small $\beta_{\mathrm{DC}} \ll 1$, and of order $B\sqrt{\beta_{\mathrm{DC}}}$ for large $\beta_{\mathrm{DC}} \gg 1$.   The AC field energy $d\mathcal{E}_{\mathrm{AC}}$ is of order $0.5\hbar\omega$.  The single-molecule time scale associated with $d\mathcal{E}_{\mathrm{AC}}$ is the Rabi period, the time it takes for the population of a two-level system to cycle once, as seen in Figure~\ref{RabiFlopp}.  In real time, this is on the order of $10$ps for the parameters in the preceding paragraph.  The time scale associated with $\omega$ is the time scale on which the small oscillations in Figure~\ref{RabiFlopp} occur, of order $0.5$ps.  The length scale of the AC field is on the order of centimeters, and so we can neglect this in light of the micron length scale of the trap.

The tunneling term has several scales.  The optical lattice near the point of confinement has a length scale given by the harmonic oscillator length $l_{\mathrm{ho,x}}^{\left(00\right)}\sim $100nm and an energy scale of $E_R\approx$1.4$\hbar$ kHz.  The energy scales of the tunneling operator proper are given by the $\left\{t_{JJ'M}\right\}$ which are of order {$10^{-1}$-$10^{-2}E_R\sim$100$\hbar$} Hz for the given recoil energy.

There are also many scales for the dipole term.  For the $B$ and $d$ specified in the first paragraph of this section and $\beta_{\mathrm{DC}}=1.9$, the characteristic length scale where the dipole-dipole energy becomes comparable to the rotational energy is
\begin{eqnarray}r_B&\equiv \left({\left|\langle\mathcal{E}; 00|\hat{\mathbf{d}}|\mathcal{E};00\rangle\right|^2/B}\right)^{\frac{1}{3}}\,,
\end{eqnarray}
approximately 348 Bohr radii (18.4nm).  Outside this region the Born-Oppenheimer adiabatic approximation is easily fulfilled~\cite{buchler:060404}.  Since the length scale of our optical lattice is of order $\mu$m, we are justified in working within the Born-Oppenheimer framework.  For the same parameters, the length scale where the off-resonant van der Waals potential $C_6/r^6\approx -d^4/(6Br^6)$ becomes comparable to the dipole-dipole interaction is \begin{eqnarray}
r_{\mathrm{vdW}}&\equiv(2\left|C_{6}\right|/\left|\langle\mathcal{E}; 00|\hat{\mathbf{d}}|\mathcal{E};00\rangle\right|^2)^{\frac{1}{3}}\,.
\end{eqnarray}
This length is very small, on the order of tens to hundreds of Bohr radii.  Outside of this region the resonant dipole potential dominates and the intermolecular force is repulsive.  This repulsion enforces the hard-core limit.  The energy scale of the dipole-dipole force is ${\left|\langle\mathcal{E}; 00|\hat{\mathbf{d}}|\mathcal{E};00\rangle\right|^2/\lambda^3}\sim$1.2$\hbar$ kHz, with higher $J$ being an order of magnitude or so lower for small $\beta_{\mathrm{DC}}$, and of the same order for large $\beta_{\mathrm{DC}}$ (see Fig.~\ref{fig:dipoles_0to10_label}).

To summarize, the scales of the problem are shown in Table~\ref{Scales}.

\renewcommand\arraystretch{1.5}

\begin{table}[t]
\begin{center}
 \begin{tabular}{|c|c|c|}
  \hline Term&Length scale&Energy scale\\
\hline\hline  Rotation&internuclear distance $\sim$ 1 \AA &$B\sim 60\hbar$ GHz $\approx2\mbox{cm}^{-1}$\\
\hline  DC field&N/A, uniform&$d\mathcal{E}_{\mathrm{DC}}\sim 120 \hbar $ GHz $\approx4\mbox{cm}^{-1}$\\
\hline AC field&$2\pi c/\omega\sim 1$cm&$\hbar\omega\sim 30\hbar$ GHz $\approx 1\mbox{cm}^{-1}$\\
\hline Kinetic&$l_{\mathrm{ho,x}}^{\left(00\right)}\sim100$nm&$E_R\sim1.46\hbar$ kHz\\
\hline  Tunneling&Lattice spacing$\sim 1\mu$m&$\left\{t_{J'JM}\right\}\sim 100\hbar$ Hz\\
\hline Resonant Dipole-Dipole&energy comparable to $B$ &$\left|\langle\mathcal{E}; 00|\hat{\mathbf{d}}|\mathcal{E};00\rangle\right|^2/\left(1\mu\mbox{m}\right)^3\sim 1.2\hbar $ kHz \\
& at $r_B\simeq348$ Bohr radii&for nearest neighbors\\
\hline
 \end{tabular}
 \caption{Comparison of energy and length scales for the Molecular Hubbard Hamiltonian of  Eq.~(\ref{eqn:discreteHamiltonian}).}
\label{Scales}
\end{center}
\end{table}

\renewcommand\arraystretch{1.0}

\subsection{Novel Features of the Molecular Hubbard Hamiltonian}
\label{ssec:novel}

The MHH, Eq.~(\ref{eqn:discreteHamiltonian}), has a number of novel features which distinguish it from the Hamiltonians typically considered in the quantum lattice and condensed matter literature~\cite{Sachdev,lewensteinM2007}.
First, the tunneling energies $\left\{t_{J,J'M}\right\}$ not only depend on the rotational level $J,M$ but even change rotational states from $J$ to $J'$.  This is due both to the polarizability tensor's dependence on rotational level, and to the dressed basis.  This differs from other Hubbard models which consider spin degrees of freedom, as tunneling does not occur between spin states -- hopping does not cause spin transitions.  If we consider populating a single mode (e.g. $J=0$, $M=0$) in the $\Omega\to 0$ limit, then Eq.~(\ref{eqn:discreteHamiltonian}) becomes the extended Bose-Hubbard Hamiltonian, and the phase diagram is known~\cite{fisher1989,kuhner2000}.  This gives ideas of how to characterize the static phases of the MHH.  However, because the tunneling energy depends on $J$, the borders of the phase diagram will depend on the rotational state of the system.  We will discuss this property and provide an application in Sec.~\ref{sec:results}.

Second, the Hamiltonian is fundamentally time-dependent because it is a driven system.  This allows for the study of dynamic quantum phases, requiring the concept of a quantum phase diagram to be generalized to an inherently time-dependent picture.  In a case study for hard core bosonic molecules at half filling presented in Sec.~\ref{sec:results}, we show that the MHH has an emergent time scale.

\section{Methods}
\label{sec:methods}

\subsection{Time-Evolving Block Decimation}
\label{ssec:tebd}

The Time-evolving Block Decimation algorithm (TEBD) is a new method~\cite{vidal2003,vidal2004} designed to study the dynamics of entangled quantum systems.  The essential idea of TEBD is to provide a moving ``spotlight'' in Hilbert space which tracks a dynamical system.  The portion of the Hilbert space so illuminated is an exponentially small fraction of the full Hilbert space; this is justified by the fact that real, physical quantum many body systems, especially in real materials, typically explore only a small, lowly-entangled part of the total Hilbert space.

In fact, TEBD moves the full quantum many-body problem from the NP-complete complexity class to the P class through an exponential reduction in the number of parameters needed to represent the many body state.  We can understand the possibility of this reduction through an analogy to image compression.  Present digital cameras are capable of producing a roughly 3000 $\times$ 3000 array of pixels.  Downloading the images from such a camera, one notices that there are far less than 10 Megapixels worth of data per image.  Image compression algorithms such as JPEG produce images of remarkable quality with only a small fraction of the raw data.  The reason that these algorithms are so effective is that a physical image, as opposed to a random 2D pixel array, is not the ``most common'' or most probable image; it contains a great deal of structure and regularity.  In the same way, physical states in Hilbert space tend to be lowly entangled (by some entanglement measure), even though a general state in Hilbert space has a much larger probability of being highly entangled.  There is no general proof of this fact, just as there is no guarantee that an image will come out perfectly crisp after JPEG compression; it is simply a trend observed in many-body quantum systems.

To be slightly more specific, TEBD performs a partial trace over a particular bipartite splitting of the lattice, and then keeps the $\chi$ largest eigenvalues of the resulting reduced density matrix.  The cut-off parameter $\chi$ is based on the Schmidt measure~\cite{nielsenMA2000}, and so it also serves as a measure of the degree of spatial entanglement.  This idea is not unique to TEBD.  In fact, the density matrix renormalization group (DMRG) method first proposed by White~\cite{whiteSR1992} did something analogous years before.  TEBD's innovation is that at each time step it re-optimizes the truncated basis (thus the ``moving spotlight'').  The Schmidt number is just the number of non-zero eigenvalues in the reduced density matrix, and so is an entanglement measure natural to quantum many body systems.  The parameter $\chi$ is the number of non-zero eigenvalues in the reduced density matrix that TEBD retains.  It is the principal convergence parameter of the algorithm, both in entanglement and in time.  Although the time-propagation method we use is Trotter-Suzuki~\cite{1990PhLA..146..319S}, it turns out that, due to a normalization drift, $\chi$ controls convergence at long times.

With $\chi$ interpreted as an entanglement measure, we can say that TEBD treats the system not as a wavefunction in a $d^L$-dimensional Hilbert space ($L$ is the number of lattice sites), but as a collection of wavefunctions in $d^2$-dimensional two-site spaces that are weakly entangled with the environment created by the rest of the system.  To facilitate this viewpoint, we replace the $d^L$ coefficients of the full many-body wavefunction with $L$ sets of $\left(d\chi^2+\chi\right)$ coefficients corresponding to the wavefunctions of each bipartite splitting.  The most computationally expensive portion of the TEBD algorithm is typically the diagonalization of these local coefficient matrices at a cost of $\mathcal{O}\left(d^3\chi^3\right)$.  Looping over all $L-1$ bipartite splittings and evolving the system for a total time $t_f$ in time steps of length $\delta t$, one obtains an asymptotic scaling of $\mathcal{O}\left(L\frac{t_f}{\delta t}d^3\chi^3\right)$.

This scaling can be greatly improved by the presence of conserved quantities.  When a conserved quantity exists in the system we are able to diagonalize reduced density matrices corresponding to distinct values of this conserved quantity independently, which can result in significantly smaller reduced density matrices to diagonalize.  Implementing this idea, scalings of $\mathcal{O}\left(\chi^2\right)$ have been reported for fixed $d$~\cite{daleythesis}.  In addition, conserved quantities in the presence of selection rules can reduce the local dimension.  For example, in the case of the MHH, $z$-polarized electric fields disallow transitions from a particular $M$ to any other.  If we begin with all molecules in a particular $M$ state, this allows us to restrict our attention only to states with this $M$.  In our numerics we conserve both the projection $M$, and the total number $N$.  Furthermore, to match our hard core requirement, we allow only zero or one molecules per site, so that the local dimension is $d \leq R+1$, $R$ being the magnitude of the greatest angular momentum that we consider (note that the local dimension $d$, mentioned only here in Sec.~\ref{ssec:tebd}, bears no relation to the permanent electric dipole moment $d$ used throughout the rest of our treatment).

A more detailed description of TEBD can be found in Ref.~\cite{carr2008h}.  We also recommend Ref.~\cite{daley2004}, besides Vidal's original papers~\cite{vidal2003,vidal2004}.

\subsection{Quantum Measures}
\label{ssec:measures}

We use a suite of quantum measures to characterize the reduced MHH, Eq.~(\ref{eqn:reducedHamiltonian}) below.  The few-body measures we use are $\langle \hat{n}_i^{J}\rangle$, the number in the $J^{th}$ rotational state on the $i^{th}$ site, $E\equiv\langle \hat{H}\rangle$, the expectation of the energy, and $\frac{1}{L}\langle\hat{n}^{J}\rangle$, the average number in the $J^{th}$ rotational state per site ($L$ is the number of lattice sites).  The latter is a $J$-dependent filling factor. The many body measures we use include the  density-density correlation between rotational modes $J_1$ and $J_2$ evaluated at the middle site
\begin{eqnarray}
g_2^{\left(J_1J_2\right)}\left(\lfloor \frac{L}{2}\rfloor,i\right)&\equiv\langle \hat{n}_{\lfloor\frac{L}{2}\rfloor}^{\left(J_1\right)}\hat{n}_i^{\left(J_2\right)}\rangle-\langle \hat{n}_{\lfloor\frac{L}{2}\rfloor}^{\left(J_1\right)}\rangle\langle \hat{n}_i^{\left(J_2\right)}\rangle,
\end{eqnarray}
where $\lfloor q\rfloor$ is the floor function, defined as the greatest integer less than or equal to $q$.  As an entanglement measure we use the Meyer Q-measure~\cite{brennenGK2003,barnumH2003,barnumH2004}
\begin{eqnarray}
Q&\equiv\frac{d}{d-1}\left[1-\sum_{k=1}^M\mbox{Tr}\left(\hat{\rho}^{\left(k\right)}\right)^{2}\right]\,,
\end{eqnarray}
where $\hat{\rho}^{\left(k\right)}$ is the single-site density matrix obtained by tracing over all but the $k^{th}$ lattice site, and the factor outside of the bracket is a normalization factor ($d$ is the on-site dimension).  This gives an average measure of the entanglement of a single site with the rest of the system.  The Q-measure can also be interpreted as the average local impurity (recall that the $\mathrm{Tr}(\hat{\rho}^2)=1$ if and only if $\hat{\rho}$ is a pure state).

To determine what measures we can use to ascertain the static phases of our model we reason by analogy with the extended Bose-Hubbard Hamiltonian where we know that the possible static phases are charge density wave, superfluid, supersolid, and Bose metal~\cite{kuhner2000}.  The charge density wave is an insulating phase appearing at half integer fillings which has a wavelength of two sites.  Like the Mott insulating phase, it has an excitation gap and is incompressible.  While the extended Bose-Hubbard Hamiltonian has only one charge density wave phase due to the presence of only one species, the MHH has the possibility of admitting several charge density wave phases due to the presence of multiple rotational states.  As such, we define the structure factor
\begin{eqnarray}
S_{\pi}^{\left(J_1J_2\right)}&=\frac{1}{N}\sum_{ij}\left(-1\right)^{\left|i-j\right|}\langle \hat{n}_i^{\left(J_1\right)}\hat{n}_j^{\left(J_2\right)}\rangle\, ,
\label{eqn:structurefactor}
\end{eqnarray}
where $N$ is the total number of molecules.  We recognize this object as the spatial Fourier transform of the equal-time density-density correlation function between rotational states $J_1$ and $J_2$, evaluated at the edge of the Brillouin zone.  This measure is of experimental interest because it is proportional to the intensity in many scattering experiments, e.g. neutron scattering~\cite{A&M}.  Crystalline order between rotational states $J_1$ and $J_2$ is characterized by a nonzero structure factor $S_{\pi}^{\left(J_1J_2\right)}$.  The charge density wave is the phase with crystalline order but no off-diagonal long-range order as quantified by the superfluid stiffness of rotational state $J$
\begin{eqnarray}
\rho_s^{\left(J\right)}&=&\lim_{\phi\to 0} L\frac{\partial^2E^{\left(J\right)}\left(\phi,L\right)}{\partial\phi^2}
\end{eqnarray}
(note that $\rho_s$ bears no relation to the density matrix $\hat{\rho}$).  If both the structure factor and the superfluid stiffness are nonzero, the phase is called supersolid.  If both the structure factor and the superfluid stiffness are zero, the phase is called Bose Metal.  Finally, if the structure factor is zero and the superfluid stiffness is nonzero, the phase is superfluid.  In one dimension the entire superfluid phase is critical, and so there is no order parameter~\cite{kuhner2000}.

\section{Case Study: Hard Core Bosonic Molecules at Half Filling}
\label{sec:results}

In the following, we consider a particular case of Eq.~(\ref{eqn:discreteHamiltonian}) for dynamical study.  We choose the hard core case, which can occur naturally due to strong on-site dipole-dipole interactions, and half filling, which is an interesting point in a number of models, including the repulsive Fermi-Hubbard Hamiltonian and the extended Bose-Hubbard Hamiltonian discussed in Sec.~\ref{ssec:measures}.  For example, in the latter case, the charge-density-wave phase requires a minimum of half-filling~\cite{kuhner2000}.

If we assume that our system begins in its ground state ($J=0$, $M=0$) we need only include states which have a dipole coupling to this state.  For $z$-polarized DC and AC fields, this means we only consider $M=0$ states, yielding the reduced Hamiltonian
\begin{eqnarray}
 \nonumber \hat{H}&=-\sum_{JJ'}t_{JJ'}\sum_{\langle i,i'\rangle}\left(\hat{a}_{i',J'}^{\dagger}\hat{a}_{iJ}+\mbox{h.c.}\right)+\sum_{J}E_{J}\sum_i\hat{n}_{iJ}-\pi\sin\left(\omega t\right)\sum_{J}\Omega_J\sum_{i}\left(\hat{a}_{iJ}^{\dagger}\hat{a}_{iJ+1}+\mbox{h.c.}\right)\\
\label{eqn:reducedHamiltonian}&+\frac{1}{2}\sum_{\tiny{ J_1,J_1',J_2,J_2'}}U_{dd}^{\tiny{ J_1,J_1',J_2,J_2'}}\sum_{\langle i,i'\rangle}\hat{a}_{iJ_1}^{\dagger}\hat{a}_{iJ_1'}\hat{a}_{i'J_2}^{\dagger}\hat{a}_{i'J_2'}\,.
\end{eqnarray}
This is the specific case of the MHH that we study using TEBD.

A matter of practical concern, as apparent in Table~\ref{Scales}, is the large disparity between the timescales of the first three (Rotational, DC, and AC) and the last three (kinetic, tunneling, and Dipole-Dipole) terms.  The accumulation of error resulting from truncating the Hilbert space at each TEBD timestep causes the algorithm to eventually fail after a certain ``runaway time,'' making studies over long times intractable~\cite{gobert:036102}.  This invites a multiscale approach in the future~\cite{KevorkianJK1996,NayfehAH2000}.  In our current numerics we artificially increase the recoil energy and dipole-dipole potential to be of the order of the rotational constant in order to study Eq.~(\ref{eqn:reducedHamiltonian}) using TEBD.  In particular, we take
\begin{eqnarray}
\label{eqn:fixeddipdip}U_{dd}^{\tiny{ J_1,J_1',J_2,J_2'}}&=\frac{10B}{d^2}\langle \mathcal{E};J_1'|\hat{\mathbf{d}}|\mathcal{E};J_1\rangle\langle \mathcal{E};J_2'|\hat{\mathbf{d}}|\mathcal{E};J_2\rangle\,,
\end{eqnarray}
\begin{eqnarray}
\nonumber t_{J}&=10B\left[\eta\left(1+2\frac{\Delta \alpha}{\bar{\alpha}}\frac{J\left(J+1\right)}{\left(2J+1\right)\left(2J+3\right)}\right)\right]^{1.051}\\
\label{eqn:ersatzHieght}&\times\exp\left[-2.121 \sqrt{\eta\left(1+2\frac{\Delta \alpha}{\bar{\alpha}}\frac{J\left(J+1\right)}{\left(2J+1\right)\left(2J+3\right)}\right)}\right]\,,
\end{eqnarray}
where the dimensionless variable $\eta$ becomes an ersatz ``lattice height.''  To see the scaling more explicitly, we compare the above with the actual expressions for the MHH parameters
\begin{eqnarray}
U_{dd}^{\tiny{ J_1,J_1',J_2,J_2'}}&=\frac{8}{\lambda^3}\langle \mathcal{E};J_1'|\hat{\mathbf{d}}|\mathcal{E};J_1\rangle\langle \mathcal{E};J_2'|\hat{\mathbf{d}}|\mathcal{E};J_2\rangle\\
&=\left(\frac{2mE_Rd^{4/3}}{\hbar^2\pi^2}\right)^{\frac{3}{2}}\langle \mathcal{E};J_1'|\hat{\mathbf{d}}|\mathcal{E};J_1\rangle\langle \mathcal{E};J_2'|\hat{\mathbf{d}}|\mathcal{E};J_2\rangle/d^2\, ,\\
t_{JM}&\approx 1.397E_R\left(\frac{\left|\mathbf{E}_{\mathrm{\mathrm{opt}}}\right|^2\bar{\alpha}}{3E_R}\left[1+2\frac{\Delta\alpha}{\bar{\alpha}}\frac{J\left(J+1\right)-3M^2}{\left(2J-1\right)\left(2J+3\right)}\right]\right)^{1.051}\\
&\times\exp\left(-2.121\sqrt{\frac{\left|\mathbf{E}_{\mathrm{\mathrm{opt}}}\right|^2\bar{\alpha}}{3E_R}\left[1+2\frac{\Delta\alpha}{\bar{\alpha}}\frac{J\left(J+1\right)-3M^2}{\left(2J-1\right)\left(2J+3\right)}\right]}\right)\,.
\end{eqnarray}
If we now scale $E_R$ to be $10B/1.397$ and set $d$ such that $\left[{2mE_Rd^{4/3}/\left(\hbar^2\pi^2\right)}\right]^{\frac{3}{2}}=10B$ for this $E_R$, we recover Eqs.~(\ref{eqn:fixeddipdip}) and~(\ref{eqn:ersatzHieght}) provided we make the
definition
\begin{eqnarray}
\eta&\equiv-\left|\mathbf{E}_{\mathrm{opt}}\left(\mathbf{x}\right)\right|^2
\bar{\alpha}/\left(3E_R\right)=V_{x}^{\left(JM\right)}\bar{\alpha}/
\left(3E_R\alpha_{JM}^{\left(t\right)}\right)\,.
\end{eqnarray}
Since this dimensionless parameter plays the same role as the quasi-1D lattice height scaled to the recoil energy did in the actual MHH, we refer to it as the lattice height.  For the polarizability tensor, we choose $\Delta\alpha/\bar{\alpha}=165.8/237$, corresponding to LiCs~\cite{deiglmayr:064309}. This rescaling does not change the qualitative static and dynamical features of Eq.~(\ref{eqn:reducedHamiltonian}); it only makes Eq.~(\ref{eqn:reducedHamiltonian}) treatable directly by TEBD, without multiscale methods.

First, we point out that if we consider populating a single rotational state (e.g. $J=0$, $M=0$) in the $\Omega\to 0$ limit, then Eq.~(\ref{eqn:reducedHamiltonian}) becomes the extended Bose-Hubbard Hamiltonian, and the phase diagram is known~\cite{fisher1989,kuhner2000}.  Because the tunneling energy is different for different rotational states (see Eq.~(\ref{eqn:JayTunn})) and this difference depends only on the properties of the polarizability tensor, we can relate the borders of the phase diagram for different rotational states to properties of the polarizability tensor.  The MHH thus gives a means to measure the polarizability tensor, a standing issue in experiments~\cite{jinPrivateCommunication2008}.  Our calculations in Sec.~\ref{sec:mhh} can be used to compare directly to the phase diagram from the literature~\cite{fisher1989,kuhner2000}.  In fact, this aspect of our work, unlike the simulations below, is not restricted to 1D.

However, our main focus at present is on the dynamics of the MHH.  In the following numerical study, we explore dynamics as a function of the physical characteristics of the lattice, namely, number of sites $L$ and effective lattice height $\eta$.  Specifically, we study $L=9$, $10$, and $21$ lattice sites with $N$=4, 5, and 10 molecules, respectively, and $\eta$ ranging from 1 to 10.  We fix the dipole-dipole term as in Eq.~(\ref{eqn:fixeddipdip}), and fix the DC field parameter to be $\beta_{\mathrm{DC}}=1.9$.  While $\beta_{\mathrm{DC}}=1.9$ may not correspond to a physically realizable situation, its exploration provides insight into the MHH.

%
 \begin{figure}[htbp]
 \begin{minipage}[t]{0.49\linewidth}
\subfigure[Site-averaged population vs.~rotational time for 9 sites. Note the general theme; a gradual decrease (increase) of the maxima (minima) of oscillations.]{\includegraphics[width=\linewidth]{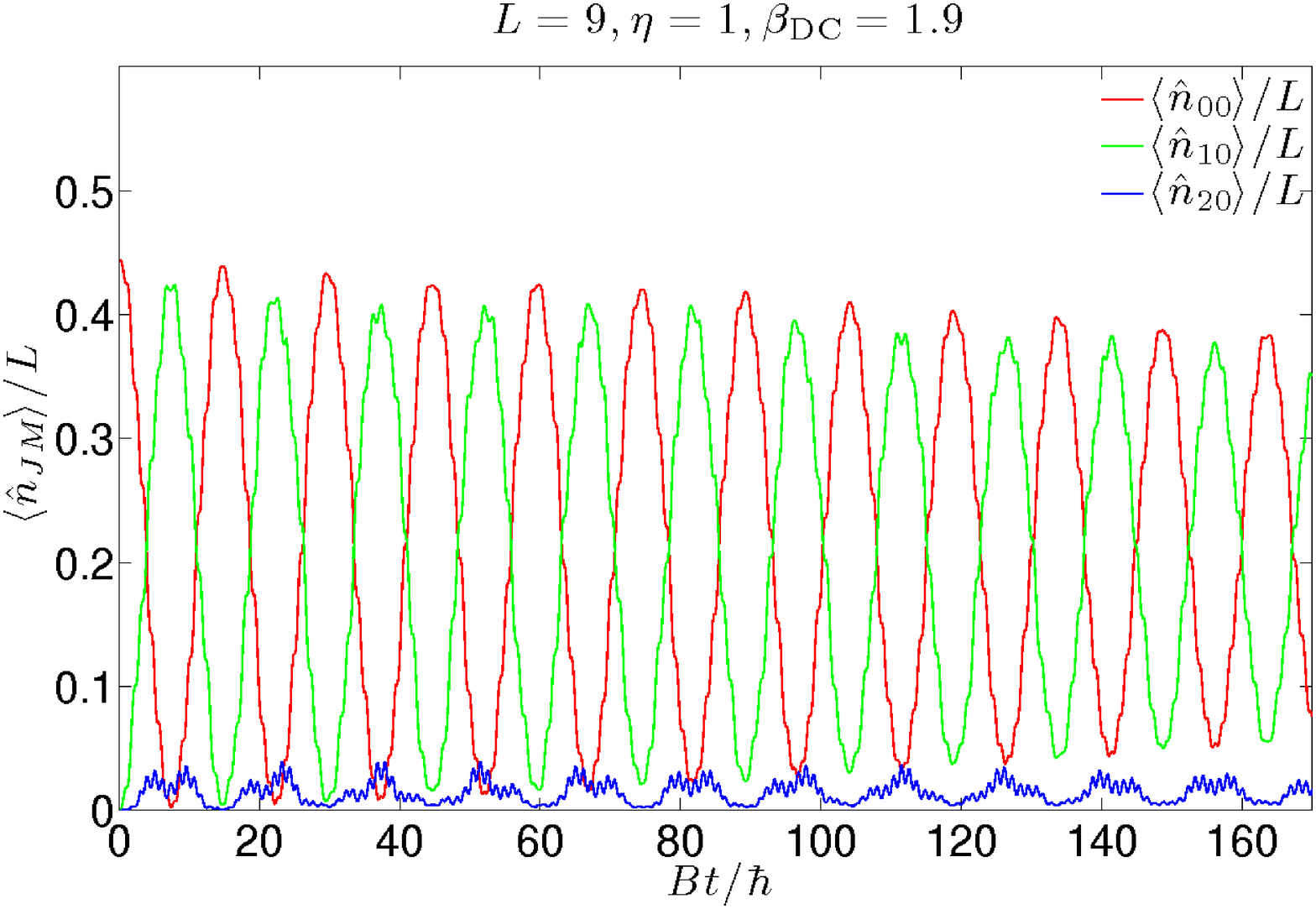}
\label{fig:Nvst911p9}
}   
    \end{minipage}
    \hspace{0.05\linewidth}
 \begin{minipage}[t]{0.49\linewidth}
\subfigure[Squared modulus of Fourier transform of site-averaged $J=0$ population vs.~rotationally scaled frequency for $L=9$ sites.  The arrow denotes the Rabi frequency $\Omega_{00}$.]{\includegraphics[width=0.8\linewidth]{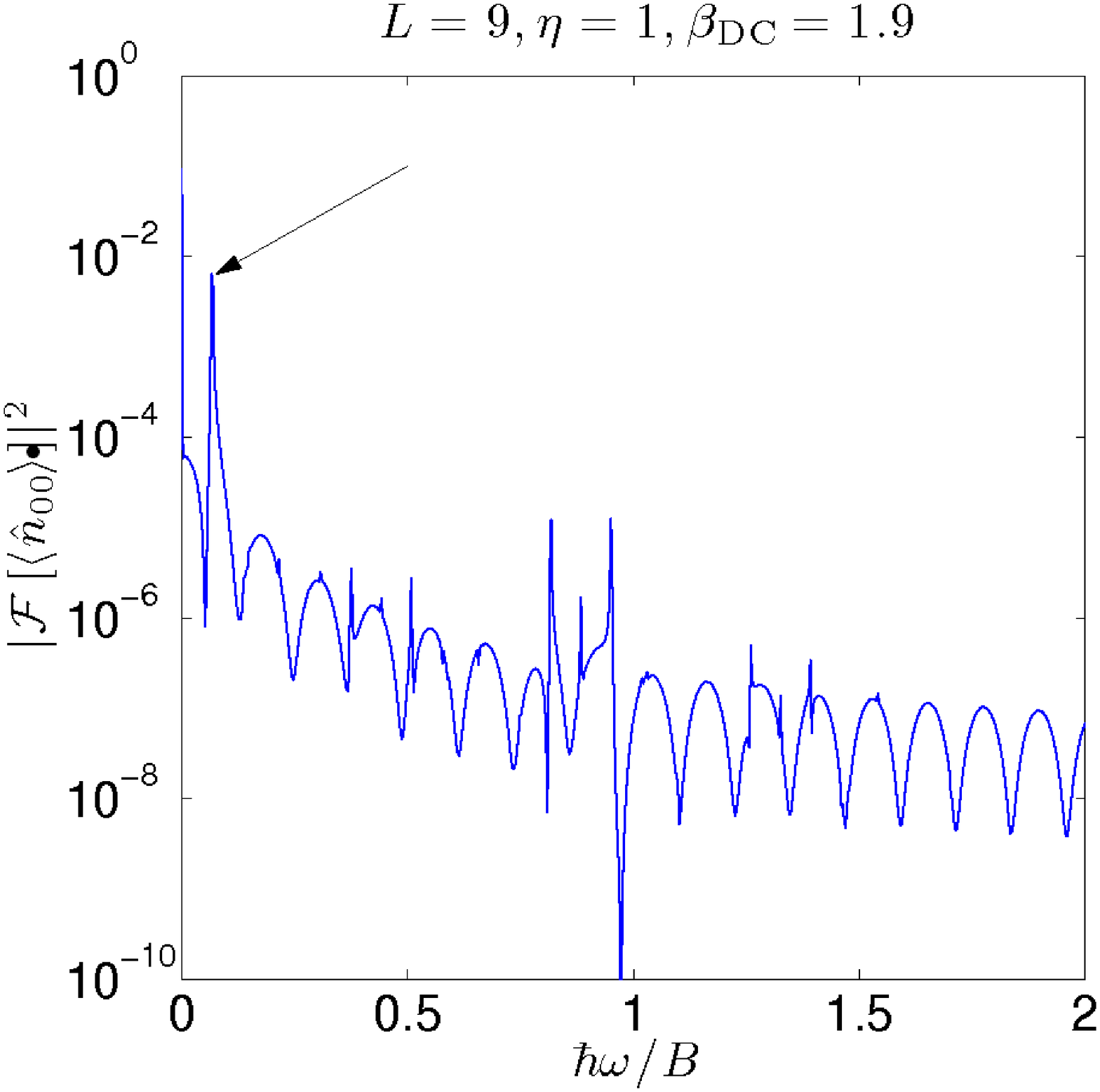}
\label{fig:PS911p9}
}   
    \end{minipage}
\begin{minipage}[t]{0.49\linewidth}
\subfigure[Site-averaged population vs.~rotational time for 10 sites.  Note that there is no significant difference between an odd and even number of sites.]{\includegraphics[width=\linewidth]{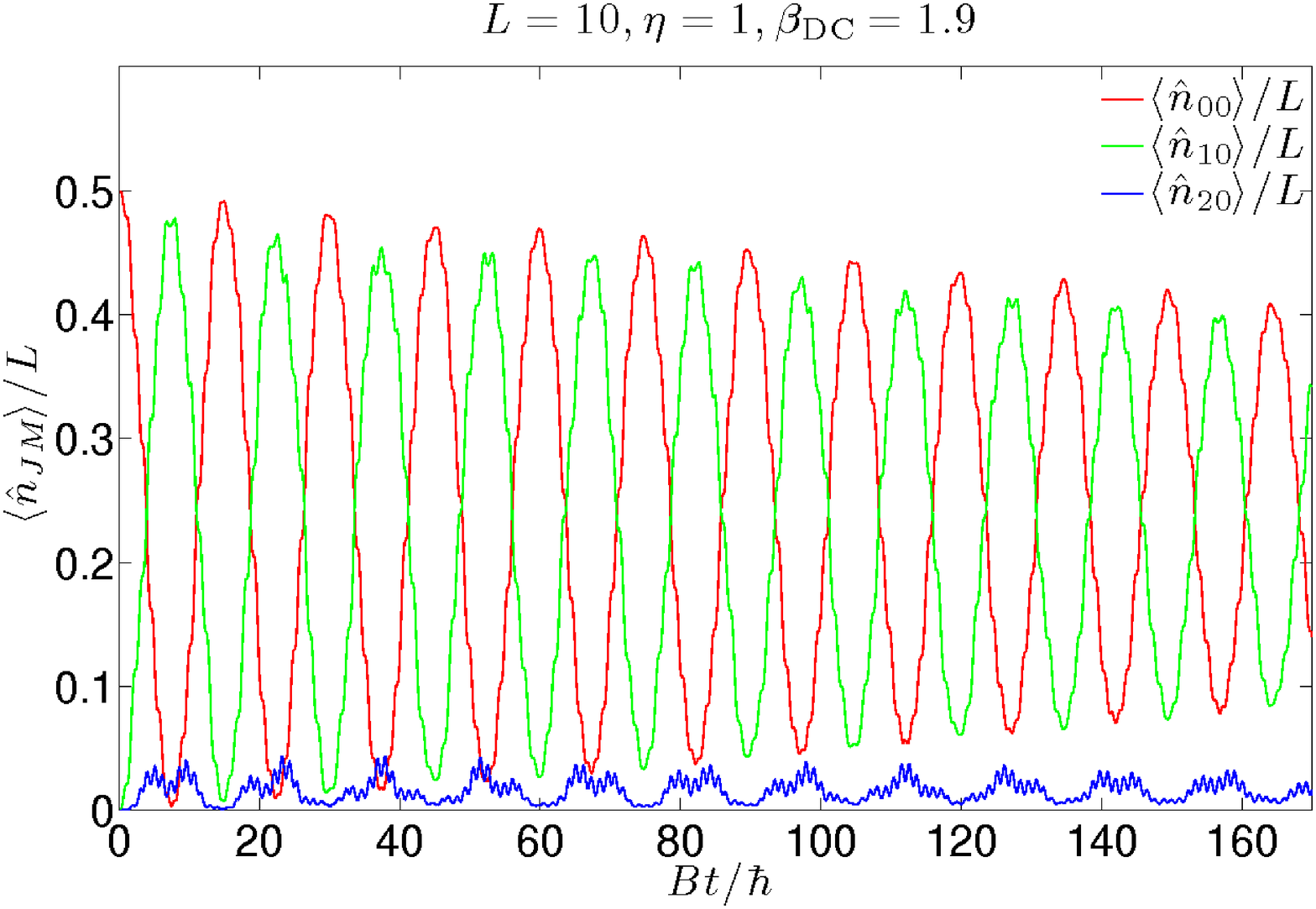}
\label{fig:Nvst1011p9}
}    
    \end{minipage}
    \hspace{0.05\linewidth}
 \begin{minipage}[t]{0.49\linewidth}
\subfigure[Squared modulus of Fourier transform of site-averaged $J=0$ population vs.~rotationally scaled frequency for $L=10$ sites.]{\includegraphics[width=0.8\linewidth]{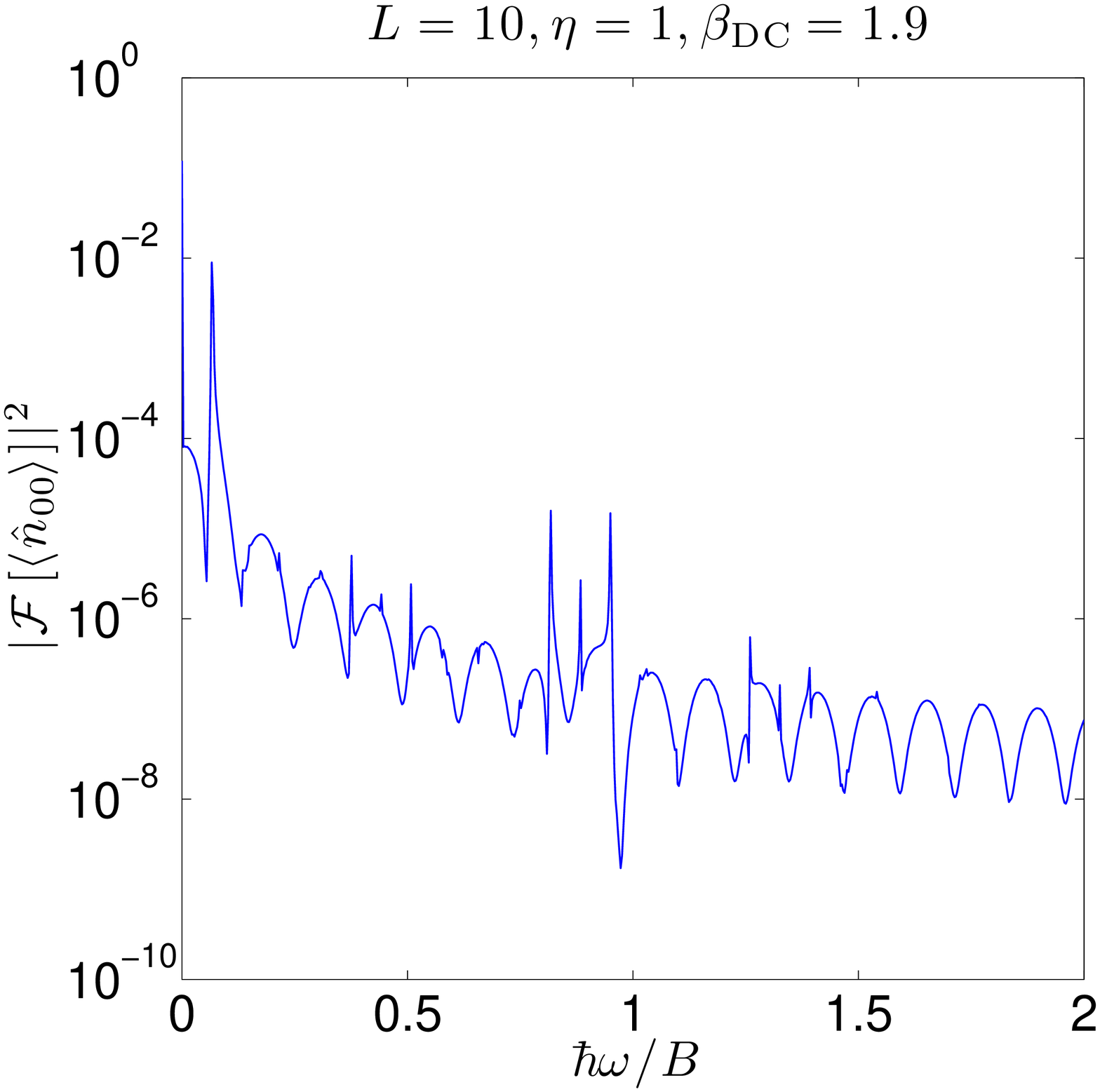}
\label{fig:PS1011p9}
}   
    \end{minipage}
\begin{minipage}[t]{0.49\linewidth}
\subfigure[Site-averaged population vs.~rotational time for 21 sites.  Note that there is no significant difference between this and the smaller system sizes.]{\includegraphics[width=\linewidth]{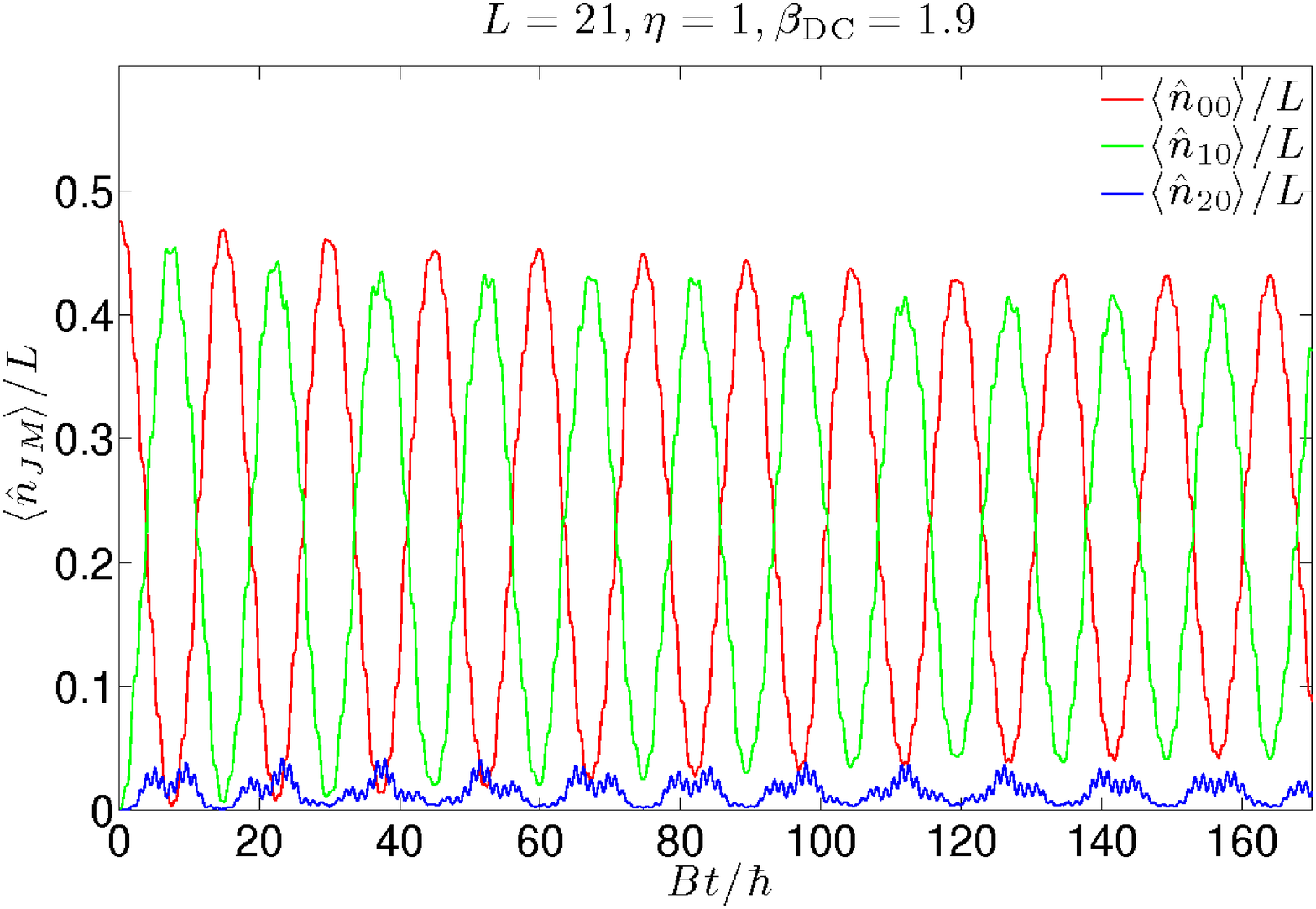}
\label{fig:Nvst2151p9}
}    
    \end{minipage}
    \hspace{0.05\linewidth}
 \begin{minipage}[t]{0.49\linewidth}
\subfigure[Squared modulus of Fourier transform of site-averaged $J=0$ population vs.~rotationally scaled frequency for $L=21$ sites.]{\includegraphics[width=0.8\linewidth]{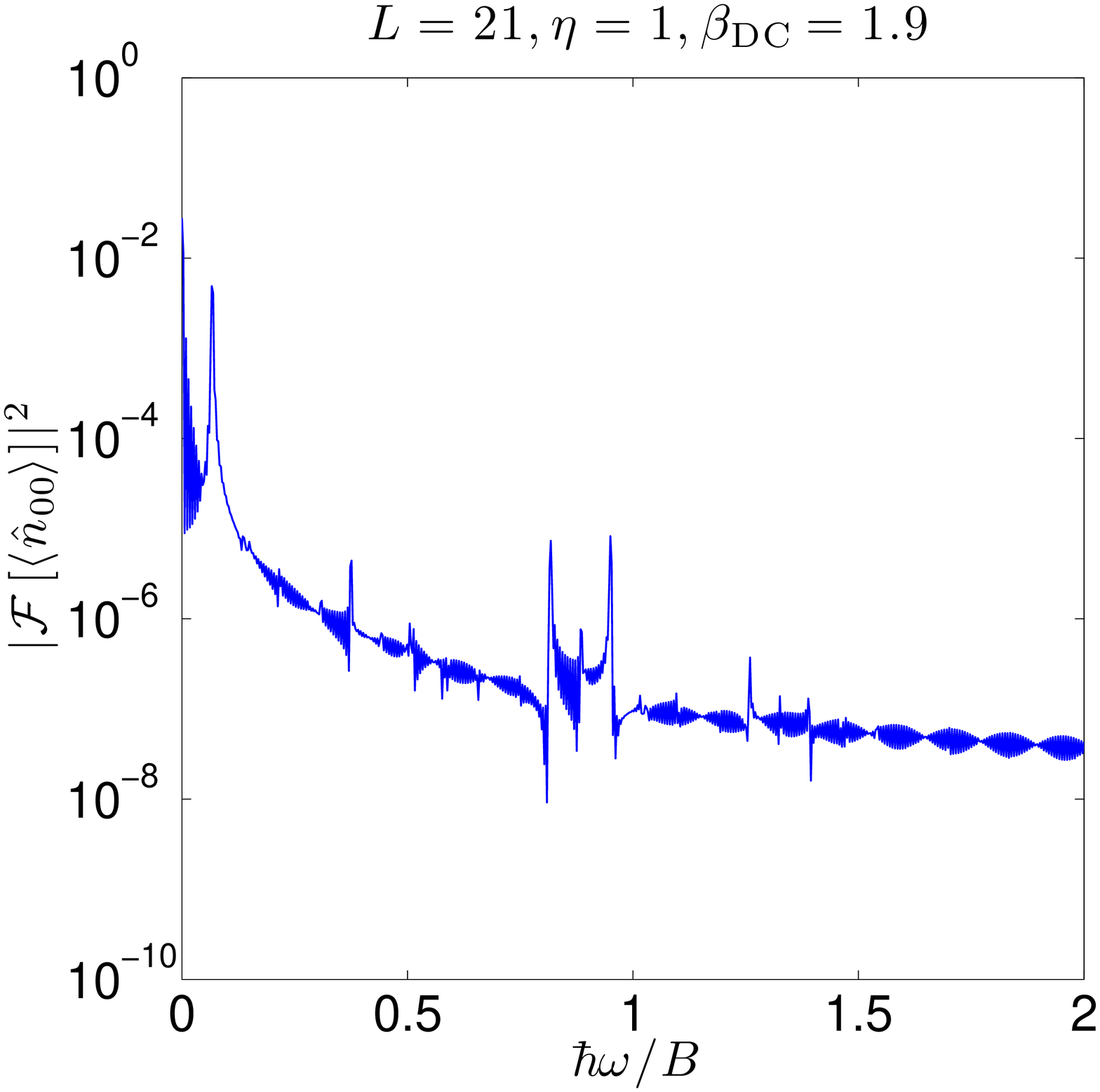}
\label{fig:PS2111p9}
}   
    \end{minipage}
\caption{Dependence of site-averaged number on lattice size $L$. For this set of parameters, the site-averaged $J=0$ and $J=1$ populations appear to asymptotically approach quarter filling.  The $J=2$ mode is populated slightly by off resonant AC couplings.  The peak near the left side of the Fourier transform plots is the Rabi frequency $\Omega_{00}$, denoted by an arrow.}
 \label{fig:Nvst}
       \end{figure}
%
\begin{figure}[htbp]
\begin{minipage}[t]{0.49\linewidth}
\subfigure[Structure factors vs.~rotational time for 9 sites.  Note the similar asymptotic behavior to the populations in Fig.~\ref{fig:Nvst911p9}.]{\includegraphics[width=\linewidth]{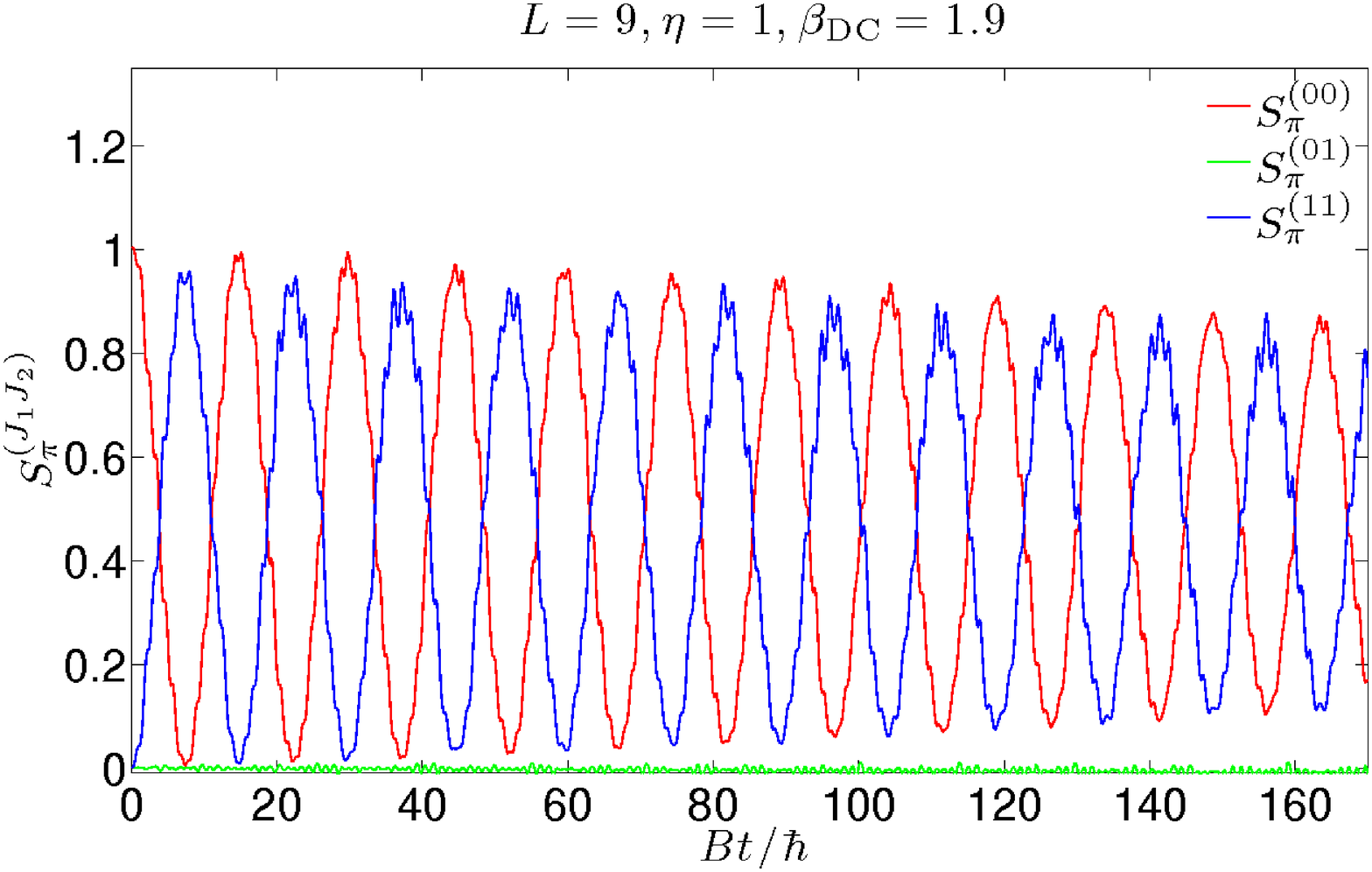}
\label{fig:Svst911p9}
}   
    \end{minipage}
    \hspace{0.05\linewidth}
 \begin{minipage}[t]{0.49\linewidth}
\subfigure[Squared modulus of Fourier transform of $S_{\pi}^{\left(00\right)}$ vs.~rotationally scaled frequency for $L=9$ sites.  Note the similarity with Fig.~\ref{fig:PS911p9} above.]{\includegraphics[width=0.8\linewidth]{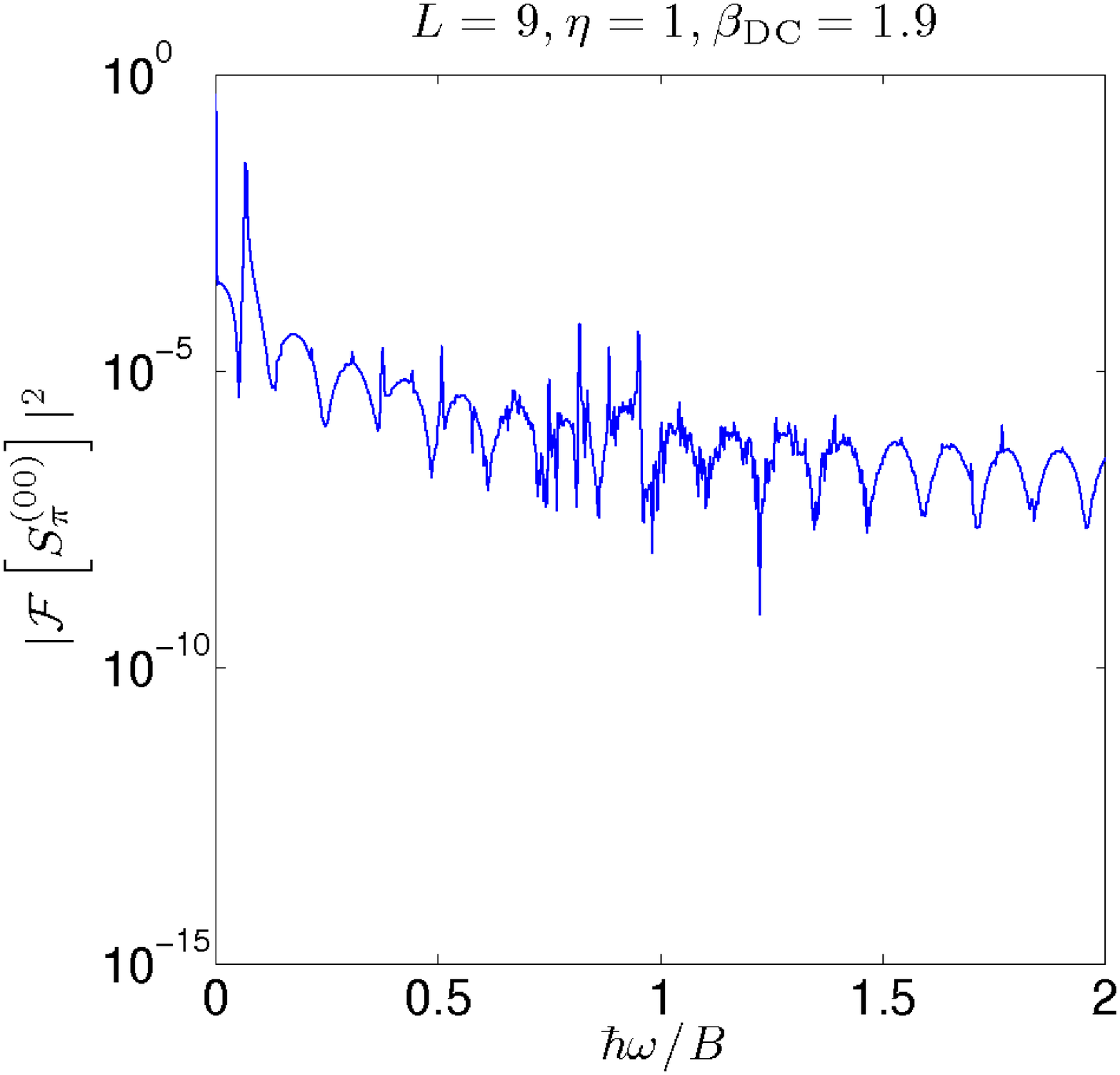}
\label{fig:SPS911p9}
}   
    \end{minipage}
\begin{minipage}[t]{0.49\linewidth}
\subfigure[Structure factors vs.~rotational time for 10 sites.  There is no significant difference in the $S_{\pi}^{\left(00\right)}$ and $S_{\pi}^{\left(11\right)}$ between even and odd $L$.  For the difference in $S_{\pi}^{\left(01\right)}$, see Fig.~\ref{fig:SPS2111p9}.]{\includegraphics[width=\linewidth]{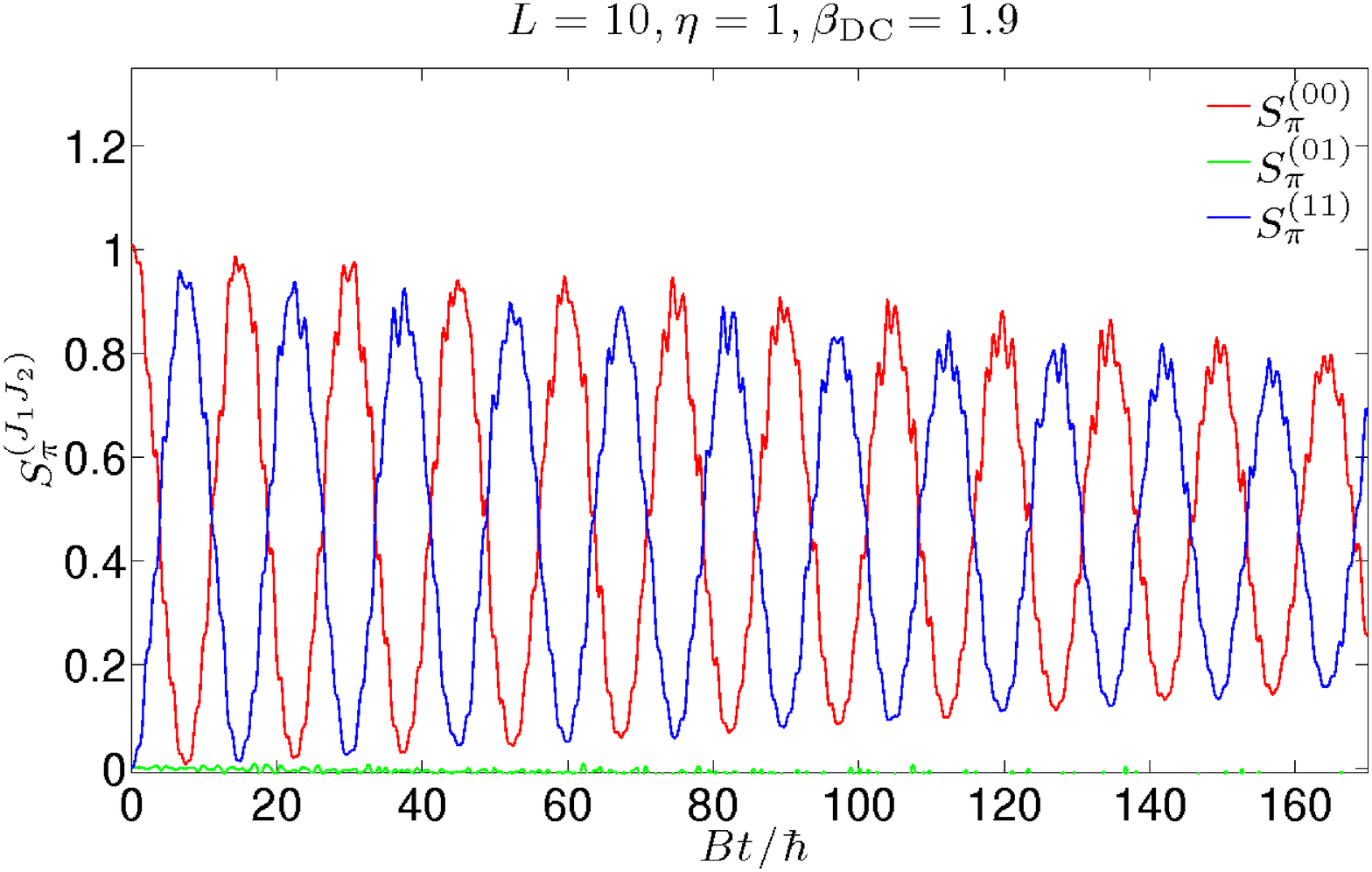}
\label{fig:Svst1011p9}
}    
    \end{minipage}
    \hspace{0.05\linewidth}
 \begin{minipage}[t]{0.49\linewidth}
\subfigure[Squared modulus of Fourier transform of $S_{\pi}^{\left(10\right)}$ vs.~rotationally scaled frequency for $L=9$ sites.  Note the absence of the Rabi frequency.]{\includegraphics[width=0.8\linewidth]{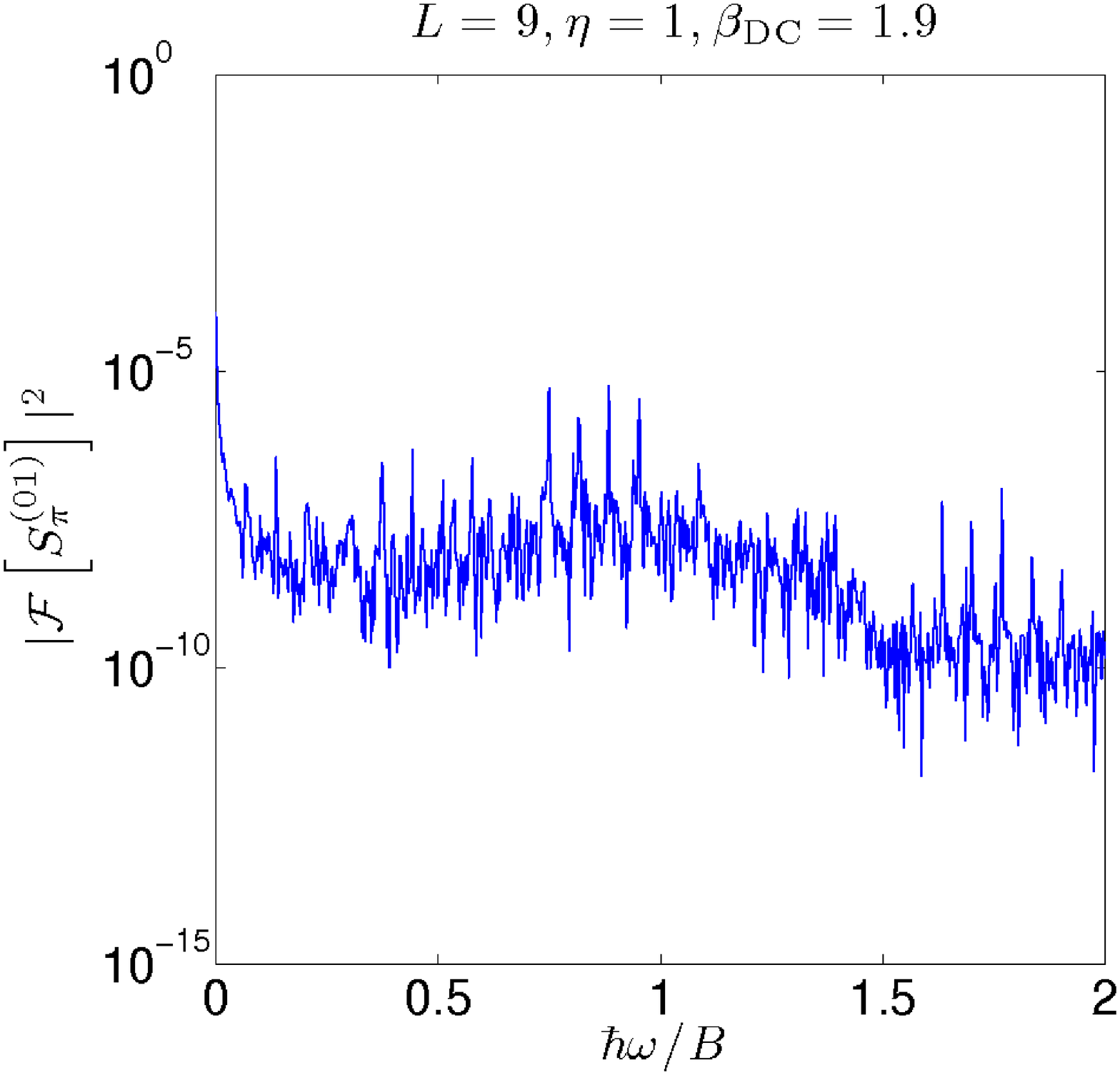}
\label{fig:SPS1011p9}
}   
    \end{minipage}
\begin{minipage}[t]{0.49\linewidth}
\subfigure[Structure factors vs.~rotational time for 21 sites.  Note the lack of significant difference with the smaller odd system size.]{\includegraphics[width=\linewidth]{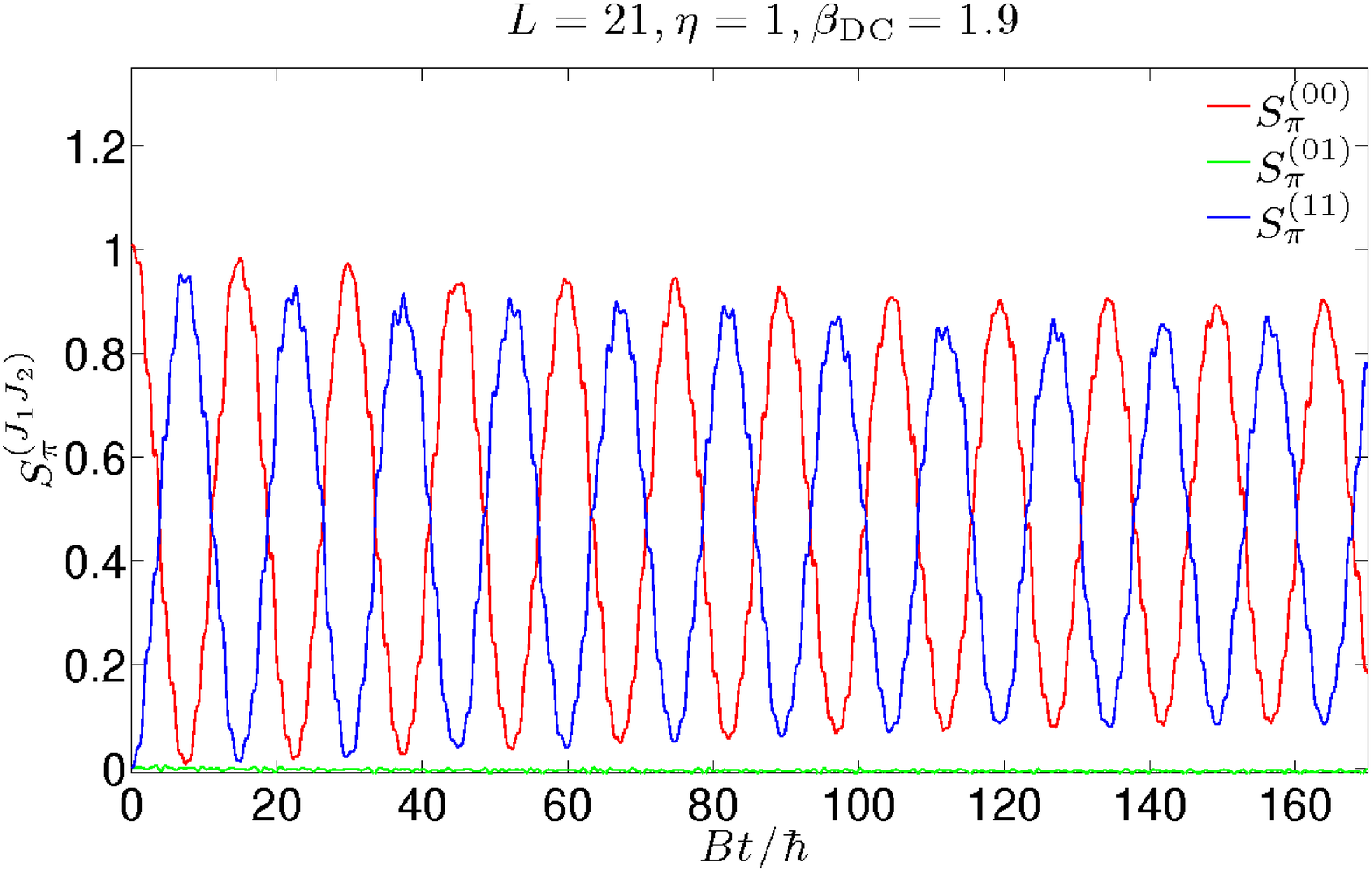}
\label{fig:Svst2151p9}
}    
    \end{minipage}
    \hspace{0.05\linewidth}
 \begin{minipage}[t]{0.49\linewidth}
\subfigure[Comparison of the $S_{\pi}^{\left(01\right)}$ correlation structure factor for odd and even numbers of sites.  Note that the even site (exactly half filling) structure factor grows faster and larger than the odd site (slightly less than half filling) structure factor.]{\includegraphics[width=0.8\linewidth]{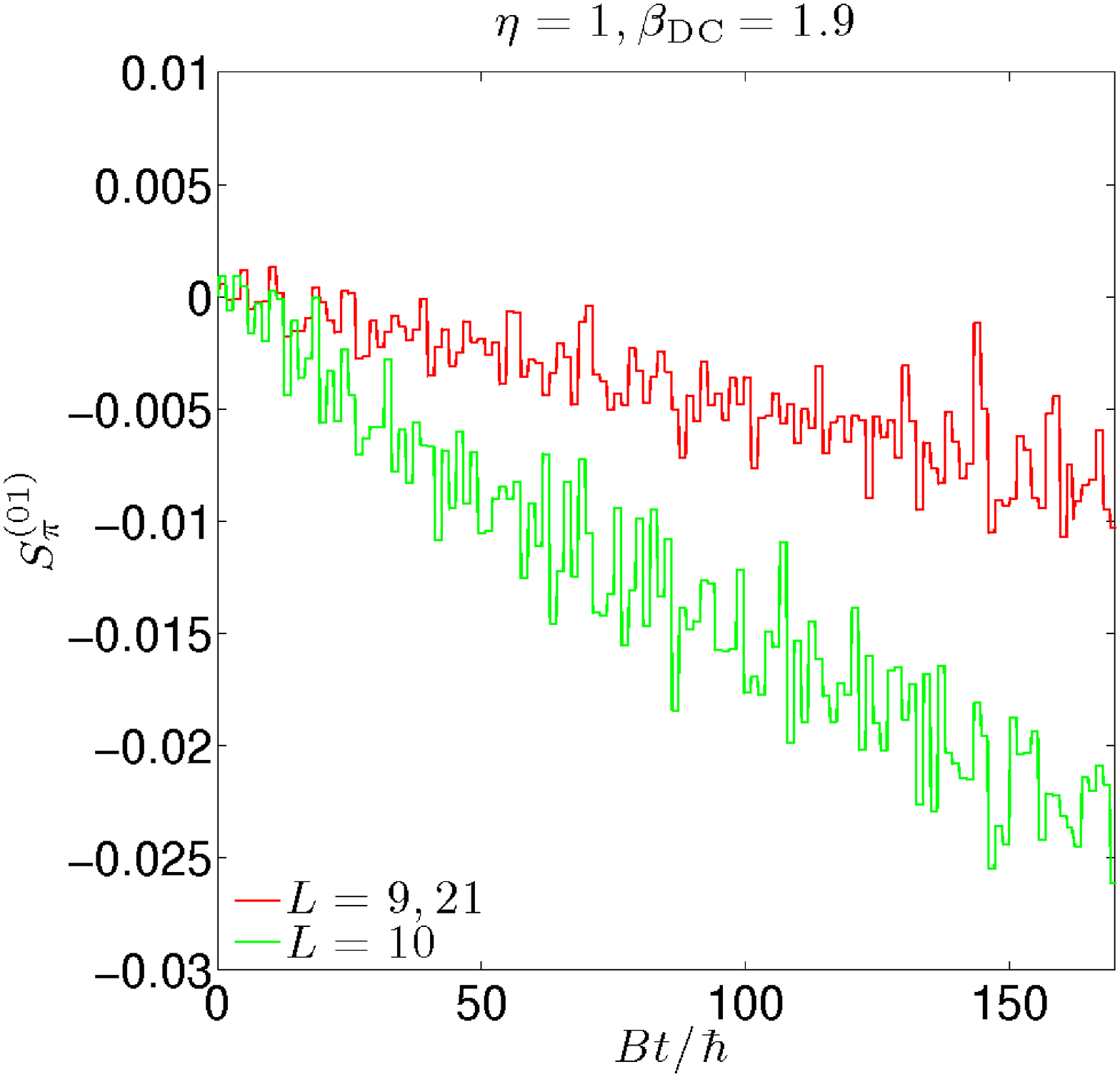}
\label{fig:SPS2111p9}
}   
    \end{minipage}

\caption{Dependence of structure factors within and between rotational states $J$ on the number of lattice sites.  We do not consider the off-resonant $J=2$ and higher rotational states because they have a very small occupation; $J=2$ is shown explicitly in Fig.~\ref{fig:Nvst}.}
 \label{fig:Svst}
       \end{figure}

\begin{figure}[htbp]
 \begin{minipage}[t]{0.49\linewidth}
\subfigure[Site-averaged population vs.~rotational time for 21 sites with $\eta=5$.  Note that the $J=0$ and $J=1$ states now appear to converge to different fillings.]{\includegraphics[width=0.9\linewidth]{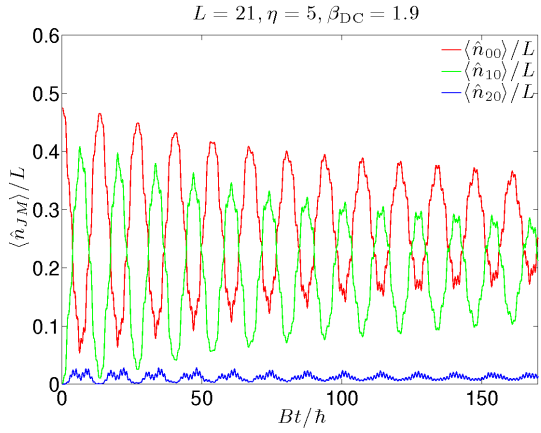}
\label{fig:N21eta5}
}   
    \end{minipage}
    \hspace{0.07\linewidth}
 \begin{minipage}[t]{0.49\linewidth}
\subfigure[Squared modulus of Fourier transform of $\langle \hat{n}_{00}\rangle$ vs.~rotationally scaled frequency for $L=21$ sites and $\eta=5$.  Note the presence of several new frequencies not observed in the $\eta=1$ case (Fig.~\ref{fig:PS2111p9}).  In particular, $\Omega_{00}$, $2\Omega_{00}$, and $3\Omega_{00}$, are denoted by arrows.]{\includegraphics[width=0.8\linewidth]{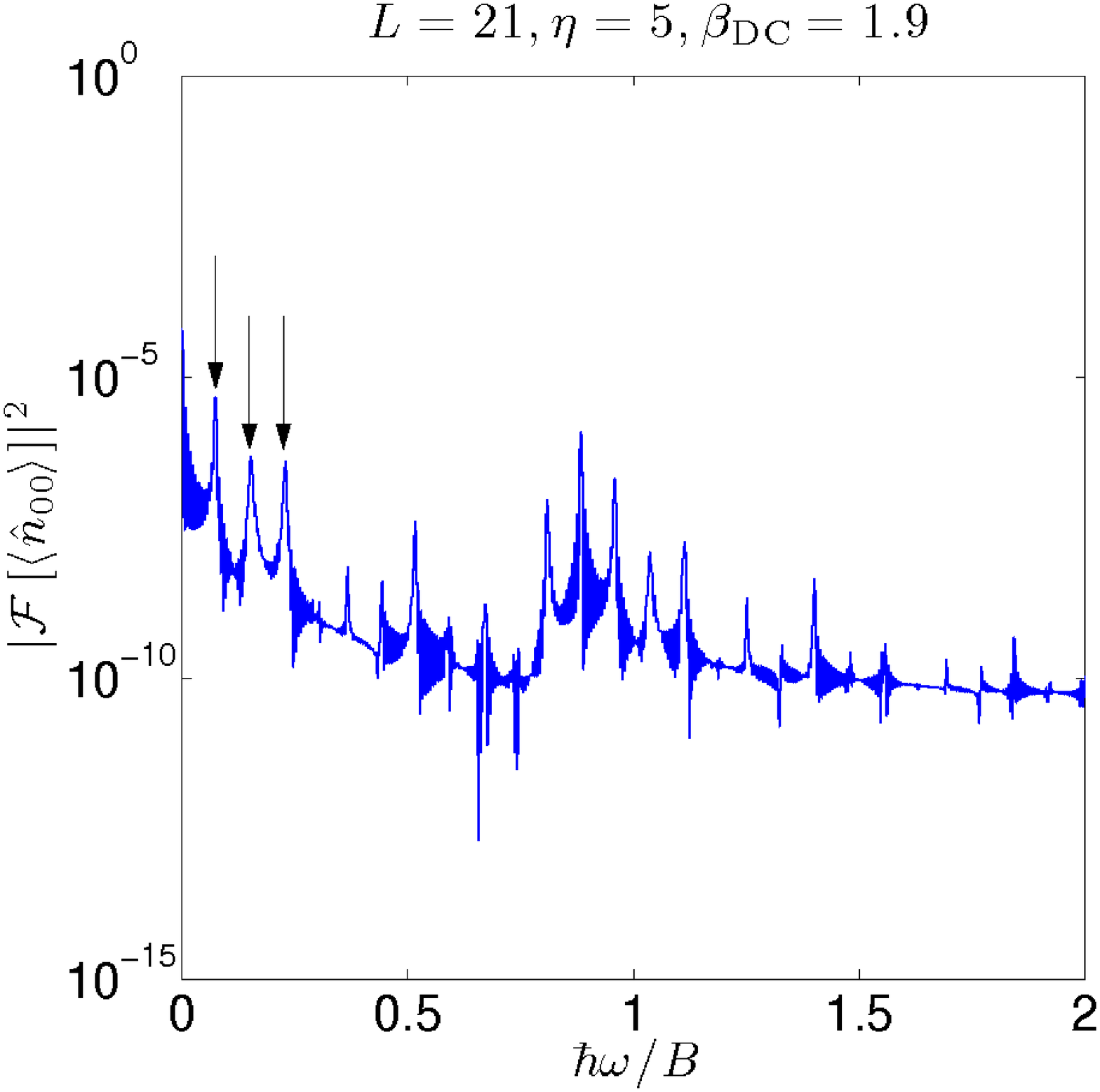}
\label{fig:NPS21eta5}
}   
    \end{minipage}

 \begin{minipage}[t]{0.49\linewidth}
\subfigure[Site-averaged population vs.~rotational time for 21 sites with $\eta=10$.  Note the similarity to the $\eta=1$ case (Fig.~\ref{fig:Nvst2151p9}) and the difference from the $\eta=5$ case(Fig.~\ref{fig:N21eta5})--the asymptotic behavior is \emph{not} a monotonic function of the lattice height.]{\includegraphics[width=0.9\linewidth]{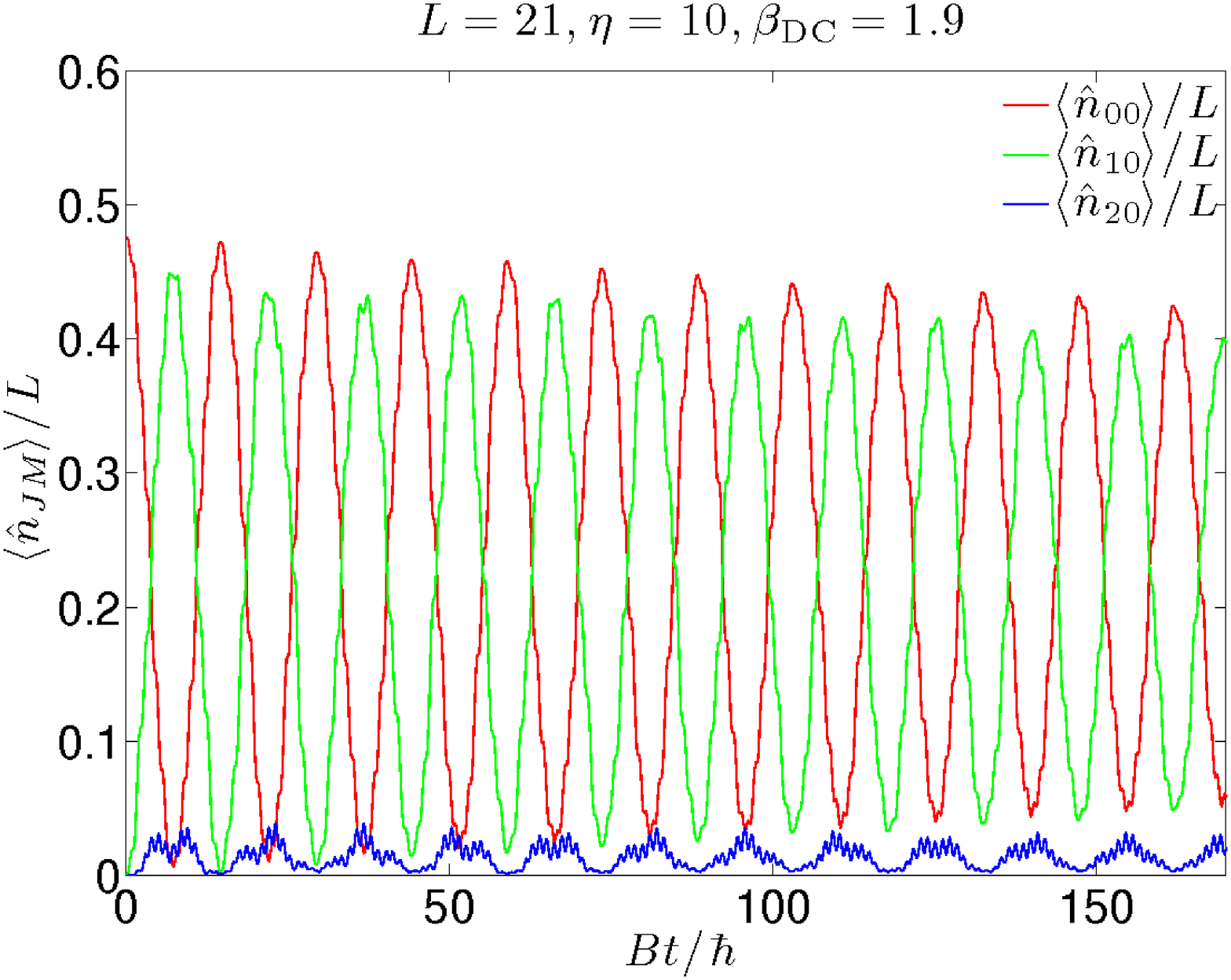}
\label{fig:N21eta10}
}    
    \end{minipage}
    \hspace{0.07\linewidth}
 \begin{minipage}[t]{0.49\linewidth}
\subfigure[Squared modulus of Fourier transform of $\langle \hat{n}_{00}\rangle$ vs.~rotationally scaled frequency for $L=21$ sites and $\eta=10$.  Note that the frequencies that emerged during $\eta=5$ have persisted.]{\includegraphics[width=0.8\linewidth]{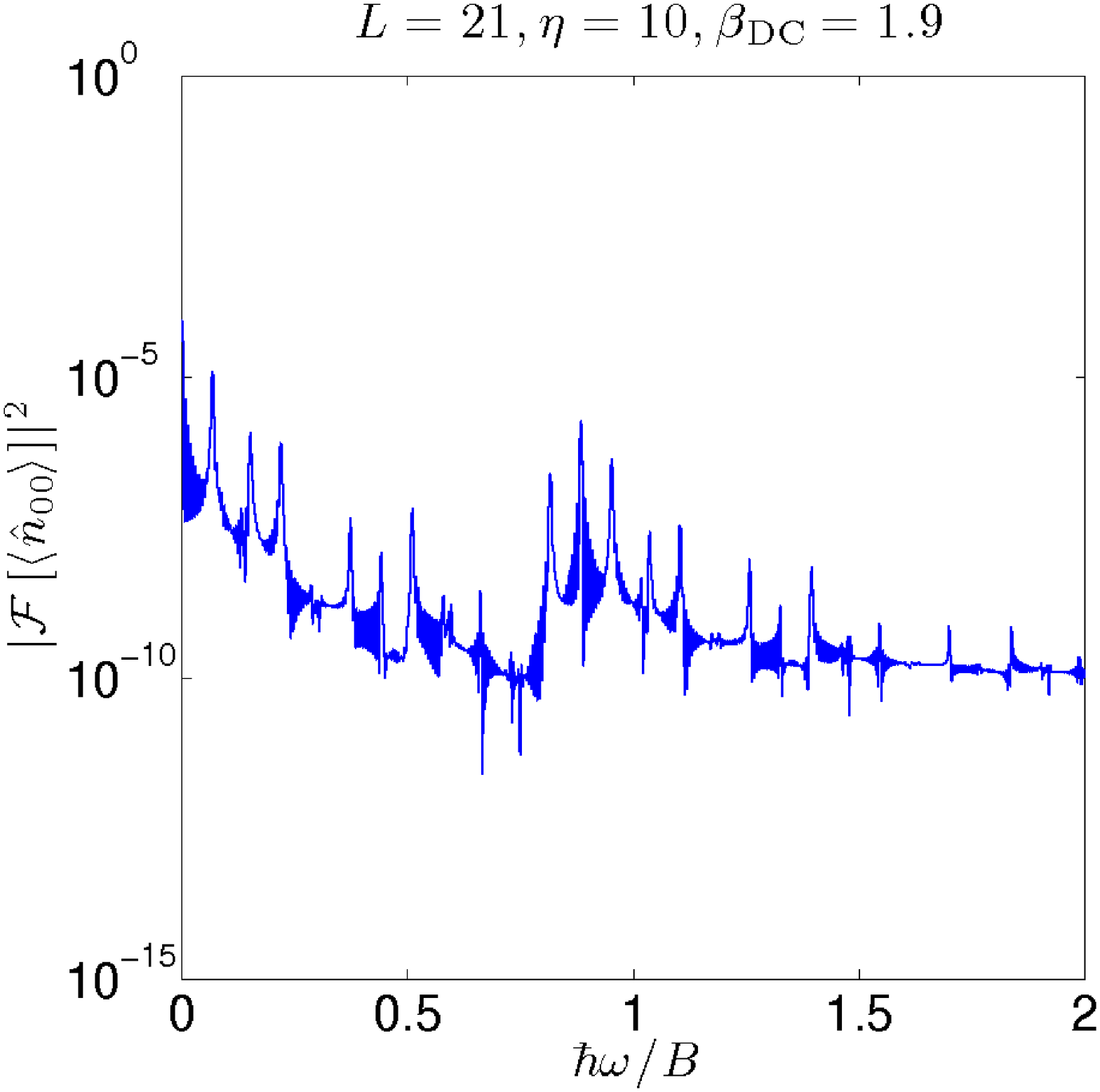}
\label{fig:NPS21eta10}
}   
    \end{minipage}

\caption{Dependence of the asymptotic behavior of rotational state populations on the lattice height $\eta$.}
 \label{fig:PAsymp}
       \end{figure}
%

\begin{figure}[htbp]
 \begin{minipage}[t]{0.49\linewidth}
\centering
\subfigure[Structure factors vs.~rotational time for 21 sites with $\eta=5$.]
{\includegraphics[width=0.9\linewidth]{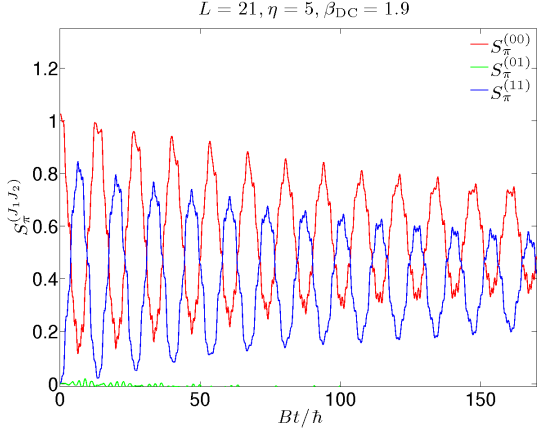}}
\label{fig:S21eta5}
 \end{minipage}
    \hspace{0.02\linewidth}
 \begin{minipage}[t]{0.49\linewidth}
\centering
\subfigure[Correlation structure factor $S_{\pi}^{\left(01\right)}$ vs.~rotational time for 21 sites with $\eta=5, 10$.]
{\includegraphics[width=0.8\linewidth]{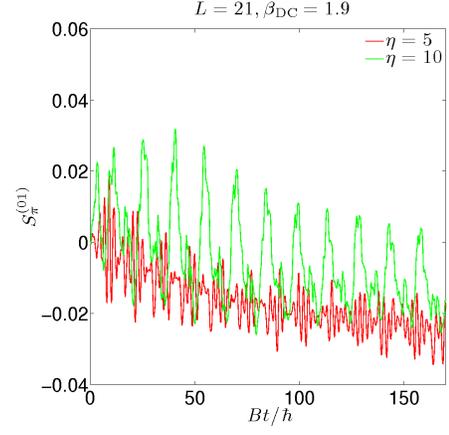}
\label{fig:S01posL21}
}   
   \end{minipage}

    \begin{minipage}[t]{0.49\linewidth}
\centering
\subfigure[Structure factors vs.~rotational time for 21 sites with $\eta=10$.  Note the similarity of $S_{\pi}^{\left(00\right)}$ and $S_{\pi}^{\left(11\right)}$ to the $\eta=1$ case (Fig.~\ref{fig:Svst2151p9}).  Note also that $S_{\pi}^{\left(01\right)}$ is now nonzero, and is periodic with the Rabi frequency $\Omega_{00}$ at short times and twice the Rabi frequency at long times (see also Figs.~\ref{fig:S01diffeta} and ~\ref{fig:S01posL21}).]
{\includegraphics[width=0.9\linewidth]{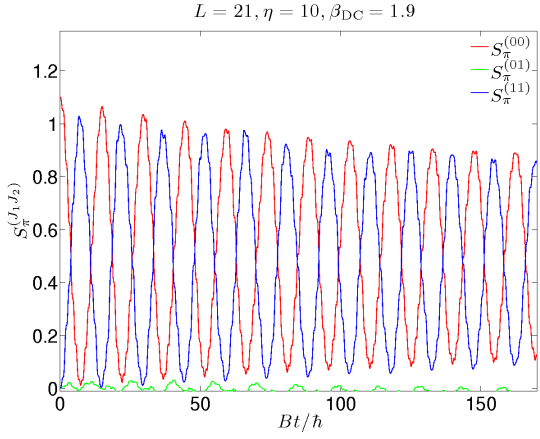}}
\label{fig:S21eta10}
 \end{minipage}
    \hspace{0.02\linewidth}
 \begin{minipage}[t]{0.49\linewidth}
\centering
\subfigure[Squared modulus of Fourier transform of $S_{\pi}^{\left(10\right)}$ vs.~rotationally scaled frequency for $L=21$ sites and $\eta=10$.  Many new frequencies appear, in particular the Rabi frequency and double the Rabi frequency, denoted with arrows.]
{\includegraphics[width=0.8\linewidth]{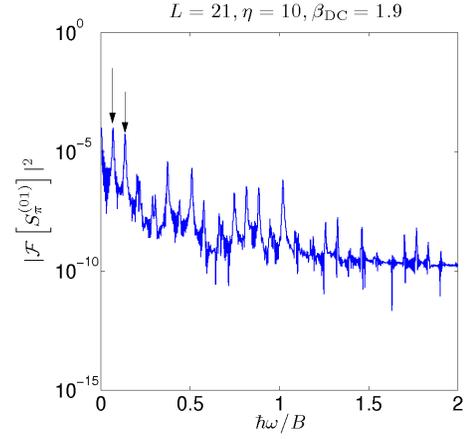}
\label{fig:S01diffeta}
}   
   \end{minipage}
\caption{Dependence of the asymptotic behavior of structure factors on the lattice height $\eta$.}
 \label{fig:SAsymp}
 \end{figure}
The Rabi oscillations between the $J=0$ and the $J=1$ states damp out exponentially in the rotational time $t_r\equiv Bt/\hbar$ as
\begin{eqnarray}
\label{fitN0}\langle \hat{n}_0\rangle&=&a_0-b_0\,e^{-t_r/\tau}\cos\left(c_0t_r\right)\,,\\
\label{fitN1}\langle \hat{n}_1\rangle&=&a_1-b_1\,e^{-t_r/\tau}\cos\left(c_1t_r\right)\,,
\end{eqnarray}
with some characteristic time scale $\tau$, as seen in Fig.~\ref{fig:Nvst}.  We note that an exponential fit has a lower reduced chi-squared than a power-law, or algebraic fit.  We also tried fit functions where the oscillations do not decay to zero, but rather persist with some asymptotic nonzero amplitude.  We find that the fit functions Eqs.~(\ref{fitN0}) and (\ref{fitN1}) above fit the data better as quantified by the convergence properties of the algorithms used, as discussed in~\ref{sec:convergence}.

The time scale $\tau$ also describes the decay of physically measurable quantities, for example the structure factors as defined in Eq.~(\ref{eqn:structurefactor}) and illustrated in Fig.~\ref{fig:Svst}.  We show the emergent time scale $\tau$ for various lattice heights and systems sizes in Table~\ref{table:timescales}.

\begin{table}[t]
\begin{center}
\begin{tabular}{|c|c|c|c|c|c|}
\hline $L$&$\eta$&$\tau B/\hbar$ &Asymp. S.E.&$\tau_QB/\hbar$&Asymp. S.E.\\
\hline 9&1&414.04&0.72\%&398.4&0.51\%\\
\hline 9&2&224.32&1.79\%&149.9&1.36\%\\
\hline 9&3&117.5&1.86\%&126.7&1.03\%\\
\hline 9&10&613.00&1.07\%&1079.66&14.09\%\\
\hline 10&1&259.96&0.76\%&240&0.6454\%\\
\hline 10&4&140.70&1.19\%&72.04&0.60\%\\
\hline 10&10&526.21&0.88\%&396.46&1.018\%\\
\hline 21&1&756.18&3.13\%&110.68&0.96\%\\
\hline 21&5&177.53&1.62\%&75.18&0.902\%\\
\hline 21&10&716.21&2.96\%&244.09&2.82\%\\
\hline
\end{tabular}
\caption{Emergent time scales $\tau$ and $\tau_Q$ and their fit asymptotic standard errors for various lattice heights and system sizes.}
\label{table:timescales}
\end{center}
\end{table}

Examining Fig.~\ref{fig:Nvst}, one observes that the driven system approaches a dynamical equilibrium that is a mixture of rotational levels.  The time scale with which the system relaxes to this equilibrium, $\tau$, cannot be determined from the single-molecule physics, and so we refer to $\tau$ as an \emph{emergent time scale}.  For the low lattice height $\eta=1$, the populations of the first two rotational states appear to oscillate around and asymptotically converge to roughly quarter filling, with $J=1$ being lower due to contributing to population of $J=2$ via an off-resonant AC coupling (Fig.~\ref{fig:Nvst911p9}).  For $\eta=5$, the asymptotic equilibrium is an uneven mixture of rotational states that favors occupation of the $J=0$ state (Fig.~\ref{fig:N21eta5}), and the emergent time scale for reaching this equilibrium is shorter than it was for $\eta=1$ by roughly a factor of four.  As the lattice height is then increased to $\eta=10$, the populations return to the trend of $\eta=1$, again converging to quarter filling with a time scale comparable to that of $\eta=1$ (Fig.~\ref{fig:N21eta10}).  This illustrates the fact that the emergent time scale $\tau$ is not, in general, a monotonic function of the parameters of the lattice.

While the dynamics of the site-averaged rotational state populations are superficially similar for $\eta=1$ and $\eta=10$, the underlying physics is not identical, as can be seen by comparing Figs.~\ref{fig:PS2111p9}, \ref{fig:NPS21eta5}, and \ref{fig:NPS21eta10}. These figures display the squared modulus of the Fourier transform of the site-averaged number in the $J=0$ state.  The only significant frequency observed for $\eta=1$ is the Rabi frequency $\Omega\sim 0.064B/\hbar$.  In contrast, the $\eta=5$ case has numerous other characteristic frequencies.   As we raise the lattice height to $\eta=10$, the frequencies that arose for $\eta=5$ remain, even though the overall visual trend of the site-averaged number reflects that of the single-frequency $\eta=1$ behavior.  While we do not explicitly see the new frequencies in the site-averaged number, we do see them in the structure factors.  An example is Fig.~\ref{fig:S01posL21}, which clearly displays the $2\Omega$ frequency behavior of the correlation structure factor $S_{\pi}^{\left(01\right)}$ for $\eta=10$.  This frequency, which we easily pick out in the site-averaged number's Fourier transform, can also be seen in the Fourier transform of $S_{\pi}^{\left(01\right)}$, see Fig.~\ref{fig:S01diffeta}.

\begin{figure}[htbp]
 \begin{minipage}[t]{0.49\linewidth}
\centering
\subfigure[Dependence of the population damping time scale $\tau$ on the number of lattice sites.  When we remove the dependence on the filling by dividing through by the total number, we see that there is little difference in the time scales with which systems of different size approach dynamic equilibrium.  Contrast Figs.~\ref{fig:Nvst911p9}, \ref{fig:Nvst1011p9}, and \ref{fig:Nvst2151p9}, which display a profound dependence on filling when the dependence on lattice sites has been removed.]
{\includegraphics[width=1.0\linewidth]{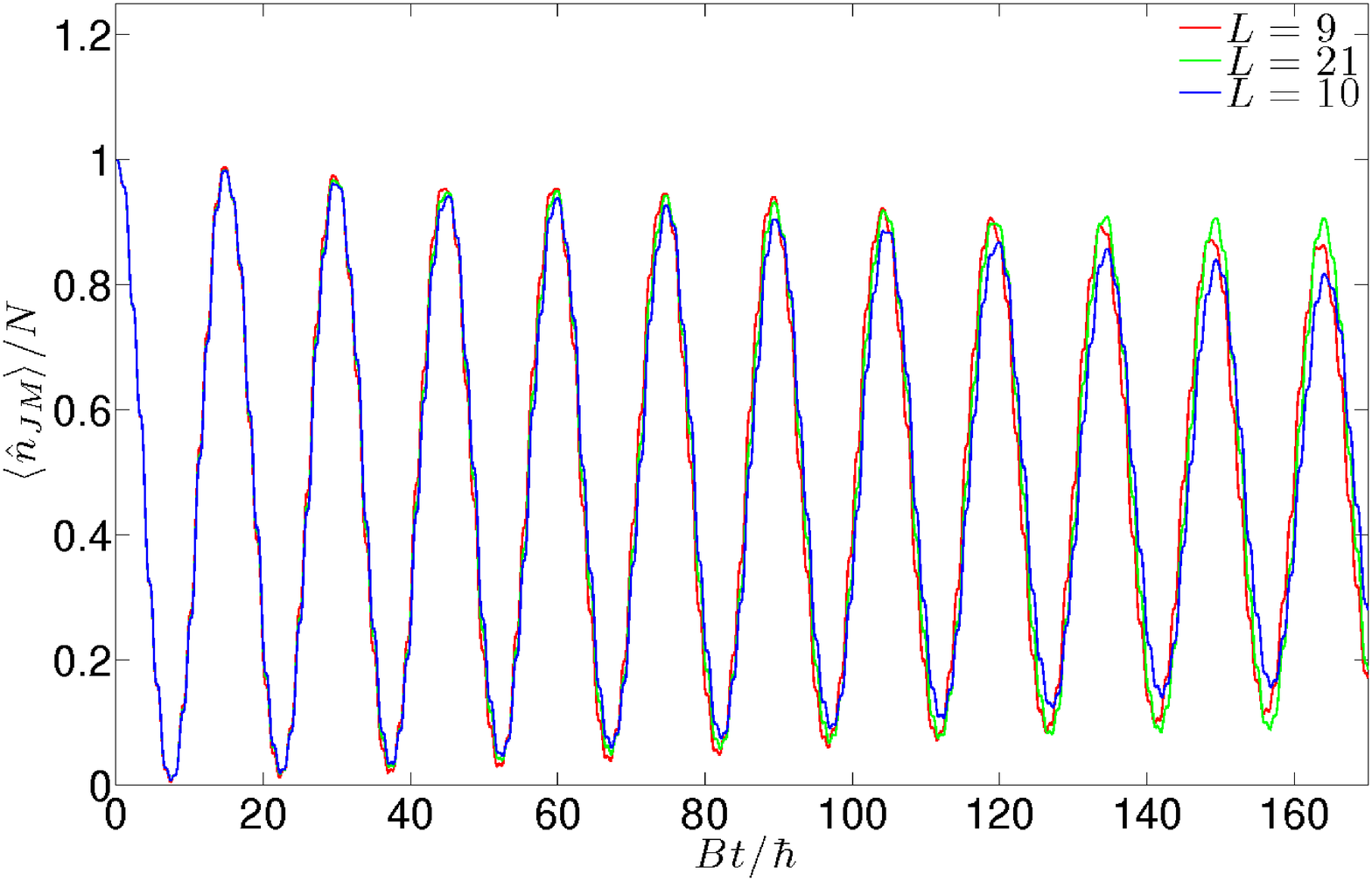}}

\label{fig:NvsL}
 \end{minipage}
    \hspace{0.02\linewidth}
 \begin{minipage}[t]{0.49\linewidth}
\subfigure[Dependence of spatial entanglement on number of lattice sites.  We see that systems of different size have different spatial entanglement in their static ground state.  The time scale of the $Q$-measure saturation, $\tau_Q$, is shorter for $L=10$ than it is for the odd $L$ cases.  This follows the general trend of $\tau$ and $\tau_Q$ responding correspondingly to changes in the Hamiltonian parameters, and so we associate this shorter time scale partially with the filling, not entirely with the system size.]
{\includegraphics[width=1.0\linewidth]{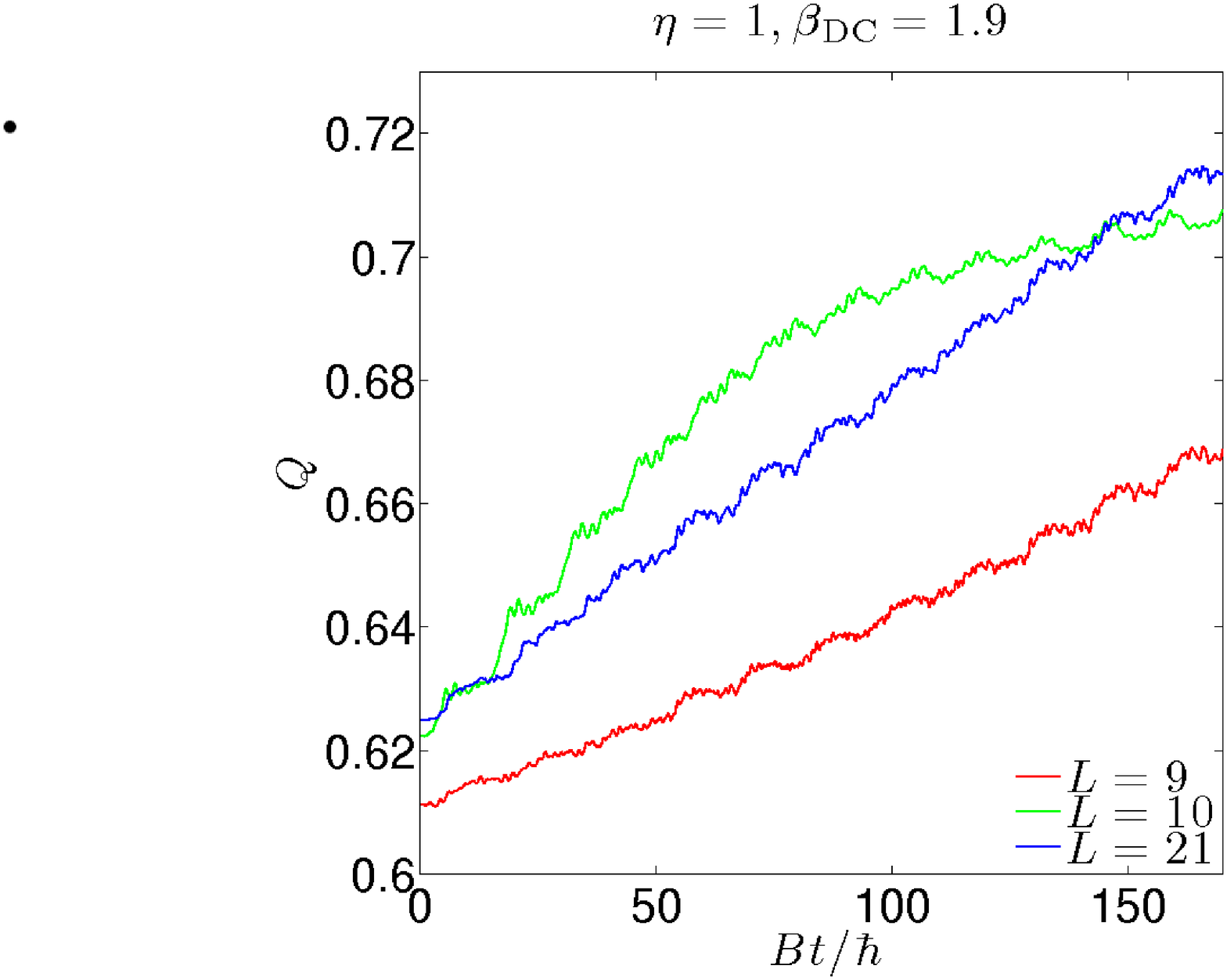}}
\label{fig:QvsSize}
   \end{minipage}
 \caption{{Dependence of emergent time scales on number of lattice sites.}}
 \end{figure}

We find that the emergent time scale $\tau$ does not depend strongly on the size of the system $L$, even though the distribution of molecules on the lattice is, in general, quite different for different numbers of sites, as can be seen by comparing Figs.~\ref{fig:Nvst911p9} and~\ref{fig:Nvst2151p9}.  Examining Fig.~\ref{fig:Nvst1011p9} and Table~\ref{table:timescales}, the $L=10$ case has a smaller $\tau$ than either of the odd $L$ cases.  We think this has to do with the filling being exactly $1/2$ and not, strictly speaking, with the number of lattice sites, as the $L=9$ and $L=21$ cases have fillings less than $1/2$.  We see this clearly by comparing Fig.~\ref{fig:NvsL} with Figs.~\ref{fig:Nvst911p9}, \ref{fig:Nvst1011p9}, and \ref{fig:Nvst2151p9}.  Fig.~\ref{fig:NvsL} displays $\langle \hat{n}_00\rangle/N$, a quantity which is independent of filling but dependent, in general, on the number of lattice sites.  There is a weak dependence on the number of lattice sites.  On the other hand, Figs.~\ref{fig:Nvst911p9}, \ref{fig:Nvst1011p9}, and \ref{fig:Nvst2151p9} display $\langle \hat{n}_{00}\rangle/L$, a quantity which is independent of the number of lattice sites but dependent, in general, on the filling.  There is a marked difference between $L=10$, which has filling of $5/10=1/2$ and the others, which have fillings$<1/2$, but there is not a significant difference between $L=9$ and $L=21$, which have fillings of $4/9$ and $10/21$, respectively.

The dependence of $\tau$ on the filling is also evidenced by the correlation structure factor $S_{\pi}^{\left(01\right)}$ in Fig.~\ref{fig:SPS2111p9}, which shows that there is a stronger correlation between the $J=0$ and $J=1$ states for exactly half filling than for fillings less than half, regardless of the system size.  Half filling is known to be important in the extended Bose Hubbard model, where it marks the introduction of the charge density wave phase.  We thus interpret this greater correlation structure factor as the appearance of a dynamic charge density wave phase \emph{between} rotational states at half filling.

This is in contrast to the usual behavior, where the structure factors $S_{\pi}^{\left(00\right)}$ and $S_{\pi}^{\left(11\right)}$ are nonzero whenever there is nonzero occupation of the particular rotational state and the structure factor $S_{\pi}^{\left(01\right)}$ is much smaller--essentially zero, see Figs.~\ref{fig:Svst911p9} and~\ref{fig:Svst2151p9}.  These results for the structure factors means that the $J=0$ and $J=1$ states tend to lie on top of one another, and not to ``checkerboard"  with a different rotational state occupying alternating sites.  This is due to the fact that the Rabi flopping time scale is much shorter than the dipole-dipole time scale, meaning that the population cycles before there is sufficient time for the molecules to rearrange to a configuration which is energetically favorable with respect to the dipole-dipole term.  However, because the population in each rotational level asymptotically reaches some nonzero value, we do see a small amount of rearrangement after many Rabi periods for any filling, corresponding to a nonzero $S_{\pi}^{\left(01\right)}$.  Note that this rearrangement does not affect the site-averaged numbers, but rather the distribution of rotational states among the lattice sites.  This asymptotic distribution emerges on time scales longer than we have considered, and is more prone to finite size effects than the site-averaged quantities, so we do not make a conjecture about it here.

\begin{figure}[htbp]
 \begin{minipage}[t]{0.49\linewidth}
\subfigure[{Dependence of spatial entanglement on lattice height.}  Note that the spatial entanglement and its associated time scale are \emph{not} monotonic functions of the lattice height.  Note also that the entanglement of the static ground state appears to be largely insensitive to the lattice height.]
{\includegraphics[width=\linewidth]{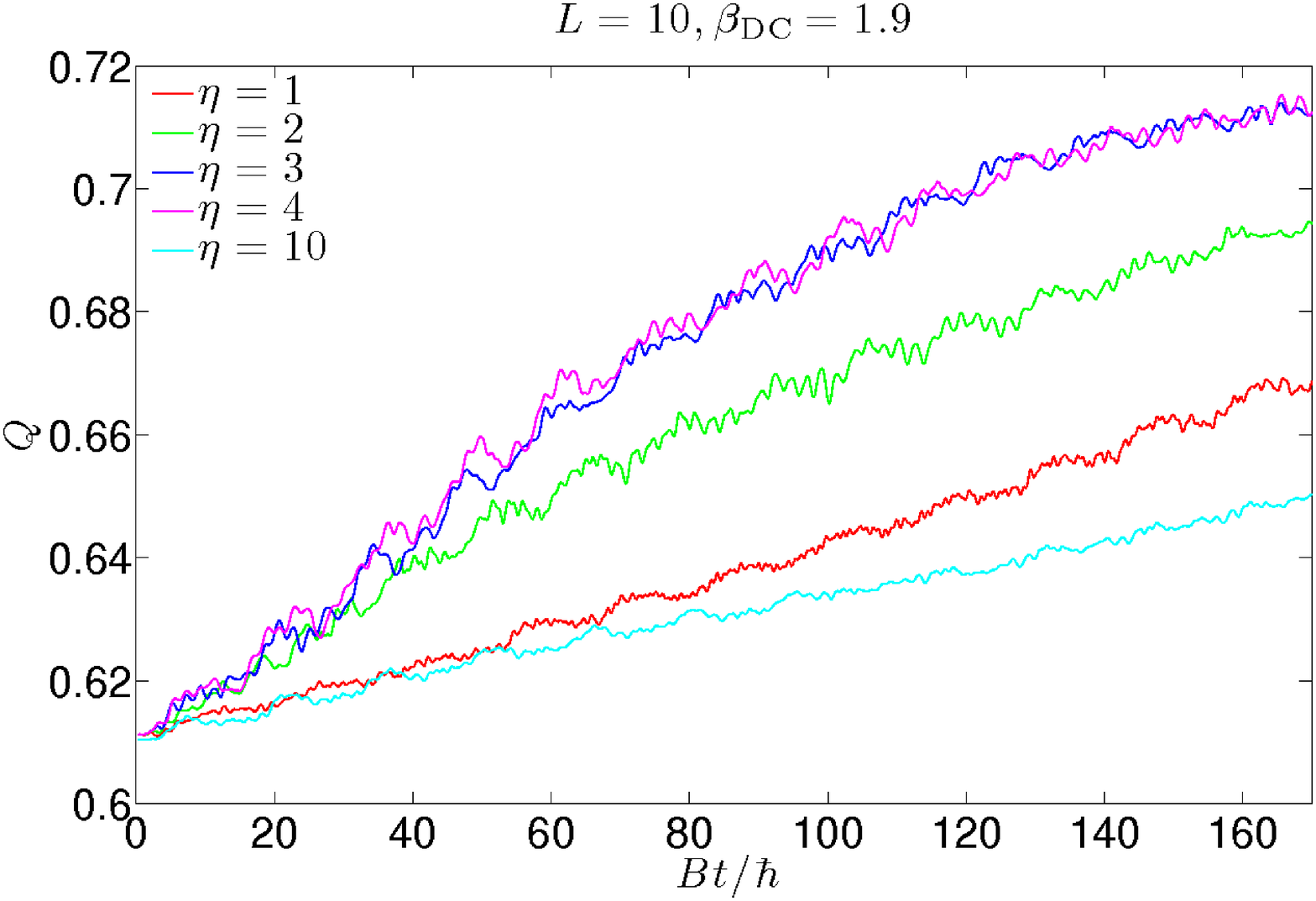}}
\label{fig:Qmeas10_1410}
 \end{minipage}
    \hspace{0.02\linewidth}
 \begin{minipage}[t]{0.49\linewidth}
\subfigure[{Dependence of the site-averaged number on the lattice height.}  Note that the emergent time scale $\tau$ is \emph{not} a monotonic function of the lattice height.  Note also that $\tau$ responds in the same way that $\tau_Q$ does to changes in the lattice height.]
{\includegraphics[width=0.95\linewidth]{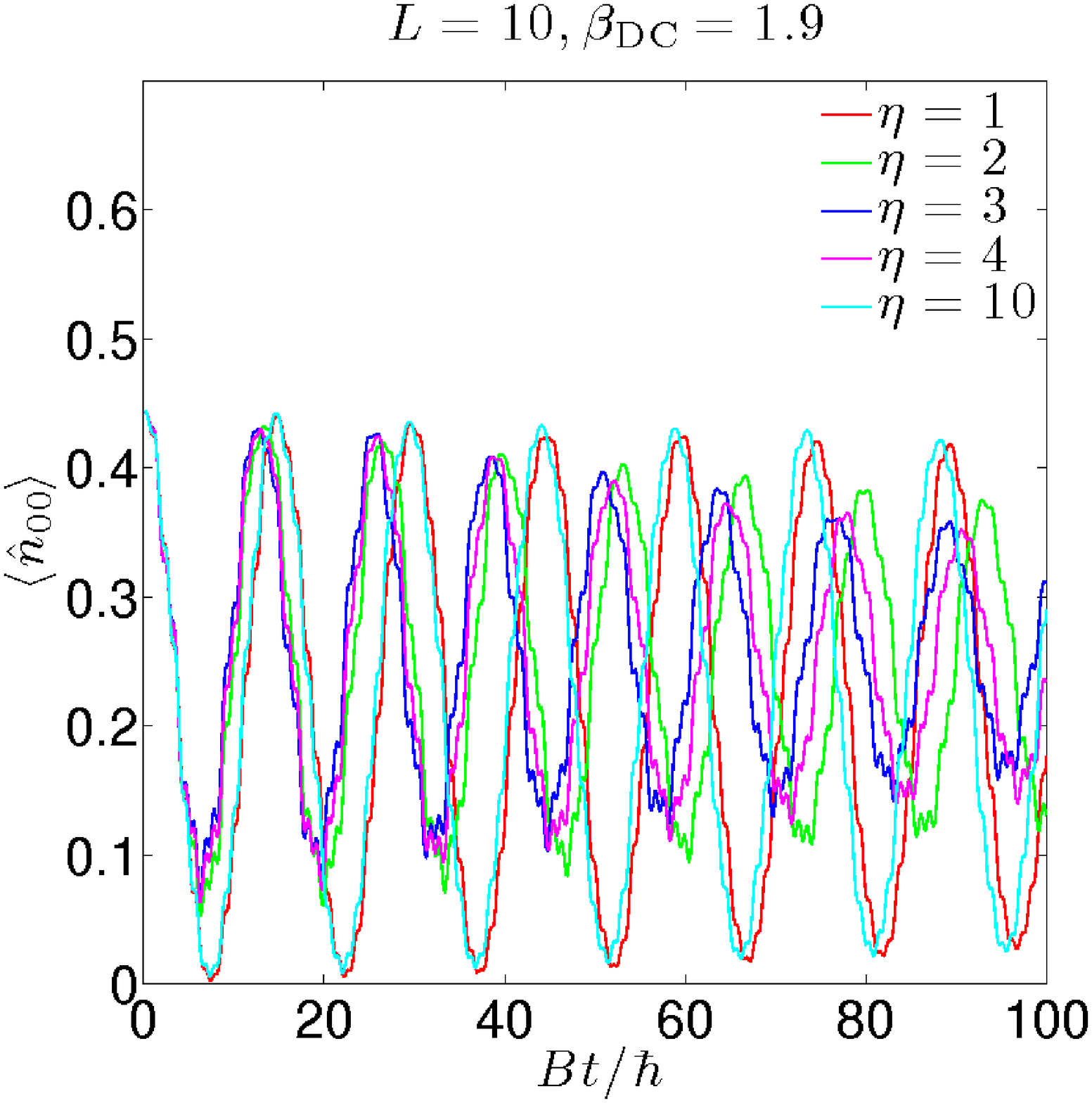}
\label{fig:NvsEta}
}   
   \end{minipage}
    \caption{{Dependence of emergent time scales on lattice height.}}
\end{figure}

We find that the $Q$-measure saturates as
 \begin{eqnarray}
\label{fitQ}Q&=&Q_{\mathrm{max}}-\Delta Qe^{-t_r/\tau_Q},
\end{eqnarray}
with a different time scale $\tau_Q$, see Fig.~\ref{fig:Qmeas10_1410} and Table~\ref{table:timescales}.  We also find that the saturation time scale of the $Q$-measure is not, in general, a monotonic function of the lattice height $\eta$, as shown in Fig.~\ref{fig:Qmeas10_1410}.

This time scale is different from the time scale $\tau$ at which the populations approach an asymptotic equilibrium, though both time scales respond similarly to changes in the Hamiltonian parameter, see Table~\ref{table:timescales}.  For example, if $\tau_Q$ gets larger as a parameter is changed then $\tau$ also gets larger, as illustrated in Figs.~\ref{fig:Qmeas10_1410} and \ref{fig:NvsEta}.  The time scale $\tau_{Q}$ displays a stronger dependence on the number of lattice sites $L$ than $\tau$, as can be seen in Figs.~\ref{fig:QvsSize} and~\ref{fig:NvsL}.  This is because $\tau$ describes a quantity that has been averaged over sites, while $\tau_Q$ does not.

\section{Conclusions}
\label{sec:conclusions}

We have presented and derived a novel lattice Hamiltonian, the Molecular Hubbard Hamiltonian (MHH). The MHH is a natural Hamiltonian for connecting theoretical studies of the dynamics of quantum phase transitions to near-term experimental setups using ultracold molecular gases.  We presented a case study of this new Hamiltonian for hard core bosonic molecules at half filling.  Starting from an initial condition of half filling in the $J=0$, $M=0$ state, we found that initial large oscillations in the system
self-damp to an asymptotic equilibrium which consists of a lattice height and filling-dependent spatially entangled superposition of dressed states.  This occurs on an emergent time scale $\tau$ which can not be predicted from the single molecule theory.  We showed that $\tau$ depends non-monotonically on lattice height, weakly on lattice size, and strongly on filling (as apparent in simulations with odd and even numbers of sites).  We also discovered a separate emergent time scale $\tau_Q$ which describes how quickly the many body spatial entanglement saturates.  We demonstrated that $\tau_Q$ and $\tau$ respond similarly to changes in the Hamiltonian parameters and that $\tau_Q$ depends on the filling, the lattice size, and, non-monotonically, on the lattice height.  In addition to these emergent time scales, we studied the time-dependent structure factors and their frequency-domain Fourier transforms.

In future studies we will consider different filling factors, DC field strength to rotation ratios $\beta_{\mathrm{DC}}$, and initial conditions, as well as polarized and unpolarized spin-1/2 fermionic molecules.  In addition, we will use multiscale methods to study how the emergent time scale demonstrated above compares to experimental time scales for physical systems, and thereby make quantitative predictions for experiments.

We acknowledge useful discussions with Deborah Jin, Heather Lewandowski, and Jun Ye.  This work was supported by the National Science Foundation under Grant PHY-0547845 as part of the NSF CAREER program.

\appendix

\section{Single molecule physics}
\label{sec:single}

\subsection*{Relationship between operators in space-fixed and molecule-fixed coordinate systems}
\label{ssec:coord}

It is well known that the representation of the angular momentum operators in a molecule-fixed coordinate frame lead to the \emph{anomalous} commutation relations $\left[J_i,J_k\right]=-i\hbar\epsilon_{ijk}J_k$~\cite{RevModPhys.23.213}.   The simplest way to avoid this trouble is to transform all expressions into the space-fixed frame where the angular momentum operators satisfy the normal commutation relations $\left[J_i,J_k\right]=i\hbar\epsilon_{ijk}J_k$~\cite{Zare}.  If the molecule-fixed axes are obtained by rotation of the space-fixed axes through the Euler angles $\left\{\phi,\theta,\chi\right\}$~\cite{ZareNote} (which we collectively abbreviate as $\left(\mathbf{R}\right)$), then the component of a $k^{th}$-rank spherical tensor $T$ that has projection $p$ along the space-fixed $z$ axis, denoted $\left(T\right)_p^{\left(k\right)}$, can be expressed in terms of the molecule fixed components as
\begin{eqnarray}
\left(T\right)_p^{\left(k\right)}&=\sum_q\mathcal{D}_{pq}^{\left(k\right)}\left(\mathbf{R}\right)^{\star}\left(T\right)_q^{\left(k\right)}\,,
\end{eqnarray}
where $\mathcal{D}_{pq}^{\left(k\right)}\left(\mathbf{R}\right)^{\star}$ is the complex conjugate of the $pq$ element of the $k^{th}$-rank rotation matrix (Wigner D-matrix).  To avoid confusion, we will label all space-fixed components with the letter $p$ and all molecule-fixed components with $q$.  From the orthogonality of the rotation matrices we have the inverse relationship
\begin{eqnarray}
 \label{qtop}\left(T\right)_q^{\left(k\right)}&=\sum_p\mathcal{D}_{pq}^{\left(k\right)}\left(\mathbf{R}\right)\left(T\right)_p^{\left(k\right)}\\
&=\sum_p\left(-1\right)^{p-q}\mathcal{D}_{-p,-q}^{\left(k\right)}\left(\mathbf{R}\right)^{\star}\left(T\right)_p^k.
\end{eqnarray}

\subsection*{Rotational Hamiltonian}
\label{ssec:rot}

In the rigid rotor approximation the rotational Hamiltonian is simply
\begin{eqnarray}
\hat{H}_{\mathrm{rot}}&=B\hat{\mathbf{J}}^2\, ,
\end{eqnarray}
where we have defined the rotational constant $B\equiv 1/2\mu r_e^2$, with $\mu$ the molecule's reduced mass and $r_e$ its equilibrium internuclear separation.  Typical values of $B$ are $\sim 60\hbar$ GHz~\cite{NISTtables}.  This Hamiltonian has eigenvalues $ BJ\left(J+1\right)$ and eigenstates $|JM\rangle$, with $J$ the total angular momentum and $M$ its projection along the internuclear axis.

\subsection*{DC Field Term}
\label{ssec:dc}

The dipole moment of a polar molecule in a rotational eigenstate is zero in an average sense due to the spherical symmetry of the rotational Hamiltonian.  We break this symmetry by introducing a DC electric field along the space-fixed $z$ axis, with Hamiltonian
\begin{eqnarray}
\label{eqn:rothami}\hat{H}_{\mathrm{DC}}&=-\hat{\mathbf{d}}\cdot\mathbf{\mathcal{E}}_{\mathrm{DC}}\,,
\end{eqnarray}
where $\mathbf{\mathcal{E}}_{\mathrm{DC}}$ is the electric field amplitude.
The field defines the spherical space-fixed axis $p=0$, and the molecule-fixed internuclear axis defines $q=0$.  We transform between them using a first-rank rotation matrix as outlined above:
\begin{eqnarray}
 \hat{H}_{\mathrm{DC}}&=-\left(\hat{\mathbf{d}}\right)^{\left(1\right)}_0\mathcal{E}_{\mathrm{DC}}.
\end{eqnarray}
The matrix elements of the DC Hamiltonian in the basis which diagonalize the rotational Hamiltonian Eq.~(\ref{eqn:rothami}) are
\begin{eqnarray}
 \langle J', M'|\hat{H}_{\mathrm{DC}}| J, M\rangle&=-d\mathcal{E}\sqrt{\left(2J+1\right)\left(2J'+1\right)}\left(-1\right)^{M}\\
\nonumber &\times\left(\begin{array}{ccc} J&1&J'\\ -M&0&M'\end{array}\right)\left(\begin{array}{ccc} J&1&J'\\ 0&0&0\end{array}\right)
\end{eqnarray}
where we use the notation $\left(\dots\right)$ for the Wigner 3-$j$ symbol~\cite{ZareNote}.  Note that the symbol $d$ refers to the permanent dipole moment of a molecule, and is not to be confused with the dipole operator denoted by $\hat{\mathbf{d}}$.  We refer to the basis which simultaneously diagonalizes the Rotational and DC Hamiltonians as the ``dressed basis,'' and we denote the kets that span this basis by $|\mathcal{E};JM\rangle$, where the labels $J$ and $M$ are the zero field values of the corresponding quantum number and the symbol $\mathcal{E}$ is a reminder that these kets are superpositions of field free rotational states and DC field.

The effects of the DC field can be clearly seen by considering the dressed state wavefunctions, energies, and dipole moments to lowest order in perturbation theory in the dimensionless parameter $\beta_{\mathrm{DC}}\equiv d\mathcal{E_{\mathrm{DC}}}/B$,  the ratio of the field energy to the rotational level splitting:
\begin{eqnarray}
 \fl |\mathcal{E};J,M\rangle=|J,M\rangle-\frac{\beta_{\mathrm{DC}}}{2J}\sqrt{\frac{J^2-M^2}{4J^2-1}}|J-1,M\rangle+\frac{\beta_{\mathrm{DC}}}{2\left(J+1\right)}
 \sqrt{\frac{\left(J+1\right)^2-M^2}{4\left(J+1\right)^2-1}}|J+1,M\rangle\,,\\
\fl \Delta E_{JM}^{\left(2\right)}=\frac{d^2\mathcal{E}_{\mathrm{DC}}^2}{2B}\left[\frac{J\left(J+1\right)-3M^2}{J\left(J+1\right)\left(2J-1\right)\left(2J+3\right)}\right]\,,\\
\fl \langle \mathcal{E};JM|\hat{\mathbf{d}}|\mathcal{E}; JM\rangle/d=-\frac{\partial E_{JM}}{\partial \beta_{\mathrm{DC}}}=\beta_{\mathrm{DC}}\frac{3M^2/J\left(J+1\right)-1}{\left(2J-1\right)\left(2J+3\right)}\,,
\end{eqnarray}
where $\Delta E_{JM}^{\left(2\right)}$ is the lowest non-zero shift in the energy.

The DC field mixes states of different $J$, breaking the $\left(2J+1\right)$-fold degeneracy of the rotational Hamiltonian, and so $J$ is no longer a good quantum number.  In the case of a $z$-polarized field, $M$ remains a good quantum number, and a degeneracy persists for all states with the same $\left|M\right|$.  This mixing aligns the molecule with the field, inducing a nonzero dipole moment.  This means of orienting polar molecules, known as ``brute force" orientation, works well for molecules that both have a large dipole moment and can be efficiently rotationally cooled~\cite{RevModPhys.75.543}.  While more effective means of orienting molecules using intense laser fields are known~\cite{PhysRevLett.74.4623}, they complicate the theoretical discussion and the experimental setup, and so we do not consider them here.

\begin{figure}[htbp]
\hspace{-1.2cm} \begin{minipage}[t]{0.49\linewidth}
\subfigure[Scaled induced dipole moments vs.~scaled DC field energy.]{\includegraphics[width=1.3\linewidth]{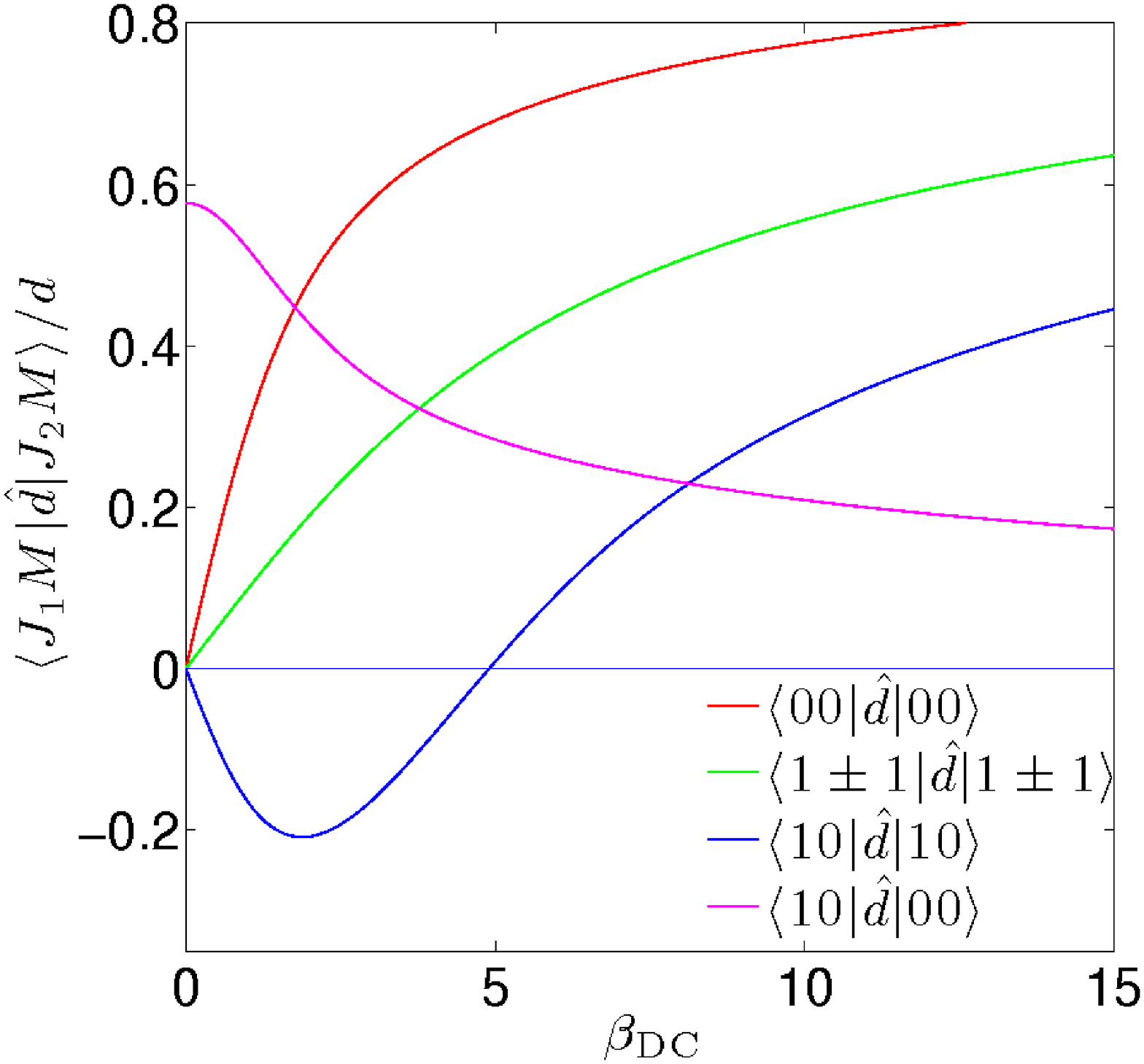}
\label{fig:dipoles_0to10_label}
}
\end{minipage}
\hspace{0.02\linewidth}
 \begin{minipage}[t]{0.49\linewidth}
\subfigure[Scaled dressed energies vs.~scaled DC field energy.]{\includegraphics[width=1.3\linewidth]{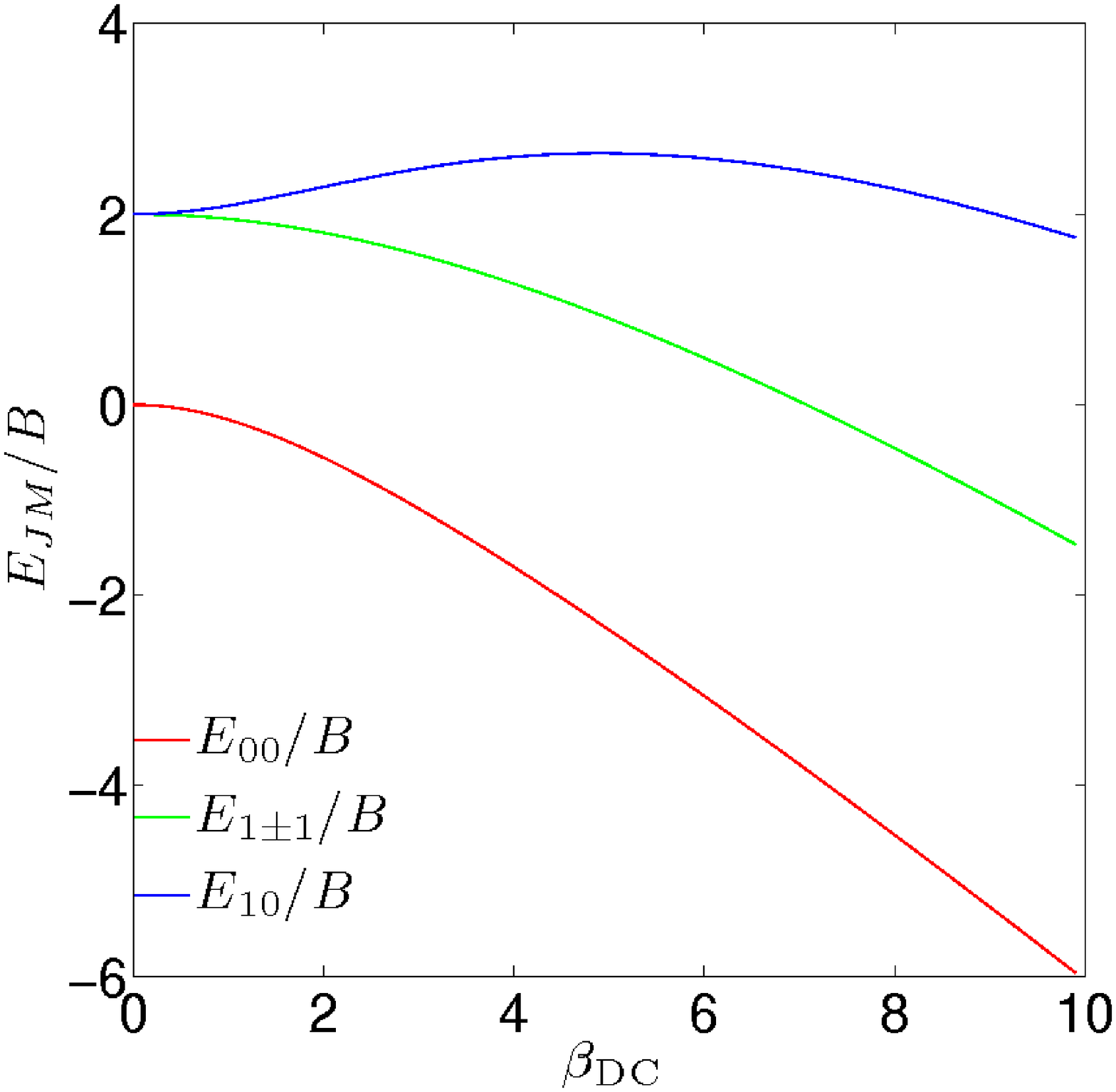}
\label{fig:dressedenergies}
}
\end{minipage}
\caption{Dressed state dipole moments and energies.  Note that the $J=1, M=0$ resonant dipole moment changes from weak-field seeking to high-field seeking at $\beta_{\mathrm{DC}}\approx 5$.      All rotational states have a field where this transition occurs, and the dipole tends monotonically towards unity after this field.  The $\langle 10|\hat{\mathbf{d}}|00\rangle$ dipole moment (and all transition dipole moments, generically) tends towards zero monotonically as $\beta_{\mathrm{DC}}$ increases.  Note also that the energetic differences between rotational levels are smallest at zero field and grow monotonically thereafter.}
    \end{figure}

\begin{figure}[htbp]
\centering
\begin{minipage}[h]{0.27\linewidth}
\subfigure[Composition of 1$^{st}$ dressed state.]{ \includegraphics[width=1.0\linewidth]{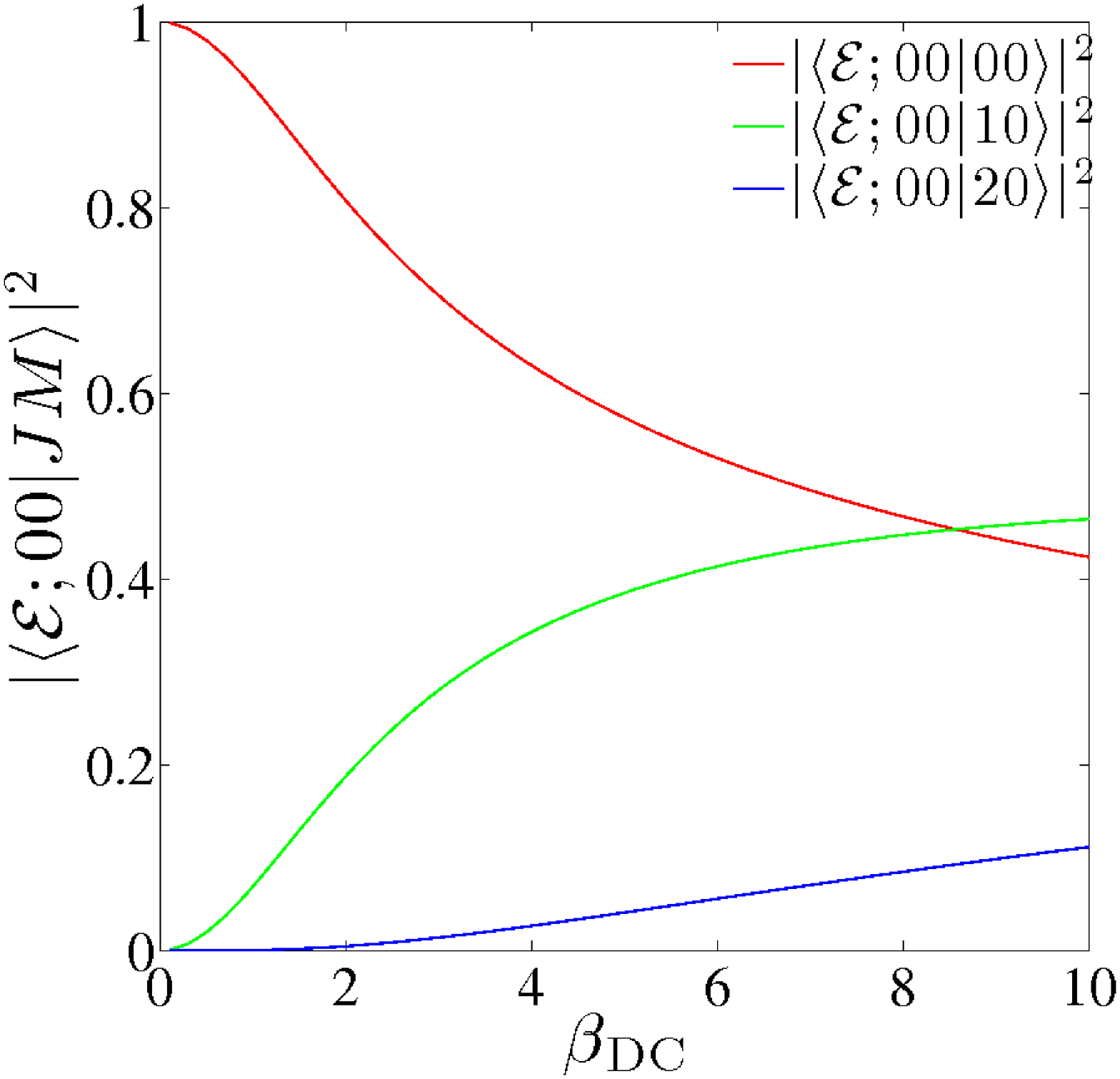}
\label{fig:CompDresStatea}
}
\end{minipage}
\hspace{0.01\linewidth}
\begin{minipage}[h]{0.27\linewidth}
 \subfigure[Composition of 2$^{nd}$ dressed state.]{ \includegraphics[width=1.0\linewidth]{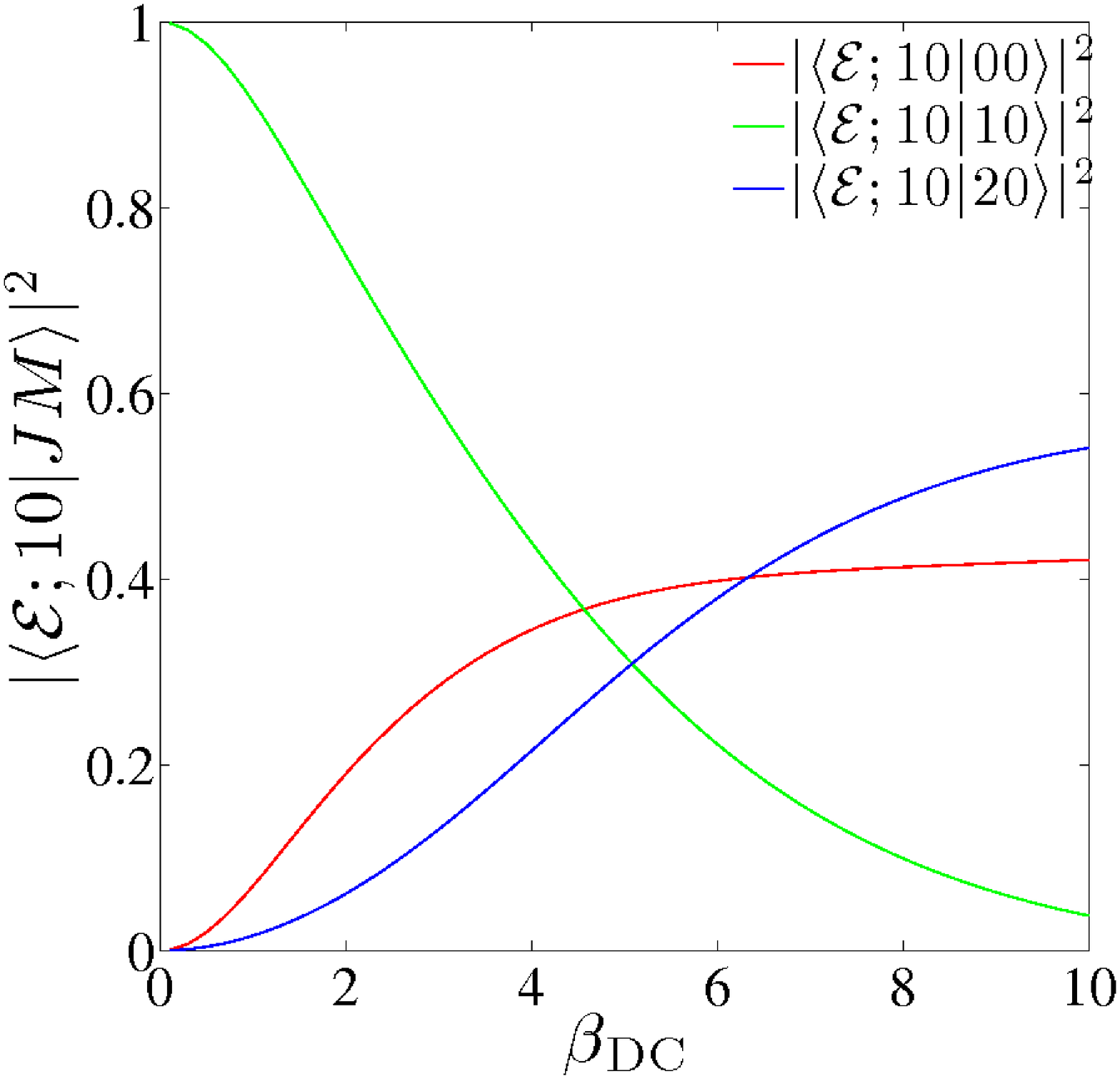}
\label{fig:CompDresStateb}
}
\end{minipage}
\hspace{0.01\linewidth}
\begin{minipage}[h]{0.27\linewidth}
\begin{center}
\subfigure[Composition of $3^{rd}$ dressed state.]{ \includegraphics[width=1.0\linewidth]{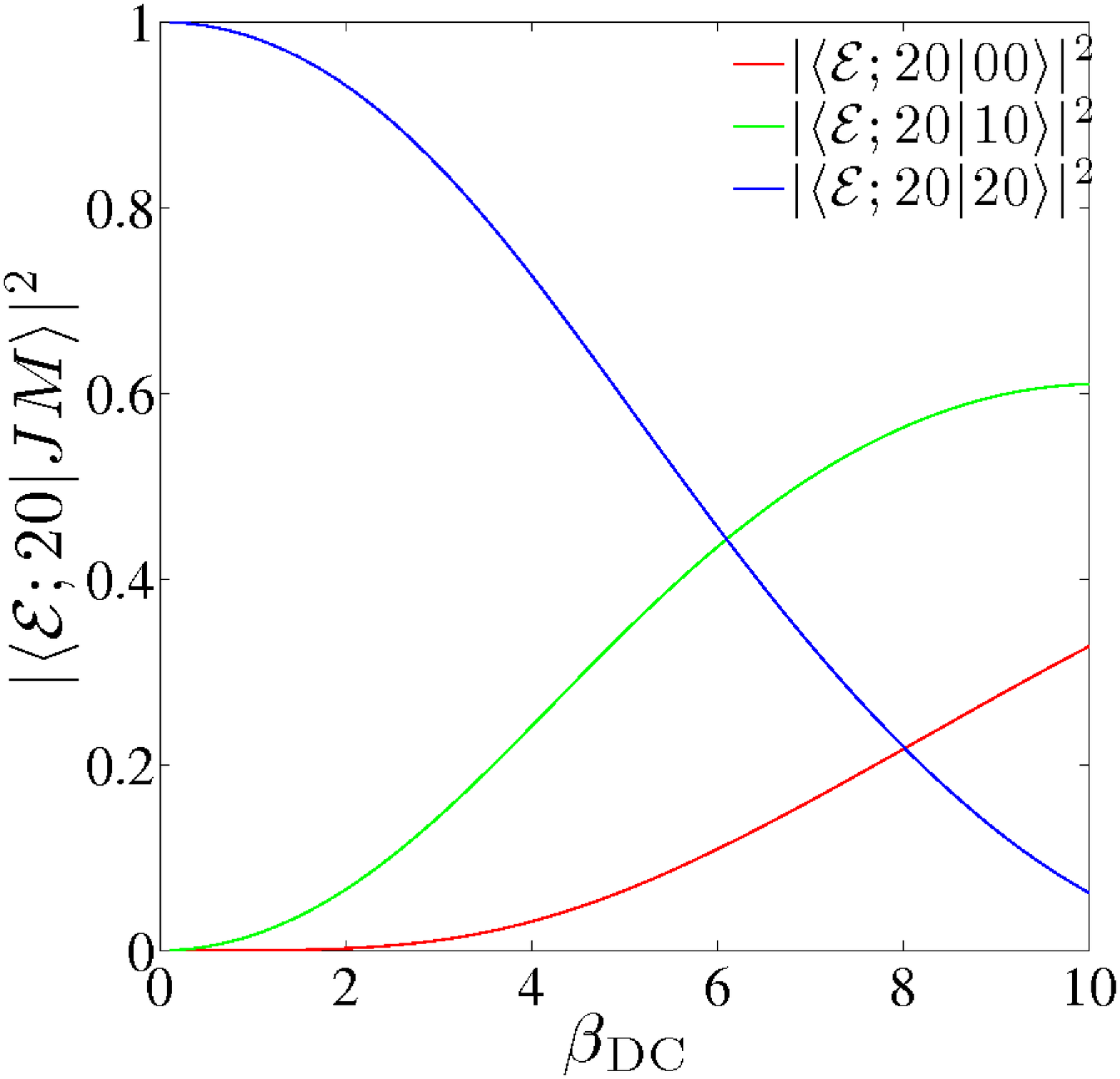}
\label{fig:CompDresStatec}
}
    \end{center}
\end{minipage}
    \caption {Compositions of dressed states vs.~scaled rotational energy.  The states become deeply mixed in large fields, and that the dressed state $|\mathcal{E};JM\rangle$ whose zero field value is $|JM\rangle$ does not always have the greatest overlap with $|JM\rangle$ for all $\beta_{\mathrm{DC}}$.   The field strength where the first dressed state changes from weak-field to high-field seeking, $\beta_{\mathrm{DC}}=5$, is also roughly the place where its overlap with the $|00\rangle$ field-free level is greater than the overlaps with all other field-free levels.}
\label{fig:CompDresState}
    \end{figure}

In larger fields the rotational levels become deeply mixed, which allows states that are weak-field seeking in low fields to become high-field seeking in high fields~\cite{avdeenkov:022707}. The actual mixing of rotational levels vs.~$\beta_{\mathrm{DC}}$ is depicted in Fig.~\ref{fig:CompDresState} for the lowest three dressed levels.  We note that there always exists a field $\mathcal{E}_R$ such that the lowest $R$ dressed states' dipole moments are all positive, as this is important to ensure the stability of a collection of dipoles.  The universal curve of the induced dipole moments (in units of $d$) vs.~$\beta_{\mathrm{DC}}$ of the first two dressed rotational manifolds are shown in Figure~\ref{fig:dipoles_0to10_label}.  The universal curve of the dressed state energies energies (in units of $B$) vs.~$\beta_{\mathrm{DC}}$ is shown in Figure~\ref{fig:dressedenergies}.  For reference, $\beta_{\mathrm{DC}}=1$ corresponds to a field of roughly 1.93$\frac{\mbox{kV}}{\mbox{cm}}$ for $B\sim 60\hbar$ GHz and $d\sim 9$ D.

Expanding the field operators in Eq.~(\ref{eqn:discreteHamiltonian}) in a Wannier basis of dressed states centered at a particular discrete position $\mathbf{r}_i$ as described in Eq.~(\ref{Wannierbasis}), we find
\begin{eqnarray}
\hat{H}_{\mathrm{rot}}+\hat{H}_{\mathrm{DC}}&=\sum_{J}\sum_{M=-J}^{J}E_{J,M}\hat{n}_{\mathcal{E},JM}\,,
\end{eqnarray}
where $E_{JM}$ is the energy of the $|\mathcal{E};J,M\rangle$ dressed state (see Fig.~\ref{fig:CompDresStatea}) and $\hat{n}_{\mathcal{E},JM}$ is the number operator associated with this same state.

If the DC field were aligned at a small angle $\theta_a$ to the $z$ field of the trap (say, in the $xz$ plane), then small dipole moments mixing $M'=M\pm1$ states would arise and the $M'=M$ dipoles would decrease slightly (we can view them as being in an effective field of $\mathcal{E}_{\mathrm{eff}}=\cos\theta_a \mathcal{E}_{\mathrm{DC}}$).  Treating the new contribution perturbatively in the small parameter $\sin\theta_a \beta_{\mathrm{DC}}$, we find the lowest order couplings to the ground state
\begin{eqnarray}
\langle \mathcal{E};00|\hat{H}_{\mathrm{DC}}|\mathcal{E};1\pm 1\rangle\simeq\frac{\sin\theta_ad\mathcal{E}}{\sqrt{6}}\left(1-\frac{49 \sin^2\theta_a}{1440}\beta_{\mathrm{DC}}^2\right)\,,
\end{eqnarray}
and associated timescale $\tau_{\theta_a}$ for occupation of $M\ne 0$ states from the ground state, \begin{eqnarray}
\label{mixing}\tau_{\theta_a}=\frac{\sqrt{6}\hbar}{\sin\theta_a d\mathcal{E}\left(1-\frac{49 \sin^2\theta_a}{1440}\beta_{\mathrm{DC}}^2\right)}\sim \frac{\sqrt{6}}{\beta_{\mathrm{DC}}\sin \theta_a}\frac{\hbar}{B}\,.
\end{eqnarray}

\subsection*{AC Field Term}
\label{ssec:ac}

An AC microwave field of frequency $\omega$ resonantly drives transitions between two DC dressed states $|\mathcal{E};J'M'\rangle$ and $|\mathcal{E};JM\rangle$ with energy difference $\left(E_{J'M'}-E_{JM}\right)/\hbar\approx \omega$ provided the induced dipole moment $\langle \mathcal{E};J'M'|\hat{\mathbf{d}}|\mathcal{E};JM\rangle$ is nonzero.  Two states separated by an energy difference $\Delta E$ that is off-resonant from the driving field (i.e. $\Delta E \gg \omega$) will also be coupled, albeit much more weakly. In our system we resonantly couple the lowest two dressed rotational levels, $|\mathcal{E};10\rangle$ and $|\mathcal{E};00\rangle$.  We consider the case of $z$ polarization, in which the effective Hamiltonian in the dressed Wannier basis is
\begin{eqnarray}
\hat{H}_{AC}\left(t\right)&=-\pi\sin\left(\omega t\right)\sum_{JM}\Omega_{JM}\left(\hat{a}_{\mathcal{E};J,M}^{\dagger}\hat{a}_{\mathcal{E};J+1,M}+\mbox{h.c}\right)\,,
\end{eqnarray}
where
\begin{eqnarray}
\Omega_{JM}&\equiv \mathcal{E}_{\mathrm{AC}}\langle \mathcal{E};J,M|\hat{\mathbf{d}}|\mathcal{E};J+1,M\rangle/\hbar \,.
\end{eqnarray}
is the Rabi frequency.  This is the frequency with which the populations of a two-level system cycles. In experiments, the AC field has spatial curvature on the order of cm which is negligible on the $\mu$m system size scale.

\begin{figure}[htbp]

\begin{minipage}[t]{0.49\linewidth}
\centering
 \subfigure[Populations of the dressed states vs.~rotational time. The small amplitude rapid oscillations occur on the time scale $1/\omega$, and are often averaged away via the rotating wave approximation.  The large amplitude oscillations occuring on the time scale $1/\Omega_{00}$ that periodically transfer the population between $|\mathcal{E};00\rangle$ and $|\mathcal{E};10\rangle$ are the characteristic ``Rabi oscillations" of a driven two-level system.]{
 \includegraphics[width=1.0\linewidth]{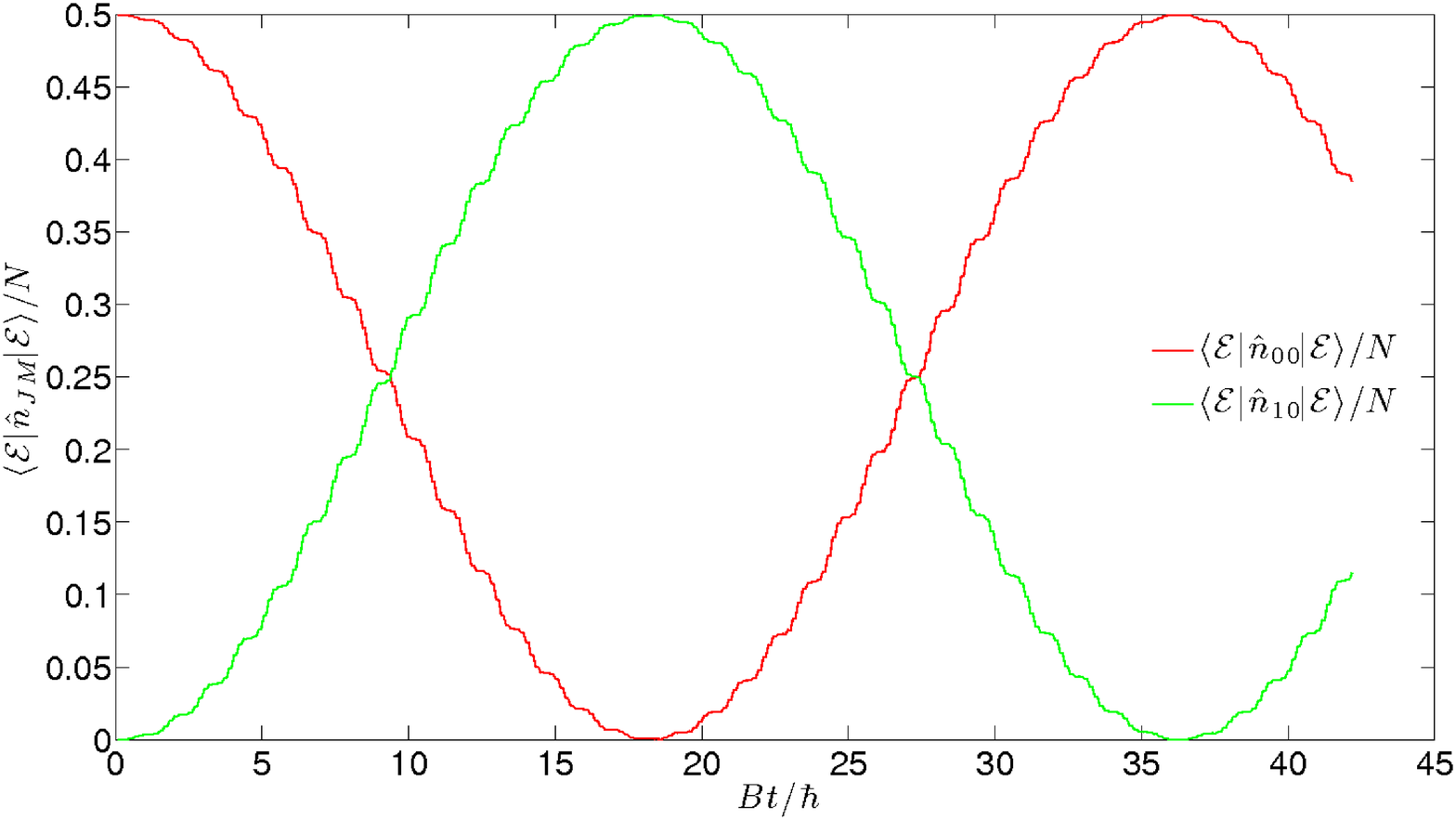}
\label{RabiFlopp}
}
\end{minipage}
\hspace{0.02\linewidth}
\begin{minipage}[t]{0.49\linewidth}
\centering
\subfigure[Populations of the field-free states vs.~rotational time.  The $|20\rangle$ state is occupied because both $|\mathcal{E};00\rangle$ and $|\mathcal{E};10\rangle$ have a nonzero projection with this state due to the mixing from the DC field, see Fig.~\ref{fig:CompDresState}.  It is apparent from comparison with Fig.~\ref{RabiFlopp} that the dressed basis greatly simplifies the AC term in the Hamiltonian.]{\includegraphics[width=1.0\linewidth]{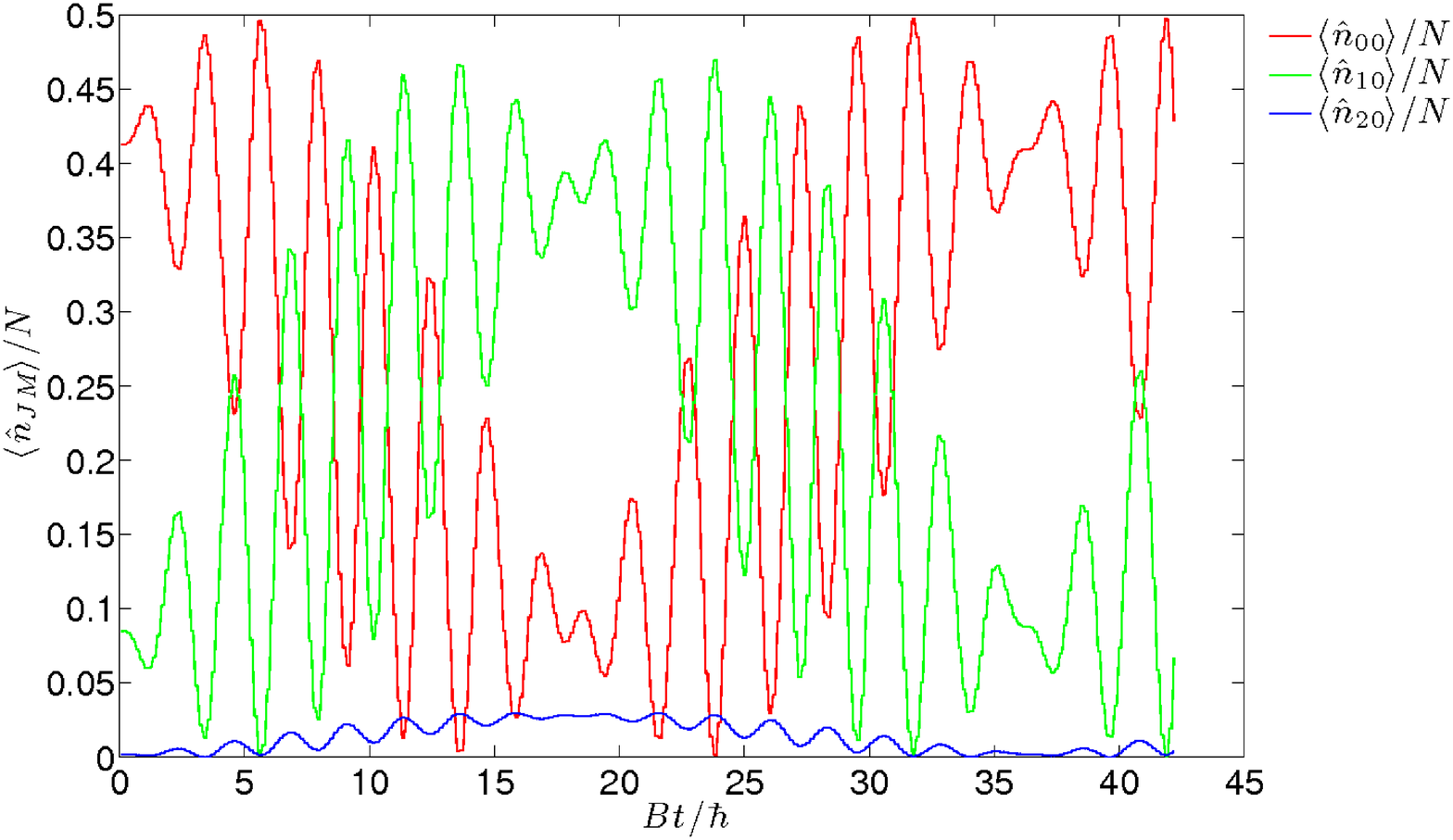}
\label{RabiFloppUndress}
}\end{minipage}
\caption{Resonant AC field induced population cycling in the dressed and field-free bases.}
    \end{figure}

In the absence of couplings between sites, the physics of the system is determined by the on-site, single-molecule physics.  The percentage population of each component in both the $|\mathcal{E};J,M\rangle$ dressed and $|JM\rangle$ field-free bases are shown below for one Rabi period.  In these plots only the $|\mathcal{E};10\rangle $ and $|\mathcal{E};00\rangle$ dressed states are considered, which is close to the actual behavior when all other states are far off-resonant.  Each site undergoes Rabi flopping independently of the others.  Figs.~\ref{RabiFlopp} and \ref{RabiFloppUndress} show this behavior for $\beta_{\mathrm{DC}}=1.900$ and $\beta_{\mathrm{AC}}\equiv{d\mathcal{E}_{\mathrm{AC}}}/{B}=0.200$, giving a Rabi period of $2\pi/\Omega_{00} = 36.5 \hbar/B$.

\section{Convergence}
\label{sec:convergence}

\subsection*{Single Molecule Considerations}

Each dressed state $|\mathcal{E};J,M\rangle$ is, in principle, an infinite linear combination of field free states
\begin{eqnarray}
\label{dressedstatedef}|\mathcal{E};J,M\rangle&=\sum_{J'=0}^{\infty}c_{J'}|J',M\rangle.
\end{eqnarray}
Numerically, we must have a finite upper bound to the sum in Eq.~(\ref{dressedstatedef}), which we call $J_{\mathrm{cut}}$.  This does not cause difficulty in practice, as the overlap of a dressed state $|\mathcal{E};JM\rangle$ with a field-free state $|J'M\rangle$ diminishes rapidly as $J'$ differs more greatly from $J$.  We find the coefficients in Eq.~(\ref{dressedstatedef}), as well as the dressed state energies and dipole moments by simultaneously diagonalizing the rotational and DC field Hamiltonians in a basis consisting of the first $J_{\mathrm{cut}}$ rotational levels.  Because TEBD scales poorly with the on-site dimension, we form as small an on-site basis as possible by keeping the eigenvectors corresponding to the $R$ lowest dressed levels.  To form a proper basis, we must renormalize these eigenvectors (which, for $z$-polarized field, does not change their orthogonality).  We now demonstrate the convergence of these two procedures

\begin{figure}[htbp]
\hspace{-2cm}
\begin{minipage}[t]{0.48\linewidth}
\subfigure{\includegraphics[width=1.4\linewidth]{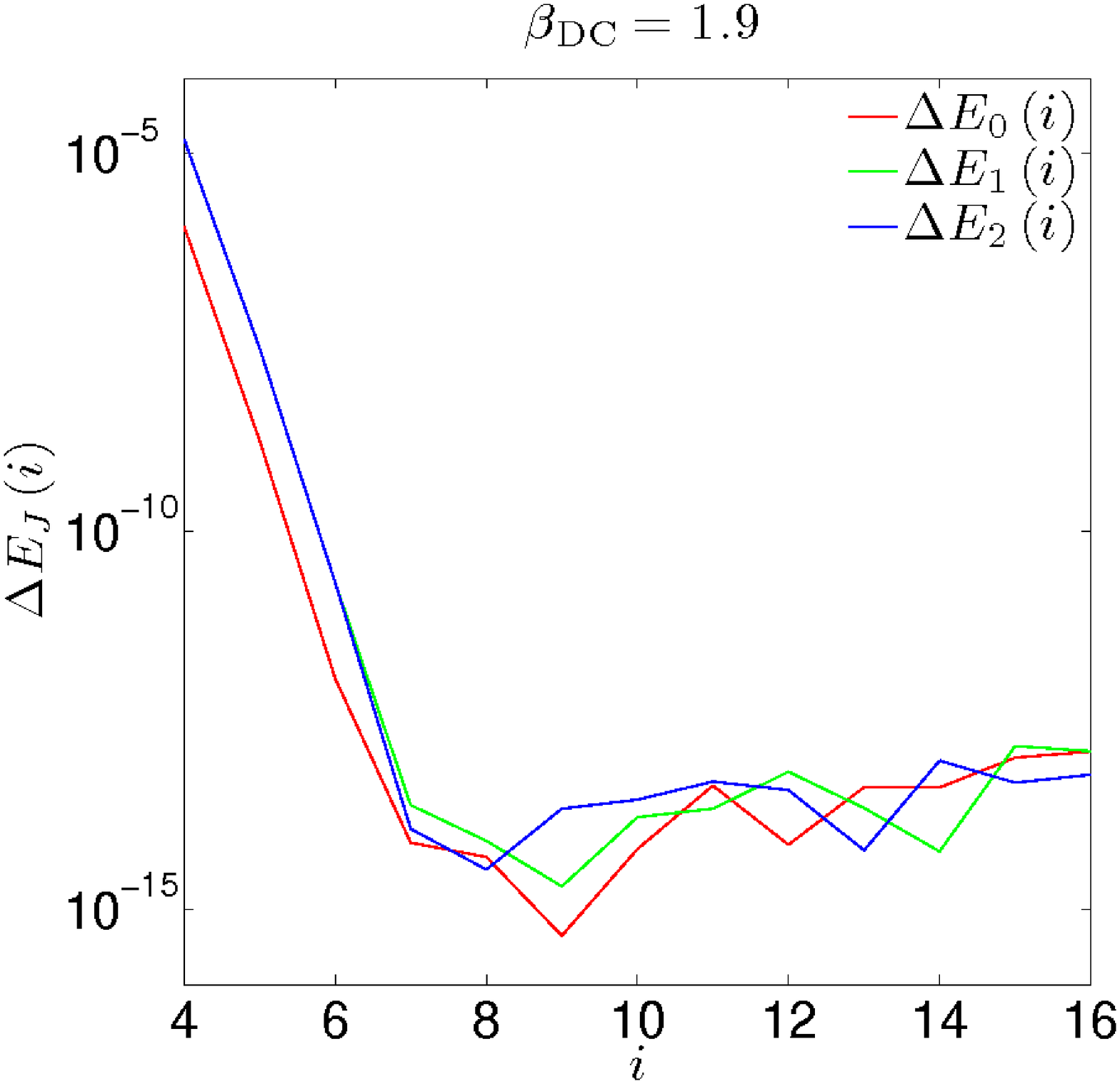}
\label{fig:E1p9}
}
\end{minipage}
\hspace{0.02\linewidth}
\begin{minipage}[t]{0.48\linewidth}
\subfigure{ \includegraphics[width=1.4\linewidth]{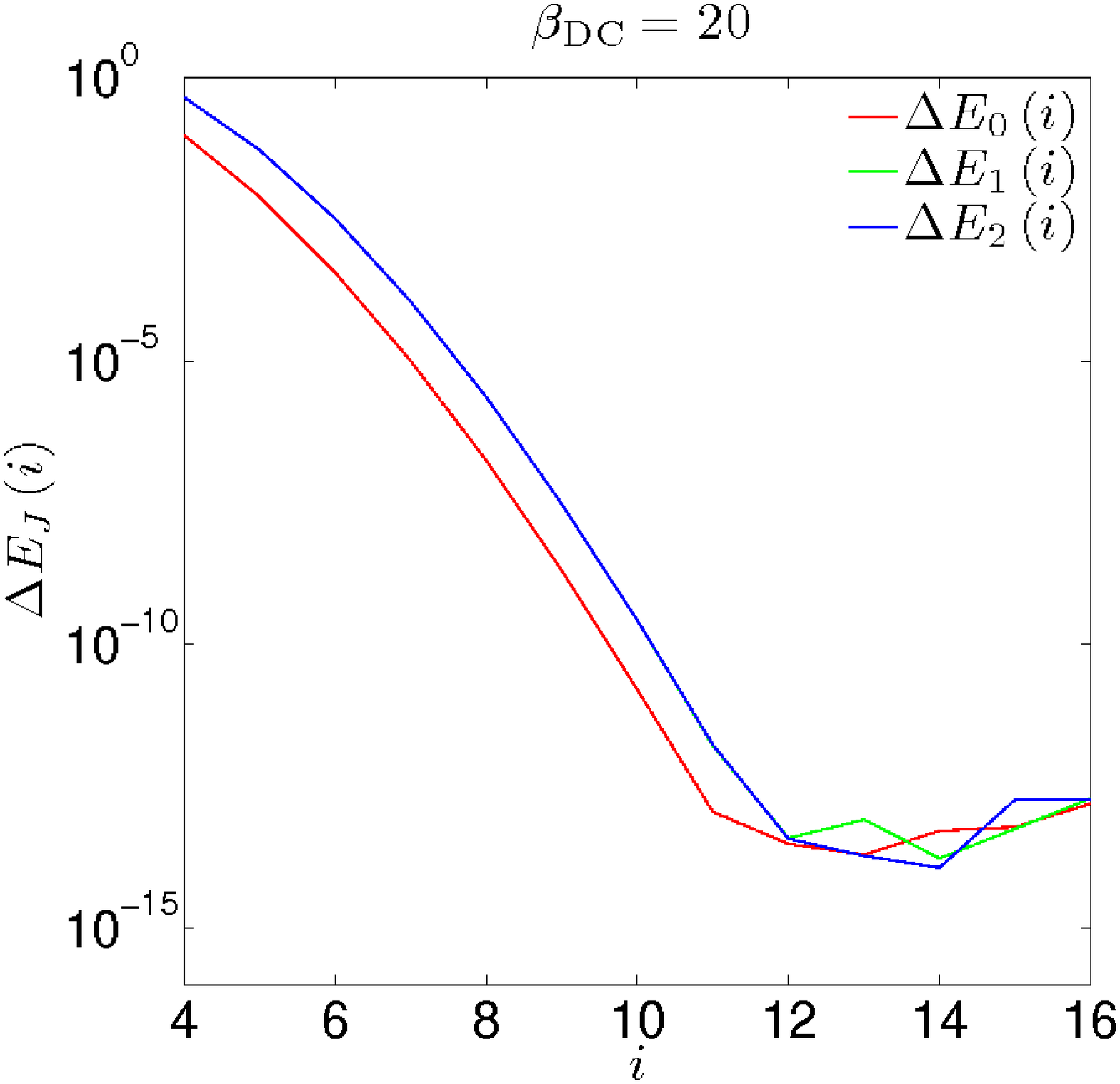}
\label{fig:E20}
}
\end{minipage}
\caption{Convergence with respect to DC dressing rotational state cutoff.  As few as 7 field-free levels are needed for the weak field $\beta_{\mathrm{DC}}=1.9$ to have the dressed state energies of interest converge to machine precision (left panel), and even a large DC field $\beta_{\mathrm{DC}}=20$ requires only 12 field-free levels for the energy to converge (right panel).}
\end{figure}

To show convergence of the first procedure, we plot the difference between the energy of the $J^{th}$ rotational state calculated for a particular value of $J_{\mathrm{cut}}=i$ and one higher value, $\Delta E_J\left(i\right)$ as a function of $i$.  The results for various field strengths are shown in Figures~\ref{fig:E1p9}-\ref{fig:E20}.  We see very fast convergence for the low fields (e.g. $ \beta_{\mathrm{DC}}=1.9$) of interest.  In our numerics we use $J_{\mathrm{cut}}=25$, which ensures convergence for any of the $ \beta_{\mathrm{DC}}$ considered.

To determine convergence with respect to the second procedure, examine Figs.~\ref{fig:CompDresStatea}-\ref{fig:CompDresStateb}, which show \begin{eqnarray}
P_J^{\left(R\right)}&\equiv 1-\sum_{i=0}^{R-1}\left|\langle\mathcal{E};J0|i0\rangle\right|^2\,,
\end{eqnarray}
the amount of the total dressed wave function norm $\left|\langle\mathcal{E};J0|\mathcal{E};J0\rangle\right|^2$ that lies outside of the first $R$ field-free rotational levels for $R=3$ and $R=4$, respectively.  For $R=4$ the renormalization of the first three rotational levels is a very small effect for the $ \beta_{\mathrm{DC}}$ we consider, and the fourth level is not populated to any appreciable extent during time evolution for any $ \beta_{\mathrm{DC}}$ (see Fig.\ref{RabiFlopp}), so we expect that keeping the $R=4$ lowest levels will give sufficient accuracy.  By direct simulation, we find six digit accuracy in the suite of quantum measures defined in Sec.~\ref{ssec:measures}; specifically, we compare $R=3$ to $R=4$.
\begin{figure}[htbp]
\hspace{-2cm}
\begin{minipage}[t]{0.48\linewidth}
\subfigure{  \includegraphics[width=1.4\linewidth]{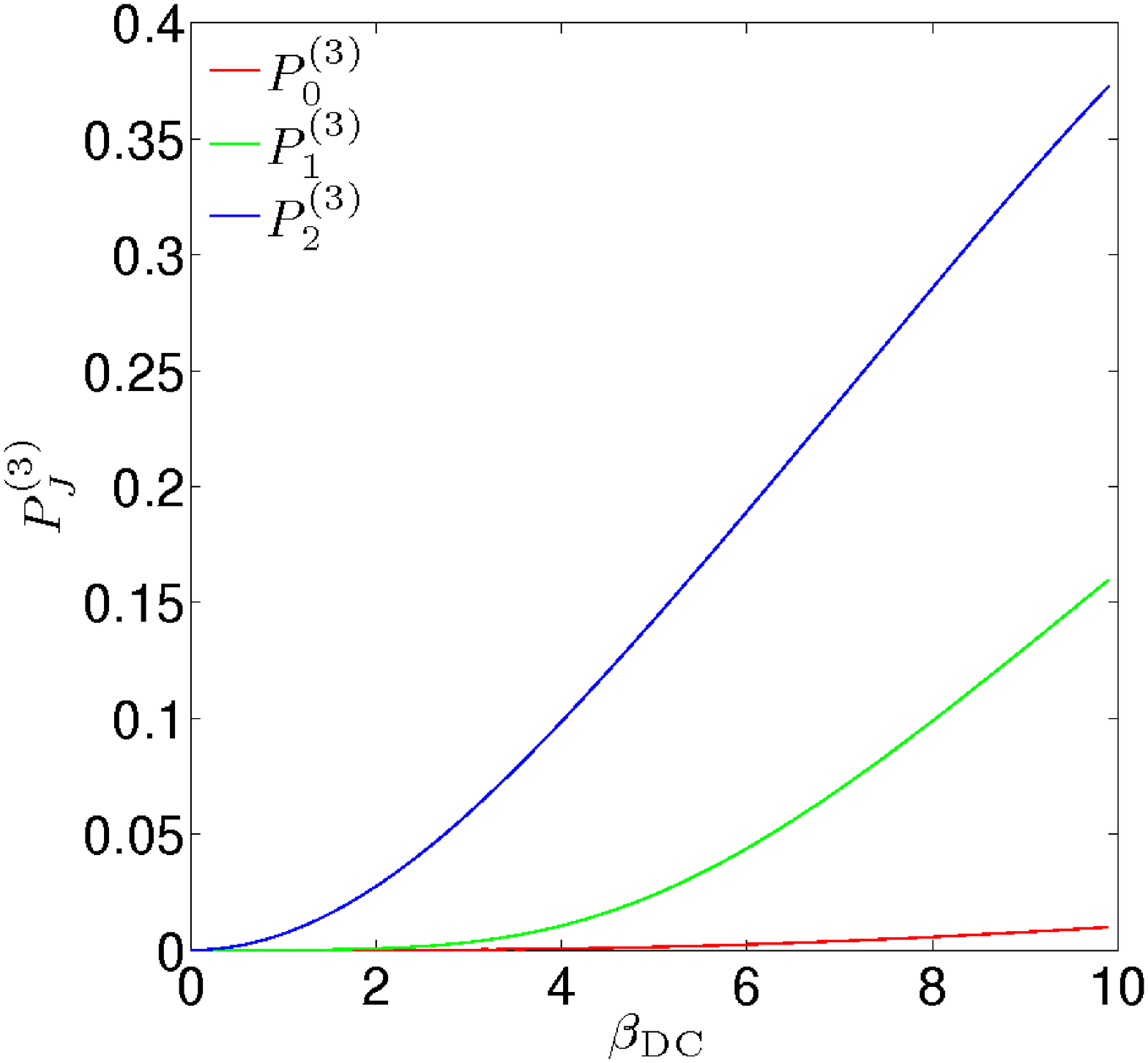}
\label{fig:CompDresStateda}
}
\end{minipage}
\hspace{0.02\linewidth}
\begin{minipage}[t]{0.48\linewidth}
\subfigure{  \includegraphics[width=1.4\linewidth]{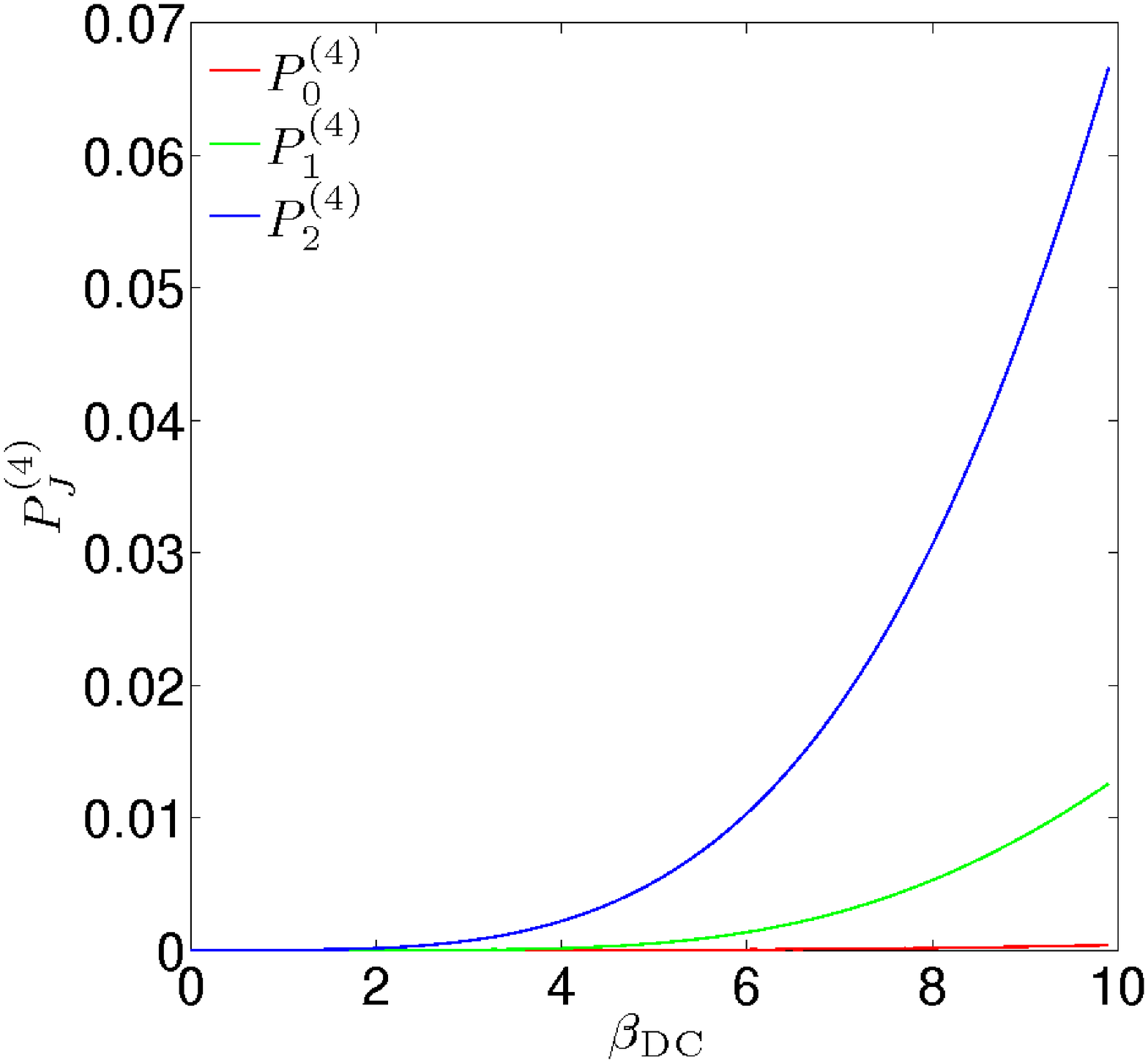}
\label{fig:CompDresStatedb}
}
\end{minipage}
    \caption {Convergence with respect to local dimension cutoff.  Dressed states with greater $J$ lose more of their norm in truncation, as mixing occurs most strongly with adjacent $J$.  Also, as the field is increased, the states become more deeply mixed, and so all states lose more of their norm.  Truncating the local basis at the $J=3$ dressed level incurs at most a 1\% loss of norm for any of the states that are appreciably populated during time evolution (right panel).}
\end{figure}

\subsection*{Many Body Considerations}

\begin{figure}[htbp]
\begin{center}
\fl
\hspace{-2.25cm}\begin{minipage}[c]{0.43\linewidth}
\subfigure{\epsfig{figure=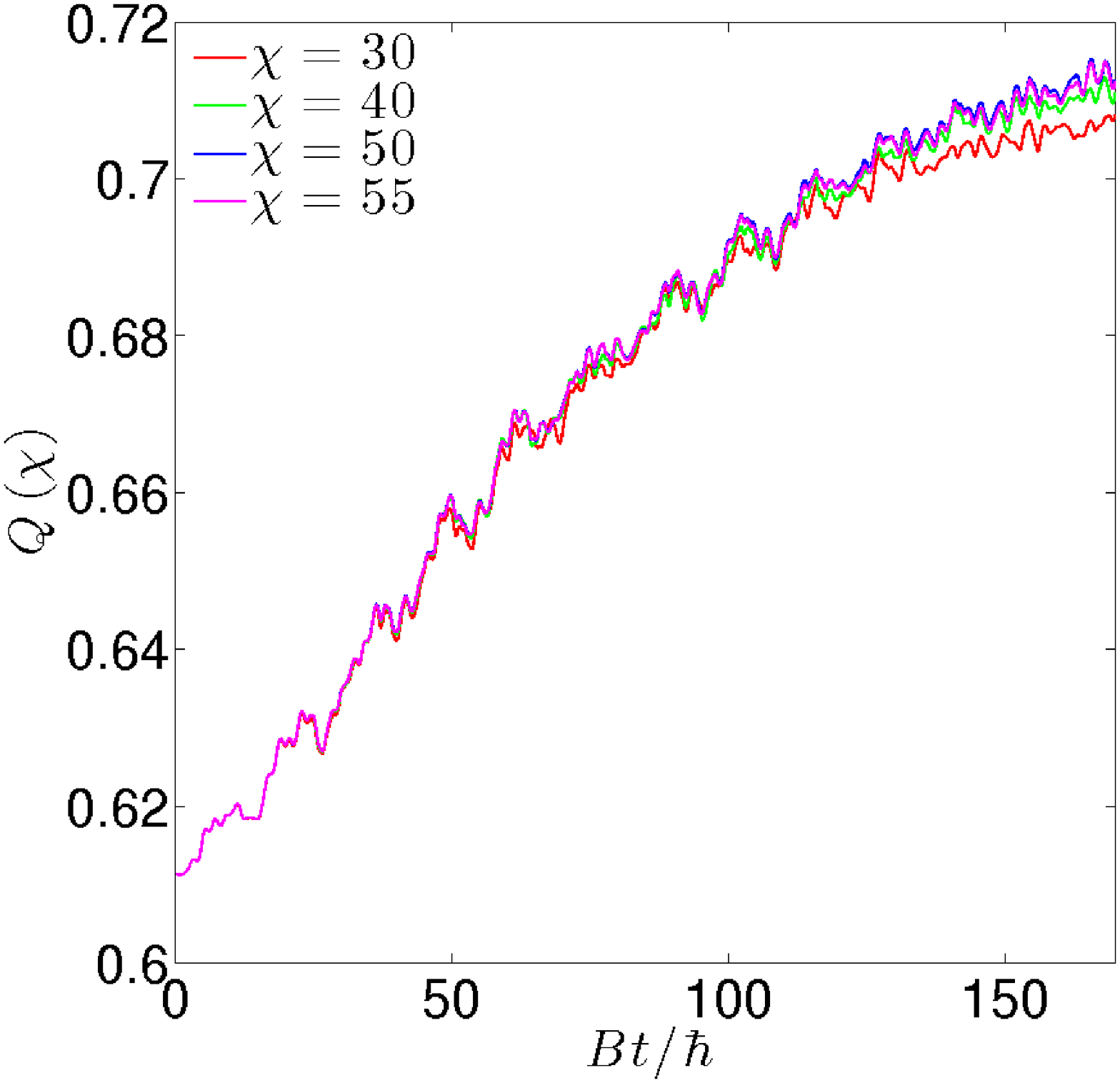, scale=0.32}}
\end{minipage}
\hspace{0.12\linewidth}
\begin{minipage}[c]{0.43\linewidth}
\subfigure{\epsfig{figure=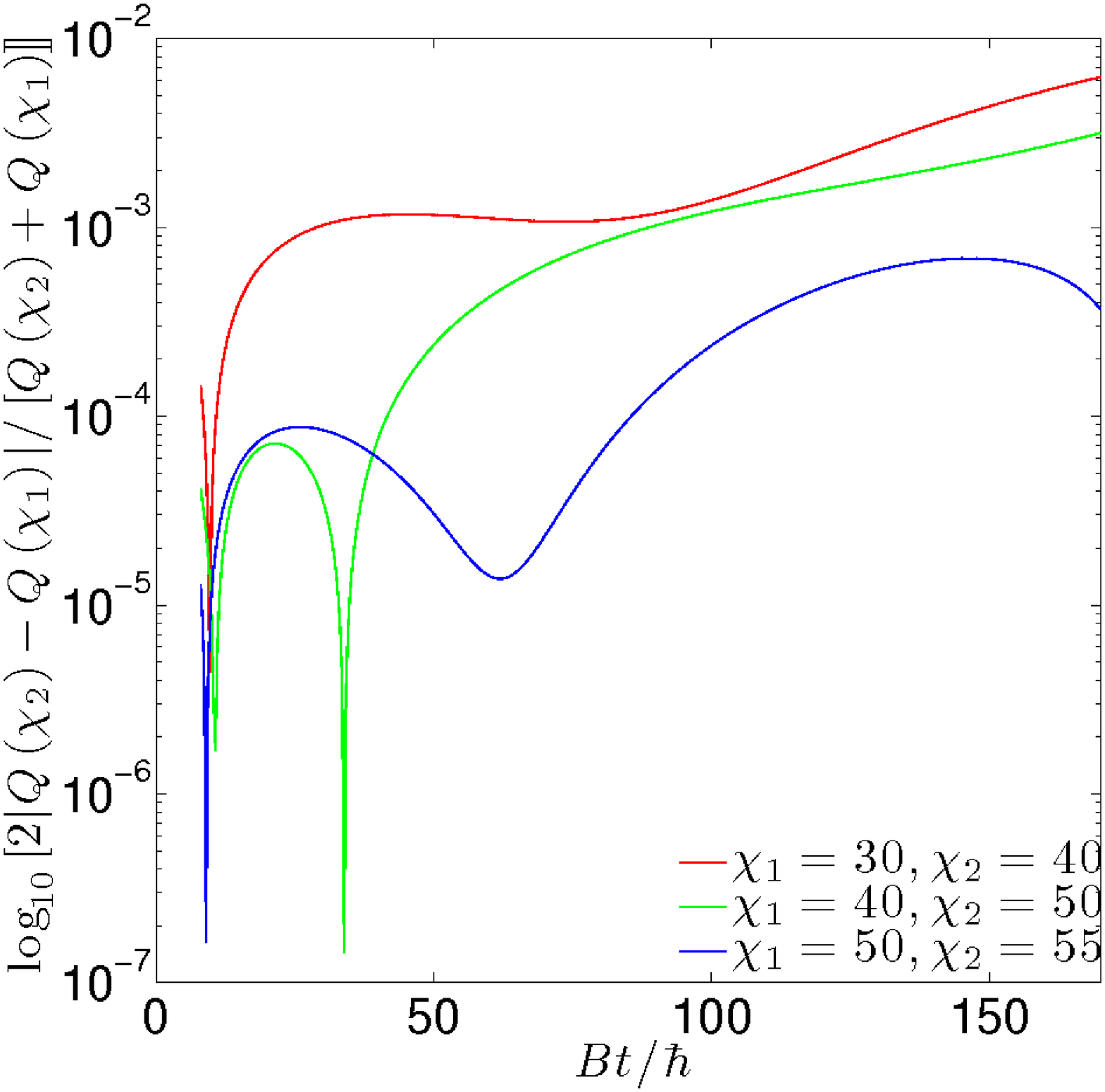,scale=0.32}}
\end{minipage}
\caption{Convergence with respect to entanglement cutoff parameter.  The left figure shows the spatial entanglement measure $Q$ for various values of the TEBD entanglement cutoff parameter $\chi$.  As $\chi$ is increased, $Q$ remains close to its true value for longer.  In the right figure we plot the log of the absolute difference in $Q$ for two values of $\chi$ divided by its arithmetic mean.  We see at least four-digit accuracy for the largest values of $\chi$ we consider.  Note also that even small values of $\chi$ are accurate for short times.}
\label{fig: Chiconv}
    \end{center}
    \end{figure}

There are also convergence issues that are inherent to the TEBD algorithm.  The first, called the \emph{Schmidt error}, is the error that arises from truncating the Hilbert space at each time step.  We can parameterize the error per step in terms of the entanglement cutoff parameter $\chi$ as
\begin{eqnarray}
\tau_l^{S}&=1-\sum_{\alpha_l=1}^{\chi}\left(\lambda_{\alpha_l}^{\left[l\right]}\right)^2
\end{eqnarray}
where $\lambda^{\left[l\right]}$ is a vector containing the eigenvalues of the reduced density matrix obtained by tracing over all sites but $l$, and $\alpha_{l}$ is the local index that entangles the site $l$ with the rest of the system, with smaller $\alpha_l$ states having greater weight.  We find that, among the measures we use, the one that is the most sensitive to $\chi$ is the $Q$-measure, which we plot for four values of $\chi$ in Fig.~\ref{fig: Chiconv}.  Increasing $\chi$ improves the accuracy over longer times, but there is always a time after which the measure begins to deviate.  This is the normalization drift alluded to in Sec.~\ref{ssec:tebd}.  The $\chi$-dependent time after which the Schmidt error dominates is referred to as the runaway time~\cite{gobert:036102}.  In the case study of Sec.~\ref{sec:results}, we used $\chi=50$ for all simulations, which gives the $Q$-measure accurately to within four decimal places over the time scales considered.

 The second intrinsic source of error in TEBD is due to the Trotter-Suzuki expansion of the propagator~\cite{1990PhLA..146..319S}.  We parameterize this error in terms of $\delta t$, the time step.  When we halve the time step from that used in the simulations above ($={2\pi}/{\left(133 \omega\right)}$), we find no change in the measures to the ninth digit.  It is clear that the Schmidt error discussed above is the chief source of error in our simulations.

To extract the emergent time scales defined in Eqs.~(\ref{fitN0}) and (\ref{fitQ}), we used two different methods.  The first is the nonlinear curve fitting routine ``fit" in gnuplot.  The second is the ``NonlinearRegression'' package in Mathematica 6.0.  Both methods use nonlinear regression, which fits the data to a specified nonlinear function of the model parameters.  The goodness of the fit is quantified by the asymptotic standard errors of the model parameters, which gives the standard deviation of each parameter.  A low percent asymptotic error means that the model parameters cannot be adjusted very far without noticeably changing the goodness-of-fit.  Both gnuplot and Mathematica returned the same values for the emergent time scales to within the stated asymptotic standard error.

\newpage

\begin{thebibliography}{10}
\bibitem{ZareNote}
We use the conventions of Zare, op. cit.
\end{thebibliography}

\end{document}